\documentclass{aa}  

\usepackage{graphicx}
\usepackage{txfonts}
\usepackage{natbib}
\bibpunct{(}{)}{;}{a}{}{,} 
\usepackage{graphicx}   
\usepackage{amsmath}    
\usepackage{epsfig}
\usepackage{animate}
\usepackage{xcolor}

\newcommand{\hii}{H\,{\sc ii}}

\begin{document}

   \title{Correlation between the gas-phase metallicity and ionization parameter in extragalactic \hii\ regions}


   \author{Xihan Ji
          \inst{1}\fnmsep\thanks{E-mail: xji243@uky.edu}
          \and
          Renbin Yan\inst{2,1}\fnmsep\thanks{E-mail: rbyan@cuhk.edu.hk}
          }

   \institute{Department of Physics and Astronomy, University of Kentucky, 505 Rose Street, Lexington, KY 40506, USA\          
   \and
    Department of Physics, The Chinese University of Hong Kong, Shatin, N.T., Hong Kong S.A.R., People's Republic of China\\
             }

   \date{Received September 27, 2021; accepted January 29, 2022}

 
  \abstract{The variations of the metallicity and ionization parameter in \hii\ regions are usually thought to be the dominant factors that produce the variations we see in the observed emission line spectra. There is an increasing amount of evidence that these two quantities are physically correlated, although the exact form of this correlation is debatable in the literature. Simulated emission line spectra from photoionized clouds provide important clues about the physical conditions of \hii\ regions and are frequently used for deriving metallicities and ionization parameters. Through a systematic investigation on the assumptions and methodology used in applying photoionization models, we find that the derived correlation has a strong dependence on the choice of model parameters. On the one hand, models that give consistent predictions over multiple emission-line ratios yield a positive correlation between  the metallicity and ionization parameter for the general population of \hii\ regions or star-forming (SF) galaxies. On the other hand, models that are inconsistent with the data locus in high-dimensional line ratio space yield discrepant correlations when different subsets of line ratios are used in the derivation. The correlation between the metallicity and ionization parameter has a secondary dependence on the surface density of the star formation rate (SFR), with the higher SFR regions showing a higher ionization parameter but weaker correlations. The existence of the positive correlation contradicts the analytical wind-driven bubble model for \hii\ regions. We explore assumptions in both dynamical models and photoionization models, and conclude that there is a potential bias associated with the geometry. However, this is still insufficient to explain the correlation. Mechanisms that suppress the dynamical influence of stellar winds in realistic \hii\ regions might be the key to solving this puzzle, though more sophisticated combinations of dynamical models and photoionization models to test are required.}


   \maketitle
%

\section{Introduction}

As one of the most widely observed and studied classes of ionized regions in galaxy studies, \hii\ regions are vital for our understanding of galaxy evolution.
By studying the emission line spectra of \hii\ regions, one can learn about their chemical compositions as well as their ionization states, which are related to the past and current evolution statuses of their host galaxies. Various methods have been developed to derive the properties of \hii\ regions based on their emission line spectra, among which the so-called strong line method is widely used in the literature. The strong line method makes use of strong optical emission lines that are easy to observe to predict the gas-phase metallicity and ionization parameter. There are in general two ways of applying this method, including empirical calibration and theoretical modeling. The former takes advantage of the observations of nearby H\,{\sc ii} regions that have very high signal-to-noise ratios (S/N), of which the metallicities can be determined by the direct method that uses faint optical auroral lines. The derived values can then be compared with the strong line ratios to obtain an empirical relation, which could later be applied to distant H\,{\sc ii} regions or star-forming (SF) galaxies without measurements of auroral lines \citep[e.g.,][]{pagel1979,pagel1980, edmunds1984, pettini2004, pilyugin2005, marino2013}. Although the empirical method does not rely on any assumption on the physical conditions of H\,{\sc ii} regions, there are unavoidable intrinsic scatters in most of the relations introduced by the variations in parameters other than metallicity (which we refer to as "secondary parameters" hereafter). In addition, the ionization parameter, which is defined as the ratio between the flux of the hydrogen ionizing photons to the hydrogen density, is almost impossible to be measured directly. It cannot be inferred from empirical calibrations and has to depend on certain assumptions on the internal structures of \hii\ regions.

Theoretical modeling, on the other hand, can self-consistently predict both the metallicity and ionization parameter, provided that the secondary model parameters are properly set and match the realistic \hii\ regions \citep[e.g.,][]{charlot2001, kewley2002, 2004ApJ...613..898T, kobulnicky2004, 2011MNRAS.415.3616D}. These parameters include the stellar spectral energy distribution (SED), the chemical abundance pattern, the dust composition, the density structure, the geometry, etc. State-of-the-art photoionization codes such as {\sc cloudy} and {\sc mappings} can be used to compute 1D models of ionized clouds based on existing atomic data and to predict emission-line ratios that are in good agreement with the observations \citep[e.g.,][and references therein]{dopita2000, 2001ApJ...556..121K, dopita2013, 2019ARA&A..57..511K}. Overall, photoionization models have been shown to be able to successfully reproduce several important emission-line ratios in observations. For example, the model predictions on [N\,{\sc ii}]$\lambda 6583$/H$\alpha$, [S\,{\sc ii}]$\lambda \lambda 6716,6731$/H$\alpha$, and [O\,{\sc iii}]$\lambda 5007$/H$\beta$ are largely consistent with observations, which explains the observed SF sequence in standard optical diagnostic diagrams \citep{stasinska2006, dopita2013}.
However, it is important to note that when the model-predicted line ratios are compared with the observations to find the metallicity and ionization parameter, the results have a strong dependence on the input parameters.
Unfortunately, many of the secondary model parameters are hard to constrain, and their importance is often overlooked in practice. This can lead to large discrepancies among the strong-line relations based on different models, as we show in this paper. A prime example of strong model-dependence can be shown by the results on the correlation between the metallicity and ionization parameter.

The metallicity and ionization parameter are often assumed to be the main contributors to the variations of line ratios among SF regions \citep[for a counterargument, however, see][]{2020MNRAS.496..339P}. 
An interesting question is whether they are correlated with each other or independent.
Surprisingly, there are a considerable number of contradictory findings in the literature on this issue. Some early works show the two quantities are clearly anti-correlated \citep[e.g.,][]{1986ApJ...307..431D, 2006ApJ...639..858M, 2006A&A...459...85N}, which is further supported by the theoretical calculation of a wind-driven bubble model for \hii\ regions by \citet{2006ApJ...647..244D} (hereafter D06). D06 argued that there are two effects driving this anticorrelation. First of all, at higher metallicities, stellar winds become more opaque and absorb more ionizing photons. In addition, the stellar atmosphere scatters photons more effectively, leading to stronger winds. This enlarges the inner shocked wind region and dilutes the ionizing flux received by the outer \hii\ region. The anticorrelation has also been confirmed by a number of more recent studies using a relatively large sample of H\,{\sc ii} regions or SF galaxies \citep[e.g.,][]{perezmontero2014, 2016A&A...594A..37M, Thomas19}.

Interestingly, some other studies found that the metallicity and ionization parameter either have no obvious correlation, or have a positive correlation in observed H\,{\sc ii} regions \citep[e.g.,][]{2011MNRAS.415.3616D, dopita2013, dopita2014, poetrodjojo2018, kreckel2019, zinchenko2019, mingozzi2020}, in contrast to the theoretical prediction of D06. \cite{dopita2014} (hereafter D14) propose that a positive correlation between the metallicity and ionization parameter only exists in starburst galaxies and not in normal SF galaxies. They argue that the underlying reason for the positive correlation in starburst galaxies could be the relation between the star-formation rate (SFR) and ionization parameter. A higher SFR could increase the ionization parameter either due to the higher mass of the star cluster or the change of the geometry of the ionized cloud. However, according to \cite{2016ApJ...827...35T} and \cite{mingozzi2020}, such a relation between the SFR and ionization parameter can also be found in normal SF galaxies. In addition, their analyses show that the correlation strongly depends on the stellar masses of the galaxies. Furthermore, \cite{mingozzi2020} found that the ionization parameter is actually more tightly correlated with the specific star-formation rate (sSFR, which is traced by the equivalent width of the H$\alpha$ line in their work) than the SFR. To make the matter even more convoluted, \cite{poetrodjojo2018} compared the slopes of the metallicity versus ionization parameter relations (MI relations) and the SFR versus ionization parameter relations (SFR-I relations) for H\,{\sc ii} regions in different galaxies. They found that the slopes of these two relations are very different for different galaxies and the overall correlations between these quantities are not significant.

The tension between the two contradictory kinds of findings is puzzling and this remains unresolved. \cite{2019ARA&A..57..511K} comment that the evidence against the expected anticorrelation mainly came from spatially resolved studies. However, as we show in this paper, the correlation between the metallicity and ionization parameter in observed H\,{\sc ii} regions or SF galaxies is less related to the scale of the observations, and it is rather a result of model assumptions and methodology. In this work, we reexamine this long-standing issue by focusing on the choice of input parameters for photoionization models that are used to fit the metallicity and ionization parameter of H\,{\sc ii} regions. We provide a self-consistent model that resolves this problem, which could help reveal the potential physical process that leads to the correlation. The layout of the paper is as follows. In Section~\ref{sec:data_models} we describe the observational data we use and the input parameters for the photoionization models. In Section~\ref{sec:results} we compare the model predictions on the correlations under different assumptions and evaluate the models based on the consistency of their predictions on different combinations of emission line ratios. We discuss the physical interpretations for the correlation indicated by our best-fit model in Section~\ref{sec:explanation} and examine the robustness of our analyses in Section~\ref{sec:discuss}. Finally, we summarize our findings and draw our conclusions in Section~\ref{sec:conclude}.

Throughout this work, we use the following abbreviations for some of the frequently mentioned emission line ratios. We denote log([N\,{\sc ii}]$\rm \lambda 6583/H\alpha$), log([S\,{\sc ii}]$\rm \lambda \lambda 6716,6731/H\alpha$), log([O\,{\sc iii}]$\rm \lambda 5007/H\beta$), log([N\,{\sc ii}]$\rm \lambda 6583/$[O\,{\sc ii}]$\lambda \lambda 3726,3729$), and log([O\,{\sc iii}]$\rm \lambda 5007/$[O\,{\sc ii}]$\lambda \lambda 3726,3729$) as N2, S2, R3, N2O2, and O3O2, respectively.

\section{Data and models}
\label{sec:data_models}

\begin{figure*}
    \includegraphics[width=0.44\textwidth]{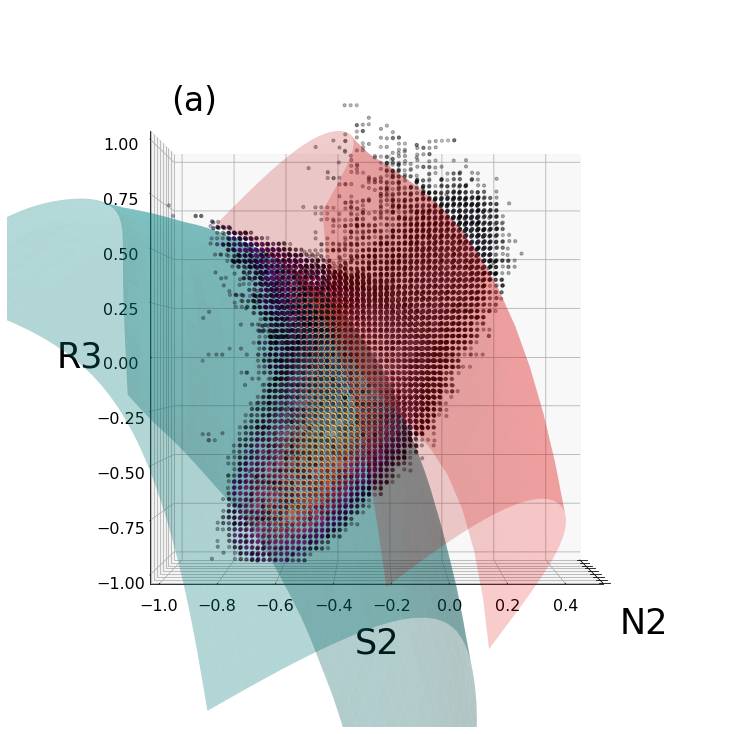}
    \includegraphics[width=0.44\textwidth]{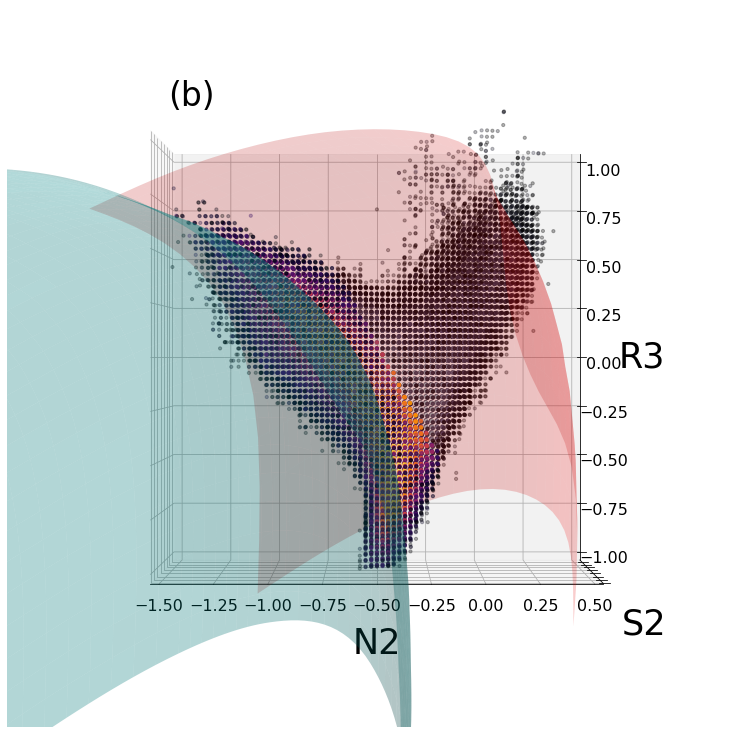}
    \centering
    \includegraphics[width=0.5\textwidth]{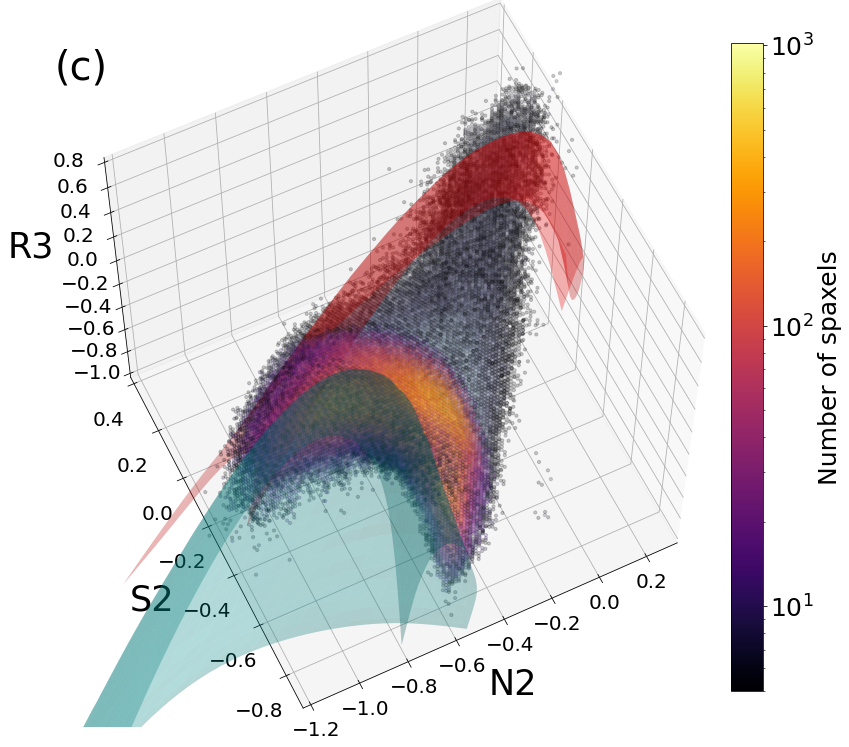}
    \caption{Density distribution of MaNGA MPL-7 spaxels in the 3D space spanned by N2, S2, and R3. Our sample H\,{\sc ii} region spaxels are colored from yellow to purple, while the rest of the spaxels are colored from white to black. Two photoionization model surfaces are shown. The cyan model and the red model are our fiducial SF model and AGN model, respectively. We show three viewing angles that lead to different 2D projections. Specifically, panel (a) and panel (b) correspond to the [S\,{\sc ii}]- and [N\,{\sc ii}]- BPT diagrams. Panel (c) shows that the two model surfaces are separate in 3D and that they are connected by a continuous mixing sequence.
    }
    \label{fig:hii_data}
\end{figure*}

In this work we use observational data from the Mapping Nearby Galaxy at Apache Point Observatory survey \citep[MaNGA,][]{bundy2015}. As one of the three major experiments of SDSS-IV \citep{blanton2017}, MaNGA was designed to obtain spatially resolved spectroscopic data of $\sim 10000$ nearby galaxies and it finished its observation in the summer of 2020. Its observing strategy and survey execution are detailed in \cite{law2015} and \cite{yan2016b}. With a median redshift of 0.03, MaNGA's targets form a primary sample made up of galaxies observed out to 1.5 $R_e$ and a secondary sample made up of galaxies observed out to 2.5 $R_e$. These targets are designed to have a nearly flat stellar mass distribution with $M_* = 10^9 \sim 10^{11} ~ M_\odot$ \citep{wake2017}. MaNGA uses the 2.5m Sloan telescope for its observation \citep{gunn2006}. Lights from galaxies are fed through IFU fiber bundles with fields of view ranging from 12$^{\prime \prime}$ to
32$^{\prime \prime}$ \citep{drory2015} to the BOSS spectrographs \citep{smee2013}. The collected spectra have a median spectral resolution of $R\sim2000$ and cover a wavelength range from 3622\AA ~to 10354\AA. These raw spectra data are then reduced and calibrated by the Data Reduction Pipeline \citep[DRP,][]{law2016, law2021a, yan2016a} and eventually fed to the Data Analysis Pipeline \citep[DAP,][]{belfiore2019, westfall2019}. The DAP uses the {\sc Penalized Pixel-Fitting} software \citep[{\sc ppxf},][]{cappellari2004, cappellari2017} at its core and fits the stellar continua and emission line spectra simultaneously. The final data products include spatially resolved models and measurements of emission lines and stellar continua. A python-based toolkit, MARVIN, further facilitates steamlined access and visulization of the DAP products \citep{cherinka2019}.

\begin{figure}
    \includegraphics[width=0.48\textwidth]{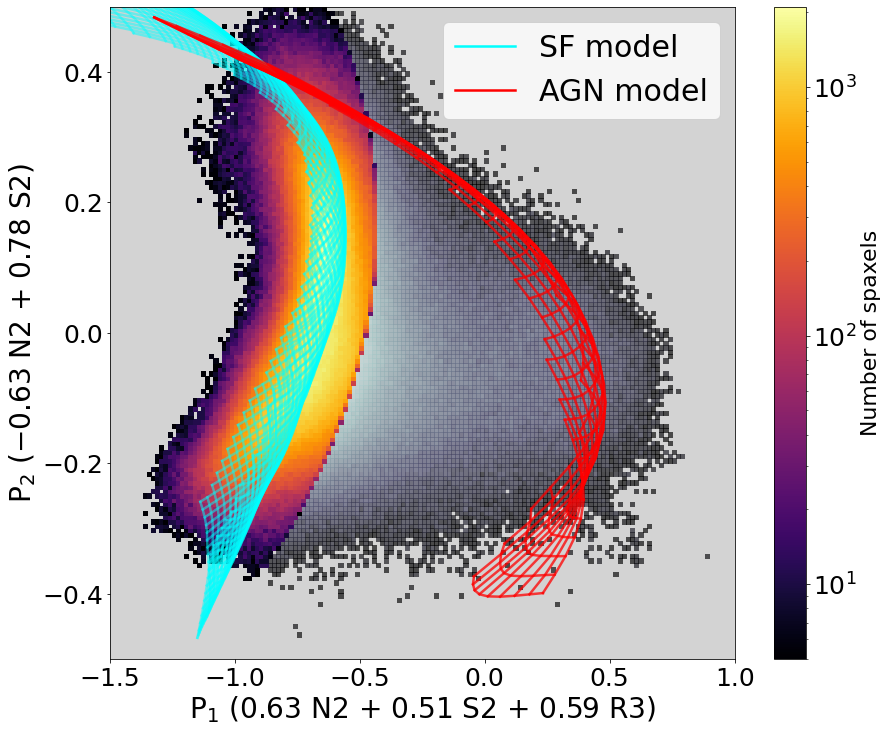}
    \caption{Density distribution of the MaNGA data in the 2D $P_1$-$P_2$ diagram, where the relevant parts of the (interpolated) SF model (cyan grid) and AGN model (red grid) appear edge-on and well separated. The sample \hii\ regions are colored from yellow to purple. 
    Here, we only plotted the parts of the model surfaces that cover the middle 98\% of the data along the $P_3$ axis (line of sight), which is perpendicular to the $P_1$ versus $P_2$ plane.
    This projection corresponds to a line of sight at ($\theta$, $\phi$) = (36$^{\circ}$, 219$^{\circ}$) in the 3D space of (N2, S2, R3), where $\theta$ and $\phi$ are the polar angle and the azimuthal angle, respectively \citep{ji2020b}.
    }
    \label{fig:hii_2d}
\end{figure}

\begin{figure}
    \includegraphics[width=0.48\textwidth]{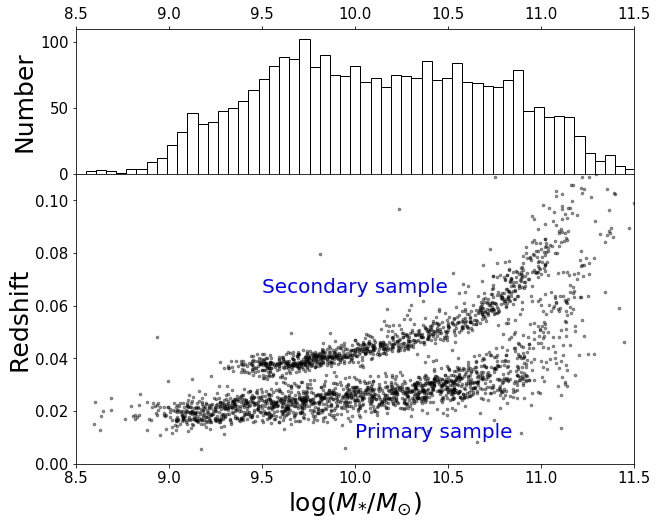}
    \caption{Redshift and stellar mass distributions of our sample galaxies. The histogram shows a relatively flat distribution in stellar mass. The sample galaxies include both primary and secondary MaNGA galaxies. The former cover spatial regions out to 1.5 $R_e$, while the latter cover spatial regions out to 2.5 $R_e$.}
    \label{fig:manga_data}
\end{figure}

Our sample is drawn from the MaNGA products in the $\rm 15^{th}$ public data release of SDSS (DR15), which includes a total of 4639 unique galaxies.
This data set is equivalent to the seventh product launch of MaNGA (MPL-7).
We selected a sample of H\,{\sc ii} regions by using the 3D optical diagnostic diagram introduced by \cite{ji2020b}, which combines the [N\,{\sc ii}]- and [S\,{\sc ii}]- Baldwin, Phillips \& Terlevich (BPT) diagrams \citep{1981PASP...93....5B, 1987ApJS...63..295V} and selects spaxels in a 3D space.
Basically, \cite{ji2020b} used two model surfaces to describe the SF and active galactic nucleus (AGN) loci in 3D space, which are shown in Figure~\ref{fig:hii_data}. The two model surfaces are well separated in 3D, allowing the decomposition of spatially mixed ionized regions into pure ionized regions, of which the line ratios are indicated by the locations of the model surfaces.
With this method, \cite{ji2020b} computed quantitatively a demarcation surface in 3D that is able to isolate SF regions with an AGN contribution of less than 10\% to the total H$\alpha$ fluxes\footnote{An H\,{\sc ii} region sample selected by using only the [N\,{\sc ii}] BPT diagram would not change our conclusion. Our selection criterion includes slightly more H\,{\sc ii} regions in the outskirts of galaxies that would have been missed by the standard BPT diagnostics.}.
The selection function is given by
\begin{equation}
    P_1 < -1.57 P_2^2 + 0.53 P_2 - 0.48,
\end{equation}
where
\begin{equation}
    P_1 = 0.63 N2 + 0.51 S2 + 0.59 R3,
\end{equation}
and
\begin{equation}
    P_2 = -0.63 N2 + 0.78 S2.
\end{equation}
Figure~\ref{fig:hii_data} shows the density distribution of our sample in the 3D space spanned by N2, S2, and R3. 
It is clear that there is a continuous mixing sequence connecting the two model surfaces, and our selected sample is well traced by the fiducial SF model surface.
The sample includes a total of 2782 galaxies and 1.65 million spaxels.
Figure~\ref{fig:hii_2d} shows the data distribution in a specific 2D projection, which we denote as the $P_1$-$P_2$ diagram, where the photoionization model surfaces appear edge-on. Since both the data and the (relevant) model surface cover a very narrow region in this projection, one can use it to visualize how well the model surfaces fit the center of the data distribution.
In addition, the projected AGN model grid is clearly separated from the SF model grid, indicating that AGN or composite regions are unlikely to be projected into the SF locus and contaminate the sample significantly. Although the two models seem to overlap with each other at the upper part of the diagram, the corresponding part of the AGN model actually describes very low metallicity AGNs, which are rare in the MaNGA sample, as indicated by the spaxels in the central regions of MaNGA galaxies \citep{ji2020b}.

Since we have tried to carry out a statistical study on the distribution of both the metallicity and ionization parameter in "typical" H\,{\sc ii} regions for this work, whether the selected sample is representative enough has a non-negligible impact on the result. For our main MaNGA sample, the mass distribution is relatively flat, as can be seen in Figure~\ref{fig:manga_data}. This ensures that the high-mass regime is well populated and helps to constrain photoionization models at high metallicities. 
We defer a more detailed discussion on the selection effect to Section~\ref{sec:discuss}.

Our fiducial photoionization model for H\,{\sc ii} regions (hereafter JY20 model) is generated by the photoionization code {\sc cloudy} \citep[v17.00,][]{ferland2017}, and it is described in detail by \cite{ji2020b}. In short, this model simulates an isobaric H\,{\sc ii} region with plane-parallel geometry. The initial hydrogen density is set to be $\rm 14~cm^{-2}$, which is derived from the median [S\,{\sc ii}]$\lambda 6716$/[S\,{\sc ii}]$\lambda 6731$ of H\,{\sc ii} regions in MaNGA \citep{ji2020a}. The ionizing SED is produced by the code {\sc starburst99} \citep[v7.01,][]{leitherer1999}, assuming a continuous star-formation history (SFH) of 4 Myr and a Kroupa initial mass function \citep[IMF,][]{kroupa2001}. When computing this SED, we used the \cite{pauldrach2001} and \cite{hillier1998} stellar atmospheres, and a standard Geneva evolutionary track. To account for secondary nitrogen, we used the N/O versus 12 + log(O/H) relation given by \cite{dopita2013}, which was derived by fitting a set of H\,{\sc ii}-region measurements derived by \cite{vanzee1998}. For the gas-phase chemical abundances, we chose the solar abundance set of \cite{grevesse2010} as the reference abundance and used the default dust depletion factors in {\sc cloudy}, which are based on measurements by \cite{cowie1986} and \cite{jenkins1987}. For each gas-phase abundance used in the model, we tried to match it with the stellar SED with the same metallicity, assuming the abundance ratio among heavy elements (including $\alpha$-elements but excluding C and N) is the same as the Sun. Since {\sc starburst99} only computes stellar metallicities up to $2~Z_{\odot}$, we extrapolated the stellar SED when the gas-phase oxygen abundance was larger than two times the solar value. Basically, we assumed that the logarithms of the stellar fluxes change linearly with the logarithmic oxygen abundances\footnote{If we do not extrapolate the SED and use the SED with mismatched abundances at high metallicities, the predicted 12+log(O/H) is $\sim 0.01$ to $0.02$ dex higher at this regime.}.

Besides this fiducial model, we also examined three other SF-ionized models in the literature, computed by \cite{levesque2010}, \cite{dopita2013}, and \cite{byler2017}, respectively (we denote these models as L10, D13, and B17 hereafter). The key input parameters of these models, including the hydrogen density, geometry, ionizing SED, solar abundance set, nitrogen prescription, and dust depletion, are listed in Table.~\ref{tab:models}. The relative importance of these parameters is further discussed in Section~\ref{sec:results}. Throughout this work, we express the gas-phase metallicity as 12 + log(O/H), which refers to the "pre-depletion" abundance unless otherwise specified. We chose 12+log(O/H) instead of [O/H] ($\equiv \log$((O/H)/(O/H)$_\odot$)) adopted by these models to avoid any confusion about the value of the solar oxygen abundance.
For the stellar metallicity, we denote it as $Z$. For all photoionization models in Table.~\ref{tab:models}, the relative stellar metallicity, $\log (Z/Z_\odot )$, and the relative gas-phase metallicity, [O/H], were set to be the same and were varied together. The ionization parameter is defined as
\begin{equation}
    U = \frac{\Phi_{0}}{n_H c} = \frac{Q_0}{4\pi r^2 n_H c},
\end{equation}
where $\Phi_{0}$ is the ionizing flux from the central stars, $Q_0$ is the number of ionizing photons per unit time, $r$ is the radius at which the ionization parameter is measured\footnote{This expression is used if the geometry is spherical. We note that $r$ is usually set to the inner radius of the \hii\ region or sometimes the Str\"omgren radius.}, $n_H$ is the hydrogen density, and $c$ is the speed of light.

\begin{table*}
        \centering
        \caption{Input parameters for the photoionization models}
        \label{tab:models}
        \begin{tabular}{l c}
            \hline
                \hline
                Parameter & Values \\
                \hline
                \multicolumn{2}{|c|}{JY20 model \citep[fiducial model in this work,][]{ji2020b}} \\
                \hline
                log(U) & $-4.0$, $-3.75$, $-3.5$, $-3.25$, $-3.0$, $-2.75$, $-2.5$, $-2.25$, $-2.0$, $-1.75$, $-1.5$\\
                $\log(Z/Z_\odot )$ & $-1.3$, $-0.7$, $-0.4$, 0.0, 0.3, 0.5\\
            $\log (n{\rm_H} / {\rm cm}^{-3})$ & 1.15\\
            Geometry & Plane-parallel\\
                Stellar SED & {\sc starburst99} model with a continuous SFH of 4 Myr\\
                Solar abundance set & \citet{grevesse2010} solar abundance set\\
                Nitrogen prescription & \citet{dopita2013} prescription\\
                Depletion factor & Default depletion set in {\sc cloudy} \citep{cowie1986, jenkins1987}\\
                \hline
                \multicolumn{2}{|c|}{L10 model \citep{levesque2010}} \\
                \hline
                log(U) & $-3.5$, $-3.2$, $-2.9$, $-2.6$, $-2.5$, $-2.2$, $-1.9$\\
                $\rm \log(Z/Z_\odot )$ & $-1.3$, $-0.7$, $-0.4$, 0.0, 0.3\\
                $\log (n{\rm_H} / {\rm cm}^{-3})$ & 2.0\\
                Geometry & Plane-parallel\\
                Stellar SED & {\sc starburst99} model with a continuous SFH of 6 Myr\\
                Solar abundance set & \cite{anders1989} solar abundance set (E. Levesque, private communication)\\
                Nitrogen prescription & unspecified\\
                Depletion factor & unspecified\\
                \hline
                \multicolumn{2}{|c|}{D13 model \citep[$\kappa \rightarrow \infty$,][]{dopita2013}} \\
                \hline
                log(U) & $-4.0$, $-3.7$, $-3.5$, $-3.2$, $-3.0$, $-2.7$, $-2.5$, $-2.2$, $-2.0$\\
                $\log(Z/Z_\odot )$ & $-1.3$, $-1.0$, $-0.7$, $-0.5$, $-0.3$, 0.0, 0.3, 0.5, 0.7\\
                $\log (n{\rm_H} / {\rm cm}^{-3})$ & 1.0\\
                Geometry & Spherical\\
                Stellar SED & (Old) {\sc starburst99} model with a continuous SFH of 4 Myr \citep[see][]{dopita2000} \\
                Solar abundance set & \citet{grevesse2010} solar abundance set\\
                Nitrogen prescription & \cite{dopita2013} prescription\\
                Depletion factor & Depletion set described in \cite{kimura2003} and \cite{dopita2005}\\
                \hline
                \multicolumn{2}{|c|}{B17 model \citep[extracted from {\sc python-fsps},][]{byler2017}} \\
                \hline
                log(U) & $-4.0$, $-3.5$, $-3.0$, $-2.5$, $-2.0$, $-1.5$, $-1.0$\\
                $\log(Z/Z_\odot )$ & $-2.0$, $-1.5$, $-1.0$, $-0.6$, $-0.4$, $-0.3$, $-0.2$, $-0.1$, 0.0, 0.1, 0.2\\
                $\log (n{\rm_H} / {\rm cm}^{-3})$ & 2.0\\
                Geometry & Spherical\\
                Stellar SED & {\sc fsps} model with a instantaneous SFH of 1 Myr\\ & (MIST evolutionary track, see \citealp{dotter2016} and \citealp{choi2016}) \\
                Solar abundance set & \citet{anders1989} solar abundance set \\
                Nitrogen prescription & \cite{dopita2000} prescription\\
                Depletion factor & Depletion set described in \cite{dopita2000}\\
                \hline
        \end{tabular}
\end{table*}

\section{Dependence of the MI correlation on model parameters}
\label{sec:results}

In this section we use photoionization models to fit  gas-phase metallicities as well as ionization parameters for our sample. The basic idea of this approach is to compare the model predictions on multiple emission line ratios with the observed values. Depending on the number of line ratios involved, the calculation could be 2D (minimal dimensions required to fit both the metallicity and ionization parameter), or it could have three or more dimensions.

Starting with the 2D diagnostics: the standard 2D BPT diagrams ([N\,{\sc ii}]-, [S\,{\sc ii}]-, and [O\,{\sc i}]-based diagrams) can provide useful constraints on the ionizing sources for the ionized regions, but they are not suitable for fitting metallicities and ionization parameters. The photoionization model grids tend to wrap around in these diagrams, which results in significant degeneracies between the fitted metallicities and ionization parameters.
Another 2D diagnostic diagram that is frequently used to fit metallicities and ionization parameters simultaneously is the [N\,{\sc ii}]/[O\,{\sc ii}] versus [O\,{\sc iii}]/[O\,{\sc ii}] diagram \citep{dopita2000}. The merit of this diagnostic diagram is that photoionization model grids do not overlap with themselves in this diagram \footnote{Sometimes O3O2 alone is used to predict the ionization parameter. We note that such usage is not recommended as the relation between O3O2 and the ionization parameter also depends on metallicity.}. However, a few caveats should be noted. First, this diagram has an extra dependence on the dust extinction correction due to the large wavelength separation of the emission lines used. Second, it relies on a good understanding of how the N/O ratio changes with the metallicity. The fact that there is a significant degeneracy between models with different N/O versus O/H relations makes this diagram alone not very useful in constraining the model parameters or providing unique fitting results.

To overcome the above difficulties associated with the 2D diagrams, one could take more line ratios into consideration and constrain the secondary model parameters before fitting the metallicity and ionization parameter. This approach effectively brings us to the regime of high-dimensional analysis. \cite{ji2020b} show that by simply combining [N\,{\sc ii}]/H$\alpha$, [S\,{\sc ii}]/H$\alpha$, and [O\,{\sc iii}]/H$\beta$ and forming a 3D diagnostic diagram (N2-S2-R3 diagnostic), one can place strong constraints on secondary parameters, including the shape of the ionizing SED and the N/O abundance pattern. 
This is achieved by simply requiring that the model surface goes through the densest part of the data surface in 3D.
The N2-S2-R3 diagnostic itself can also be used to predict the metallicity and ionization parameter. Here we would like to define the concept of "best fitting" in high dimensions. Ideally, there exists an optimal photoionization model, which gives consistent predictions over all the line ratios, should the physical properties of the studied \hii\ regions all have narrow distributions about the median values. 
Still there are scatters around the manifold of the model due to the widths of the distributions of different parameters. Following the previous practices on modeling \hii\ regions, we make the following assumptions.

First, there are two primary parameters, that is to say the metallicity and the ionization parameter, that contribute the largest line ratio variations.
Second, at a fixed metallicity and ionization parameter, the variations in other (secondary) parameters result in comparably less variations in line ratios for a statistically large sample of \hii\ regions.

By varying the metallicity and ionization parameter of the photoionization model, we obtain a 2D surface in any given line-ratio space. The variations in secondary parameters appear as small scatters around the model surface in high dimensions, but the best-fit model surface still lies in the center, or the densest part of the data distribution.
The location of the "central surface" can then be used to constrain the median values of the secondary parameters.
We expect the desired model that best fits the central surface of the data to produce identical relationships between the metallicity and ionization parameter, within the uncertainties, no matter which subset of line ratios is used to derive them (except for those subsets with significant degeneracy). Whether the model predicts the same MI correlation using different line ratios is thus a useful test for its self-consistency.

In the following part of this section, we compare the fitting results from different photoionization models. We used the Bayesian inference to estimate the joint and marginalized probability distribution functions (PDFs) of the logarithmic metallicity, [O/H], and ionization parameter, log(U), for each of the data points, assuming flat priors in both dimensions \citep[see e.g.,][for a general discussion on the Bayesian approach]{blanc2015}. Basically, for each data point, we calculated the likelihood as
\begin{equation}
    p(D|M, \theta)=
    \frac{{\rm exp}(-\chi ^2 /2)}{(2\pi )^{n/2}\rm det({\boldsymbol C})^{1/2}},
    \label{eq:likelihood}
\end{equation}
where D, M, and $\theta$ represents the data, the adopted model, and the model parameters, respectively (i.e., [O/H] and log(U)). Furthermore, $n$ is the number of emission line ratios considered, and 
$\chi ^2$ is given by
\begin{equation}
    \chi ^2 = [\boldsymbol{X_D}-\boldsymbol{X_M}]^T \boldsymbol{C}^{-1} [\boldsymbol{X_D}-\boldsymbol{X_M}],
\end{equation}
where $\boldsymbol{X_D} - \boldsymbol{X_M}$ is an $n$ dimensional vector describing the logarithmic line-ratio differences between the data point and the model point, and $\boldsymbol{C}$ is the covariance matrix \citep{hogg2010}.
Following \cite{belfiore2019} who examined the uncertainties reported by DAP using repeated observations in MaNGA, we inflated all the uncertainties in emission line fluxes by a factor of 1.25.
This is equivalent to multiplying all terms in $\boldsymbol{C}$ by 1.5625, which does not have a significant effect on our results.
The likelihood was then combined with the flat priors and normalized to give the posterior, $p(\theta | D, M)$ (we discuss the influence of nonflat priors in \S~\ref{sec:discuss}). Finally, we calculated the weighted average metallicity and ionization parameter for each data point using the marginalized posteriors\footnote{Using the metallicity and ionization parameter with the maximum posterior would result in almost identical values, as the posterior distribution is single-peaked. This is because all the photoionization models investigated in this work behave regularly in our choices of line-ratio space (i.e., not showing a large curvature around data points, or a significant degeneracy between the metallicity and ionization parameter).}
\begin{equation}
    <{\rm \log(O/H)}>=\sum _{\rm \log(O/H)_{min}}^{\rm \log(O/H)_{max}}p({\rm \log(O/H)}|D,M){\rm \log(O/H)},
\end{equation}
\begin{equation}
    <{\rm \log(U)}>=\sum _{\rm \log(U)_{min}}^{\rm \log(U)_{max}}p({\rm \log(U)}|D,M){\rm \log(U)}.
\end{equation}
It should be noted that the metallicity calculated here is the pre-depletion metallicity, that is the metallicity before any depletion onto dust grains occurs. The relation between the pre-depletion metallicity and the post-depletion metallicity depends on the depletion factors adopted by each model. For the JY20 model, this is given by
\begin{equation}
    \rm 12+\log(O/H)_{pre} = 12+\log(O/H)_{post} + 0.22.
\end{equation}
The post-depletion metallicity is the current gas-phase metallicity in the ionized cloud. Thus, one should be cautious when comparing metallicities derived from photoionization models to those derived through other methods.

For each model, we performed three sets of fitting, the first one in the N2O2 versus O3O2 space, the second one in the N2-S2-R3 space, and the last one in the 5D space composed of N2O2, O3O2, N2, S2, and R3. When fitting the data using the N2O2 versus O3O2 method, we applied extinction corrections based on the Balmer decrements, assuming an intrinsic Balmer ratio $F_{H\alpha}/F_{H\beta}=2.86$. The extinction curve we used is from \cite{fitzpatrick1999} with $R_V = 3.1$. We see whether these results agree with each other, and what is the main driver of different MI correlations.

\subsection{Comparison of model predictions based on different emission-line ratios}

\begin{figure*}
    \includegraphics[width=0.30\textwidth]{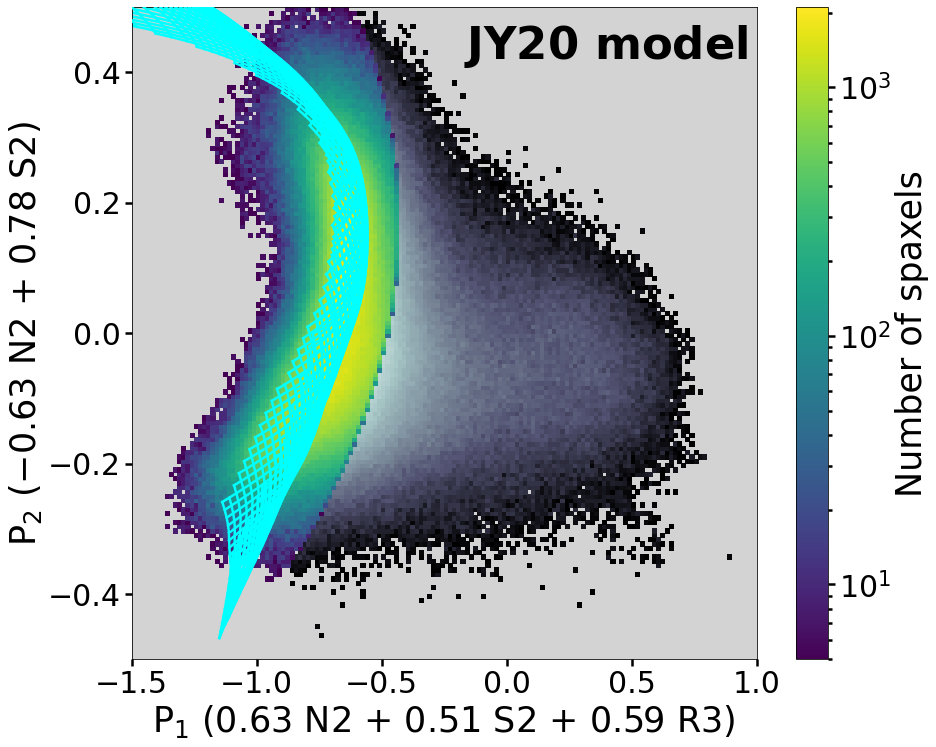}
    \includegraphics[width=0.7\textwidth]{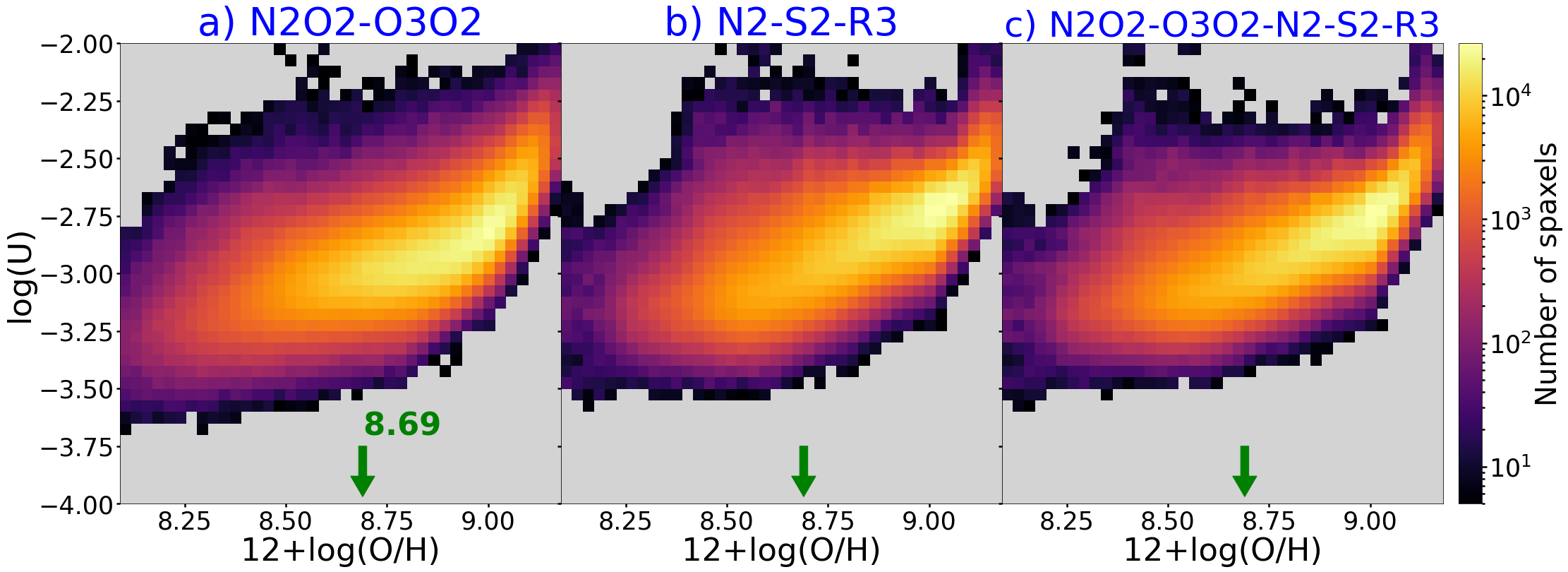}
    \includegraphics[width=0.30\textwidth]{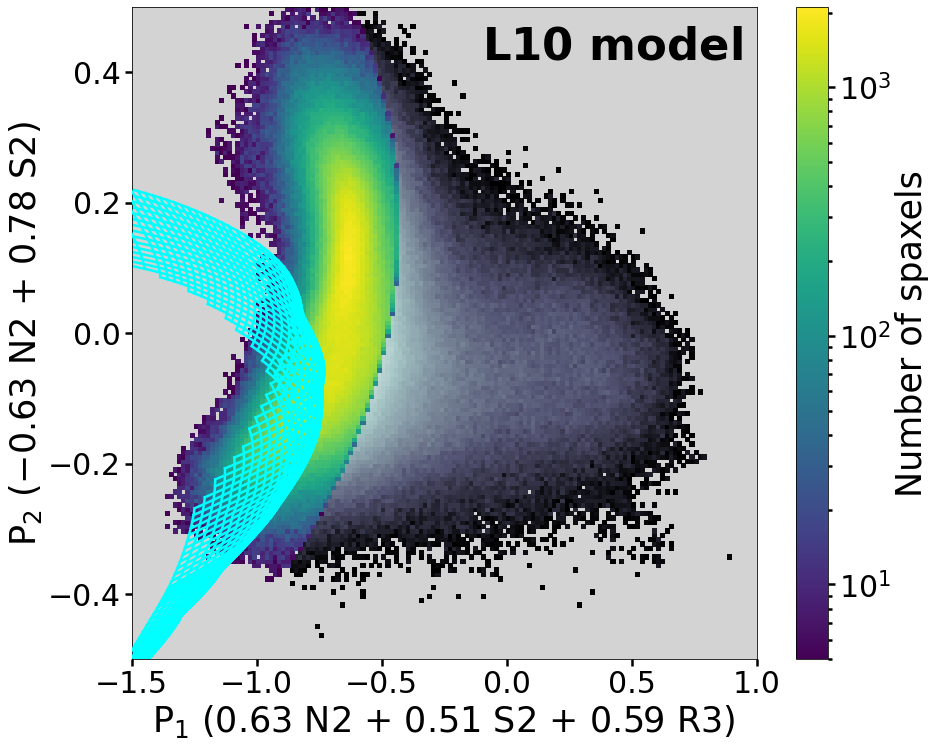}
    \includegraphics[width=0.7\textwidth]{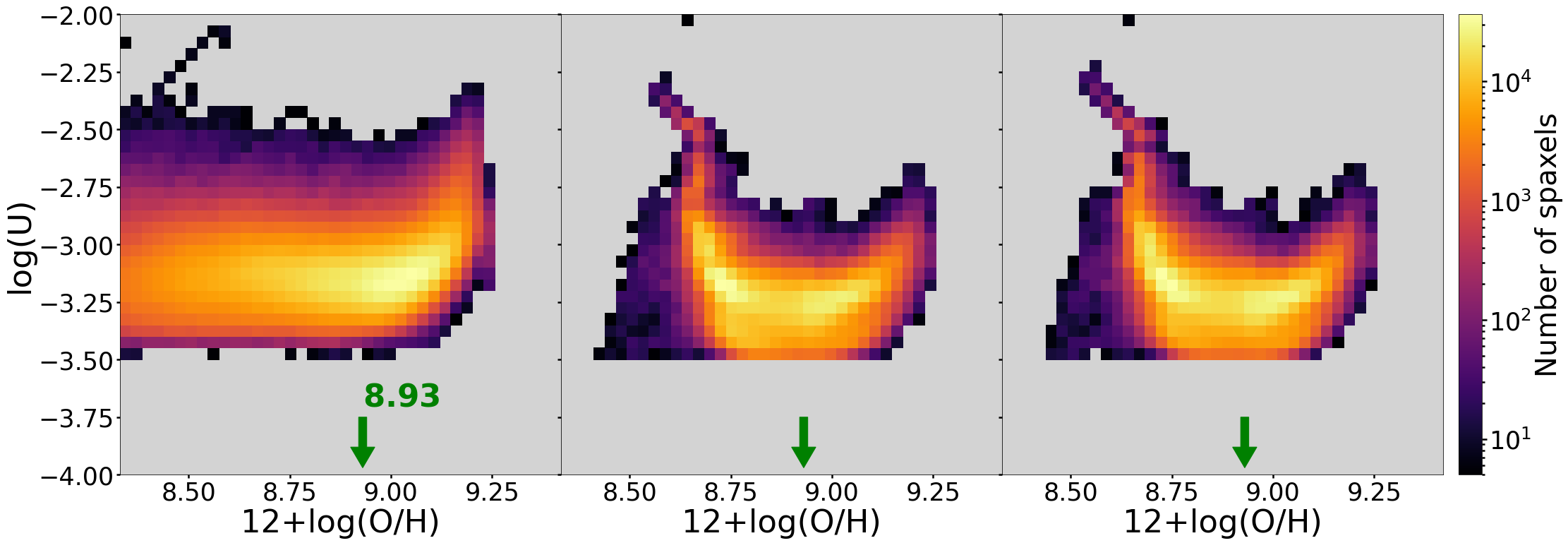}
    \includegraphics[width=0.30\textwidth]{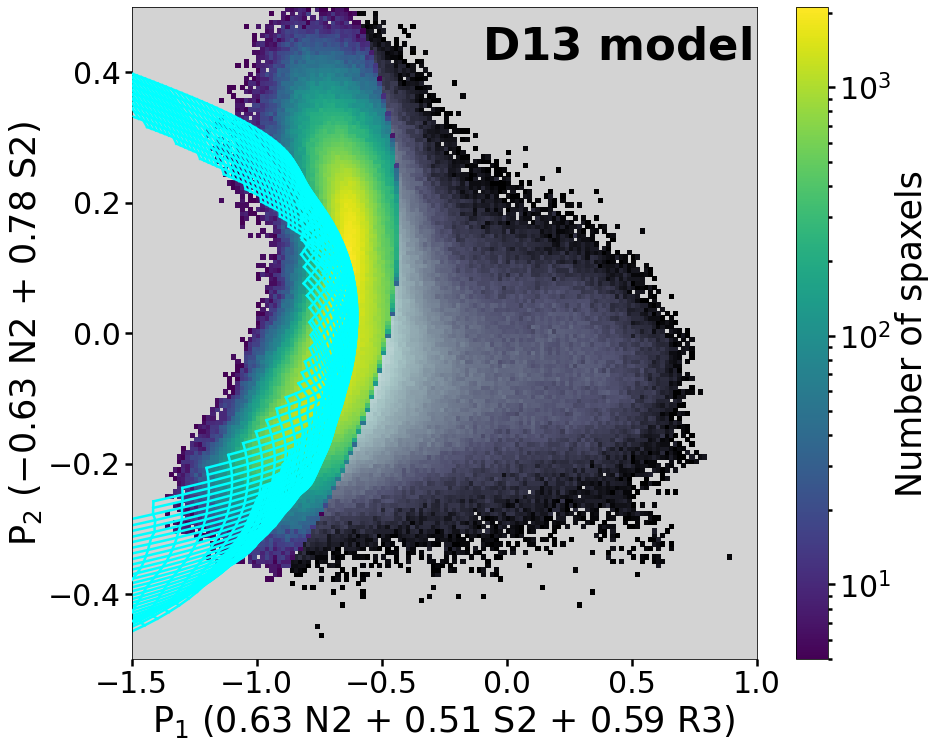}
    \includegraphics[width=0.7\textwidth]{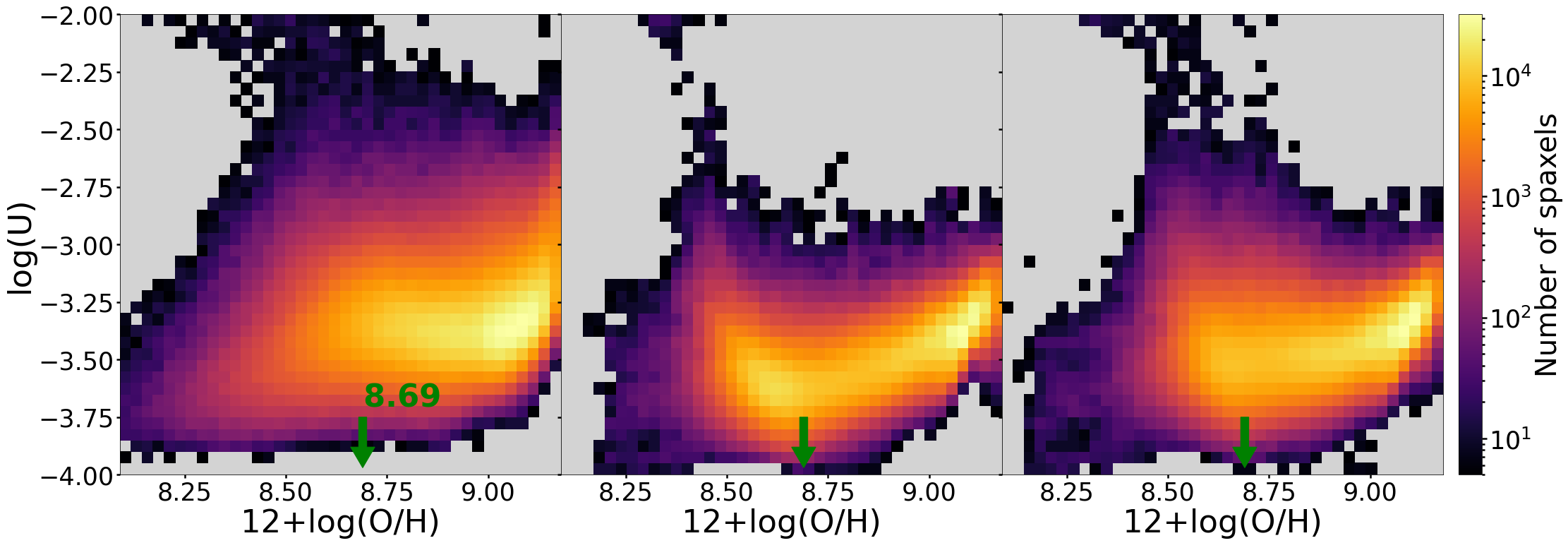}
    \includegraphics[width=0.30\textwidth]{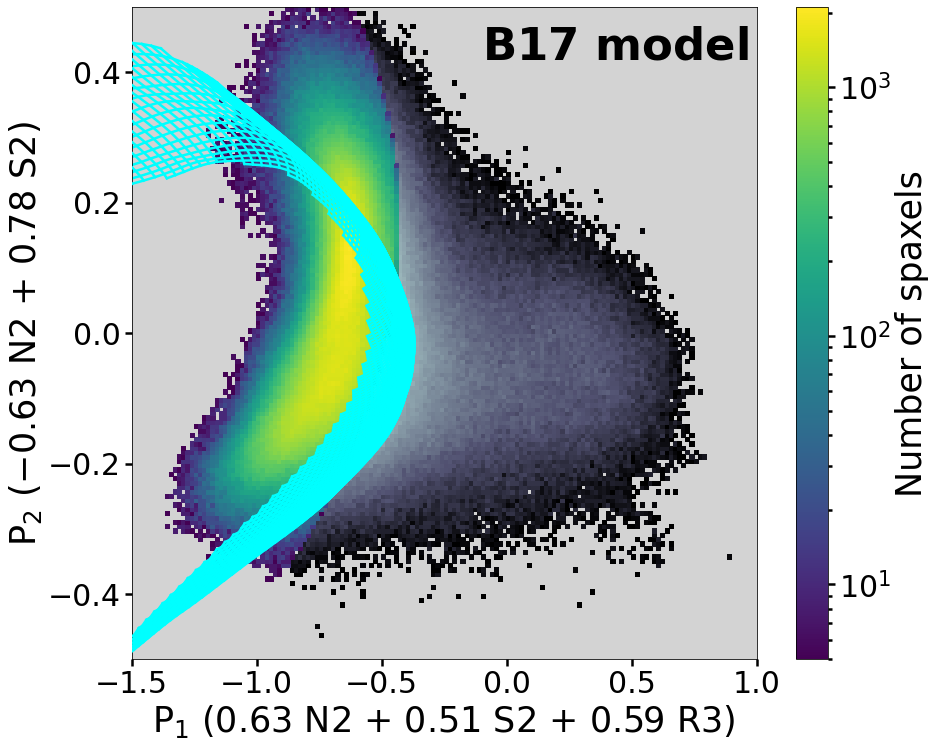}
    \includegraphics[width=0.7\textwidth]{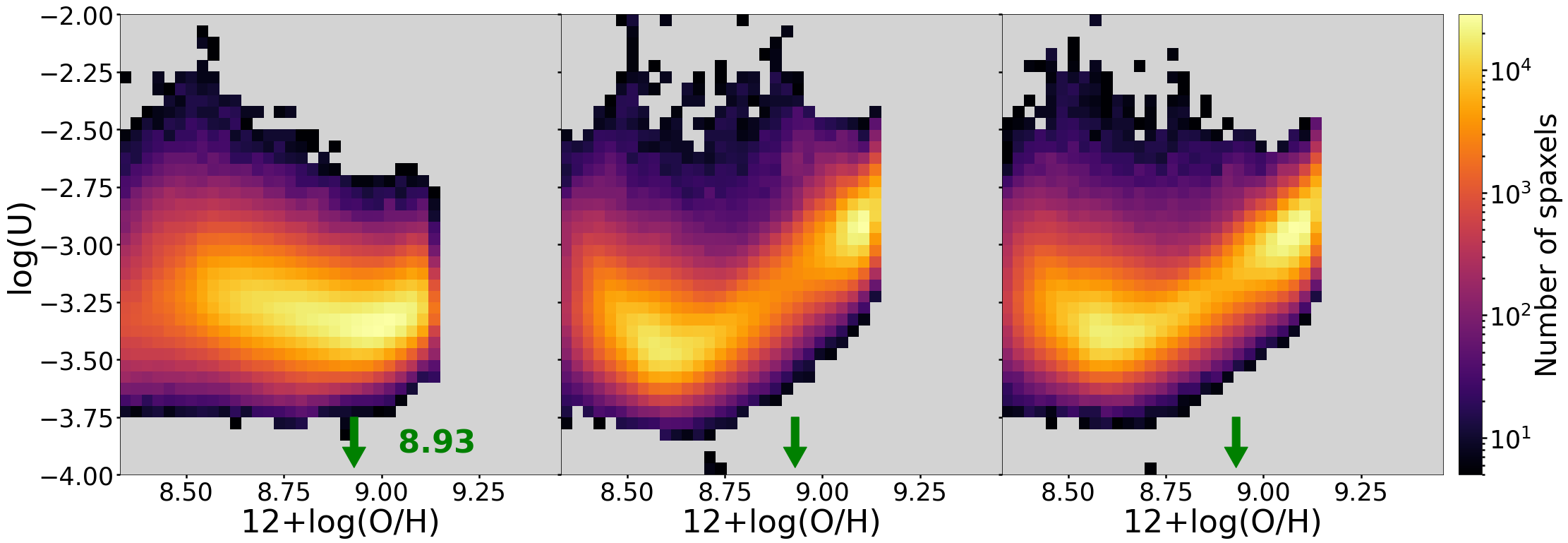}
    \caption{Different photoionization models viewed in the $P_1$-$P_2$ diagram and their predicted relations between the metallicity and ionization parameter. Leftmost column: Different SF-ionized models and MaNGA MPL-7 data in the $P_1$-$P_2$ diagram. The models were interpolated and cut so that only the parts that cover the middle 98\% of the data along the hidden $P_3$ axis are shown. The spaxels with contributions from SF ionization greater than 90\% are colored from yellow to green, while the rest of the spaxels are colored from white to black. Only the former were used to derive the metallicities and ionization parameters. Right columns [a) to c)]:  Derived metallicities and ionization parameters using different SF-ionized models and different combinations of line ratios. In each panel, the green arrow marks the solar oxygen abundance adopted by the corresponding model.}
    \label{fig:fitting}
\end{figure*}

\begin{figure*}
    \includegraphics[width=0.5\textwidth]{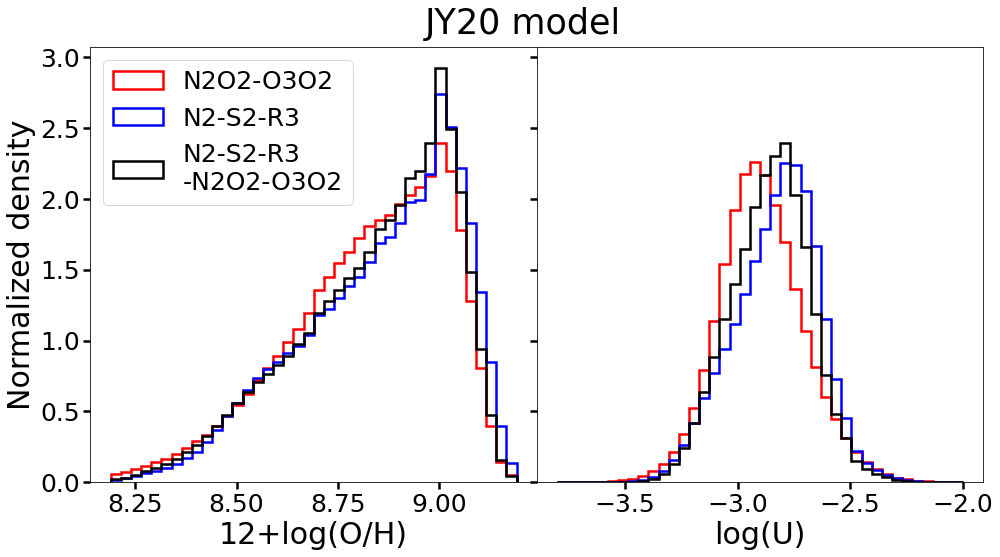}
    \includegraphics[width=0.5\textwidth]{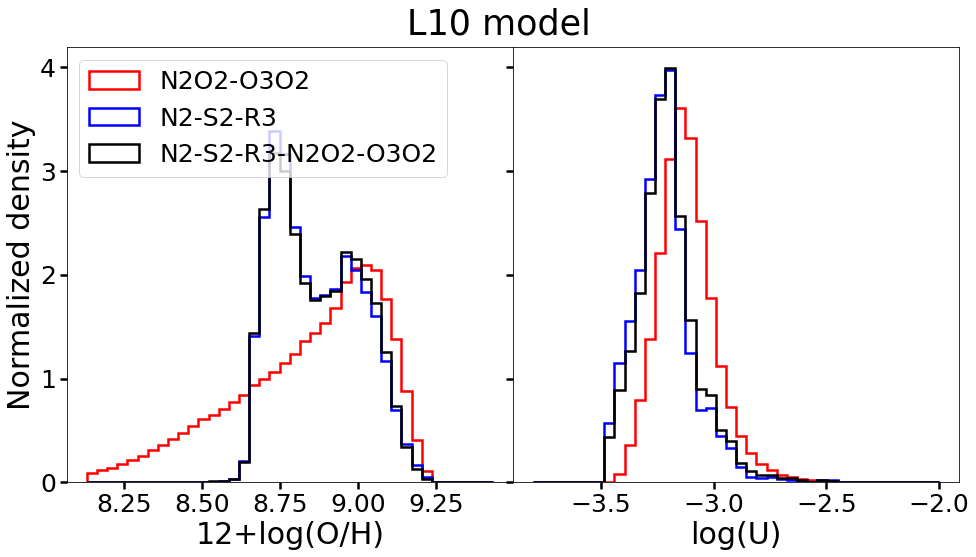}
    \includegraphics[width=0.5\textwidth]{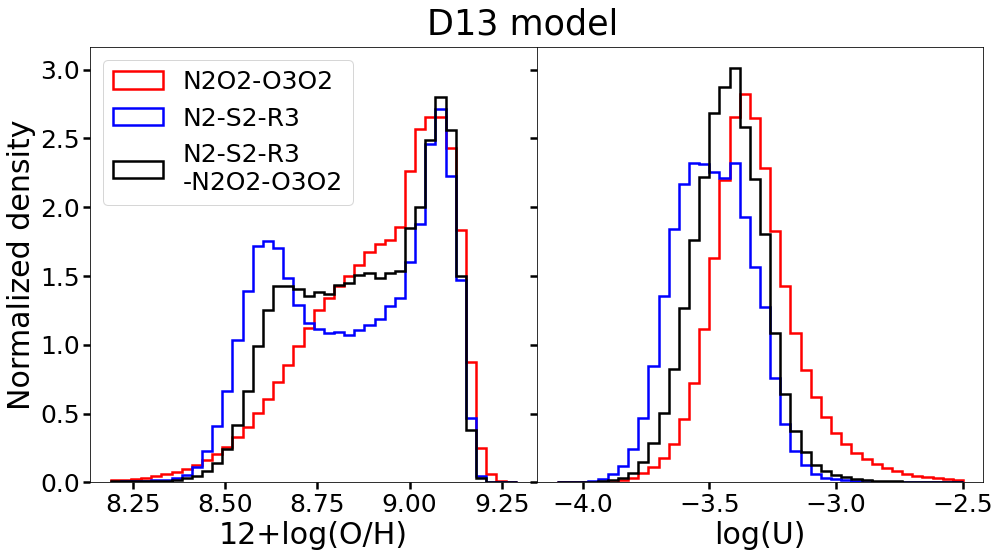}
    \includegraphics[width=0.5\textwidth]{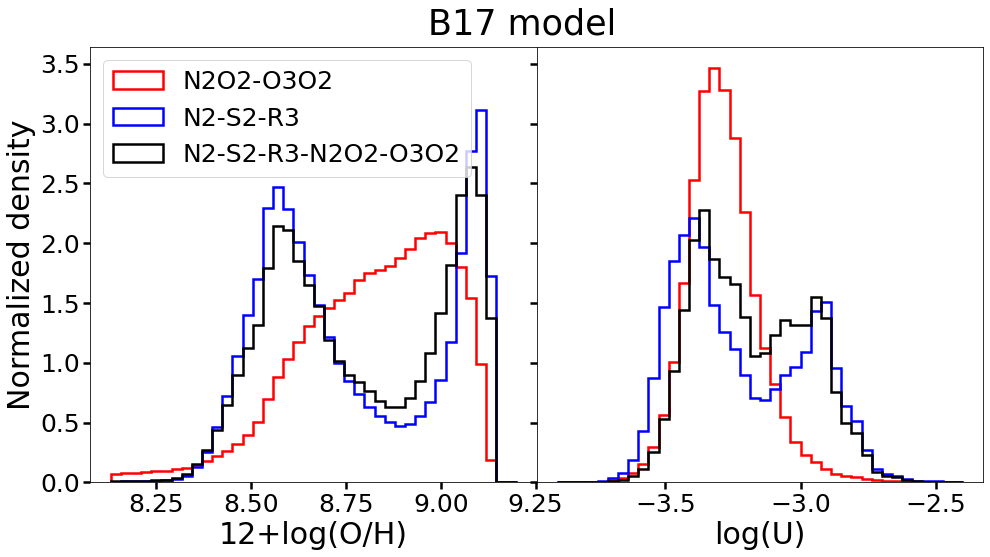}
    \caption{Histograms showing the number distributions of 12+log(O/H) and log(U) predicted by different SF-ionized models and different combinations of line ratios.}
    \label{fig:hist}
\end{figure*}

\begin{figure}
    \includegraphics[width=0.48\textwidth]{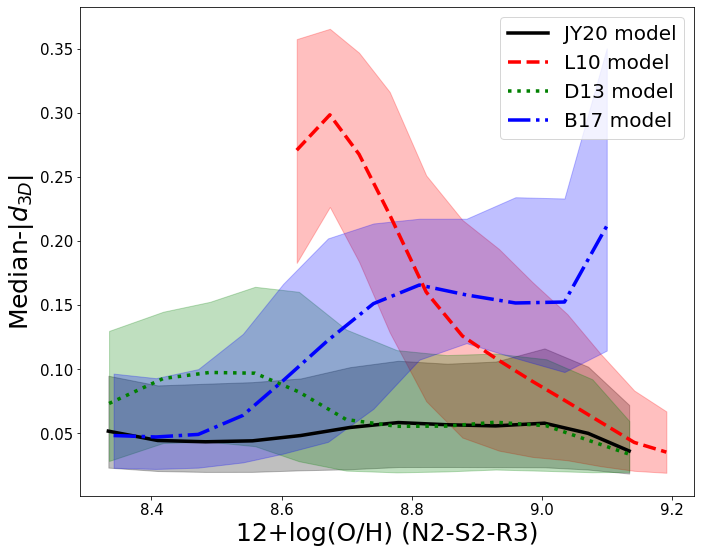}
    \caption{Median absolute Euclidean distances in the N2-S2-R3 3D space as a function of fitted metallicities for different SF-ionized models. The metallicities shown were derived based on different models using the N2-S2-R3 method, thus having different ranges. The shaded regions encompass the middle 68\% of the data.}
    \label{fig:3d_dist}
\end{figure}

In Figure~\ref{fig:fitting} we plotted all four SF-ionized models in the $P_1$-$P_2$ diagram as well as their predicted MI correlations.
This projection makes the model surfaces roughly edge-on and thus best reveals any discrepancy between the data and model surfaces in the 3D line-ratio space.
The leftmost column compares the model surfaces with the \hii\ regions (which are colored in green and yellow). All models were interpolated using the 2D interpolation function {\sc griddata} in {\sc python}. To better compare the locations of the model surfaces with the data distribution, we show only parts of the models that cover the middle 98\% of the data along the $P_3$ axis, which is perpendicular to the $P_1$-$P_2$ plane in 3D.

By eye, the JY20 model provides the best fit to the densest part of the data distribution in the $P_1$-$P_2$ diagram. For all three combinations of line ratios, the JY20 model predicts positive correlations between 12+log(O/H) and log(U). The 2D distributions of 12+log(O/H) and log(U) derived by different sets of line ratios show good consistency overall.
The only difference may be that the constraints based on the N2O2-O3O2 method gives a slightly smoother distribution than the other two methods.

In comparison, the L10 model underpredicts $P_2$ values significantly for $P_1 \lesssim 0$. Since $P_2$ anticorrelates with metallicity \citep{ji2020b}, the L10 model fails to describe the majority of the data at relatively low metallicities. As expected, for this model the fitting results from different methods show discrepancies at subsolar metallicities. When using the N2O2-O3O2 method, the L10 model predicts no correlation between 12+log(O/H) and log(U) for spaxels with subsolar metallicities. Whereas the N2-S2-R3 method gives a clear anticorrelation between 12+log(O/H) and log(U). When all line ratios are combined, the resulting distributions of 12+log(O/H) and log(U) closely resemble those predicted by N2, S2, and R3.
This could be caused by the fact that N2, S2, and R3 overall have smaller uncertainties than N2O2 and O3O2. However, we note that the position of the model surface relative to the data distribution in 5D could also be relevant.

For the D13 model, it shows better consistency with the data in the N2-S2-R3 space, but it is still offset from the dense part of the reprojected data locus at low metallicities. Using this model, the N2O2-O3O2 method predicts a very broad distribution of log(U). The resulting 12+log(O/H) and log(U) seem to have no correlation except at very high metallicities. On the other hand, the N2-S2-R3 method again predicts a negative correlation between the two quantities at subsolar metallicities and a positive correlation at supersolar metallicities. When all five line ratios are used, the overall shape of the relation between 12+log(O/H) and log(U) looks similar to the one obtained using the N2-S2-R3 method, but with a larger scatter in log(U). In summary, the fitting results of the D13 model share similar features with those of the L10 model due to their similar behaviors in the N2-S2-R3 space. They all underestimate these line ratios at (their) subsolar metallicities and show better consistency with the data distribution at supersolar metallicities, which lead to two different correlations between 12+log(O/H) and log(U) in these two regimes. In regions where these two model surfaces get close to the center of the data distribution in 3D, the derived 12+log(O/H) and log(U) are clearly positively correlated.

Finally, the B17 model gives a more complicated result as it crosses the dense part of the H\,{\sc ii} region sample at $P_2 \sim 0.2$. It underestimates N2, S2, and R3 at very low metallicities, while it overestimates them at higher metallicities. The N2O2-O3O2 predictions show an anticorrelation between 12+log(O/H) and log(U) until the metallicty reaches the solar value, after which these two quantities seem to start becoming positively correlated again. The predictions given by the N2-S2-R3 method exhibits a similar behavior, despite the turning point of the correlation lying at a subsolar metallicity rather than a solar metallicity. The 5D fitting result is again dominated by N2, S2, and R3.

We see that only the JY20 model predicts a consistent and positive MI correlation no matter which combination of line ratios is chosen. The predictions of other models all have significant dependencies on which a combination of line ratios was used, due to their obvious discrepancies with the data surface in the multidimensional line-ratio space. For these models, the predicted MI relations through the N2-S2-R3 method all have two different trends. That is, with increasing metallicities, the slopes of the relations are at first negative, and then become positive at some metallicities, which seems to be related to where the models start to get close enough to the dense part of the data in 3D.

Figure~\ref{fig:hist} shows the 1D distributions of 12+log(O/H) and log(U) predicted by different models and different sets of line ratios. Only the results from the JY20 model show good agreement among different sets of line ratios. Judging from the histograms under the JY20 model, the N2-S2-R3 method seems to yield slightly more high metallicity spaxels, which implies that our linearly extrapolated stellar SED at the highest metallicity is not perfect. Another noticeable difference is that the peak of the log(U) distribution derived by the N2-S2-R3 method is higher than that derived by the N2O2-O3O2 method by $\sim 0.1$ dex. This difference originates from the fact that the model predicted O2 value is not entirely consistent with the observed O2 value (if we assume that the model predicted N2, S2, and R3 values are all correct). Since in our case O2 is only used in combination with N2 or O3, which have much larger wavelengths, it is also possible that the extinction correction is not correctly applied. We further discuss this issue in Section~\ref{subsec:o2_corr}.

The other models all show bimodal 1D distributions in 12+log(O/H), consistent with what we have seen in the 2D distributions. Unlike the N2-S2-R3 method, the N2O2-O3O2 method always gives single-peaked metallicity distributions. For log(U), the L10 model, D13 model, and B17 model all predict systematically lower values than those given by the JY20 model. Interestingly, only the B17 model predicts an obvious bimodal distribution in log(U) when using the N2, S2, and R3 lines or all five emission lines. Given the relatively flat stellar mass distribution in our sample, it is reasonable to assume that the true distribution in either 12+log(O/H) and log(U) should be a single-peaked function; otherwise, one needs to explain why the chemical evolution model would favor a very different metal-enrichment timescale for galaxies with an intermediate metallicity. The double-peaked distributions we see can be qualitatively understood as the following. Even as the model surface significantly deviates from most of the spaxels in high-dimensional line-ratio space, the fitting algorithm is still looking for the closest part of the model surface to the data. Depending on the relative curvature of the model surface to the data distribution, data points that originally correspond to a wider distribution in 12+log(O/H) (or log(U)) might be preferentially assigned to specific parts of the model grids that are closest to the data surface, causing the double-peaked distribution.

Figure~\ref{fig:3d_dist} shows how the median absolute distances from the data points to the model surfaces in the N2-S2-R3 space changes as a function of the metallicities predicted by the models. The JY20 model surface lies closest to the majority of the data in 3D space, with the median distances $d_{3D}\lesssim 0.05$ dex. The D13 model surface is equally close to the high-metallicity data, but fits the low-metallicity data worse. The L10 model surface lies similarly close to the data at 12+log(O/H)  $\approx 9.2$, but it moves further away from the data as metallicity gets lower. Finally, the B17 model surface appears closest to the data at 12+log(O/H) $\approx 8.4$, and it starts to deviate from the data as metallicity increases. These results are consistent with what we found in Figure~\ref{fig:fitting} with the $P_1$-$P_2$ diagram. This again shows that the agreement between the model surface and the central surface of the data distribution in multidimensional line-ratio space is an important metric for model evaluation.

\subsection{Comparison of model parameters}

\begin{figure}
    \includegraphics[width=0.48\textwidth]{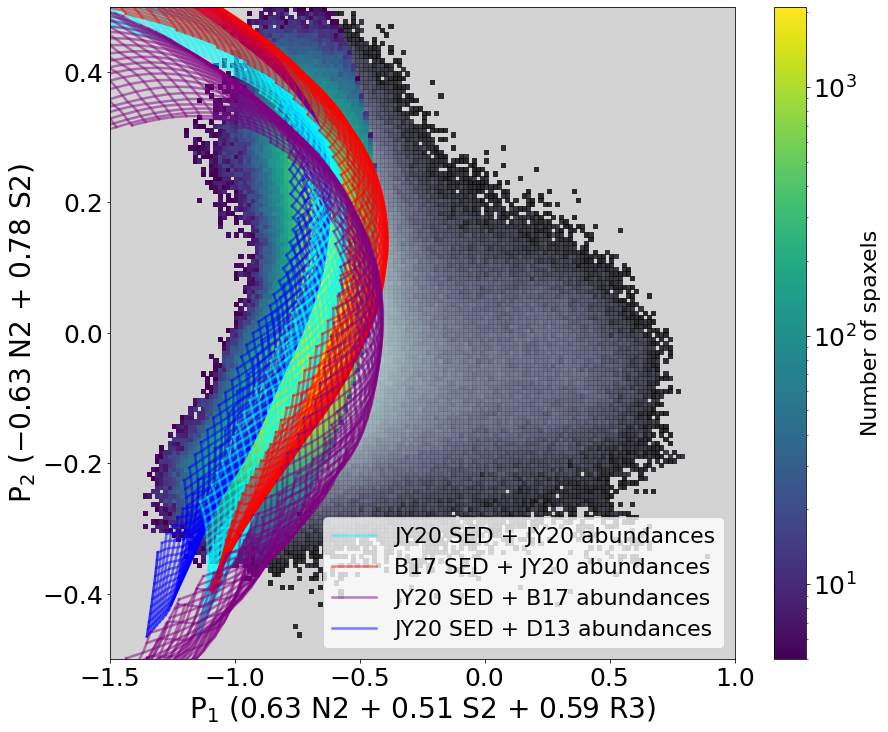}
    \caption{Comparisons of photoionization models with different input parameters in the $P_1$-$P_2$ diagram. 
    The cyan model is our fiducial model. The red model has the fiducial abundances and depletion set, but uses the SED adopted by the B17 model. The purple model uses the fiducial SED, but adopts the abundances and depletion set of the B17 model. The blue model uses the fiducial SED, but adopts the abundances and depletion set of the D13 model.
    The models are interpolated and cut so that only the parts that cover the middle 98\% of the data along the hidden $P_3$ axis are shown. The density distribution of the \hii\ region sample is plotted in green and yellow, while the rest of the MaNGA sample is plotted in black and white.}
    \label{fig:params_com}
\end{figure}

\begin{figure*}
    \includegraphics[width=0.45\textwidth]{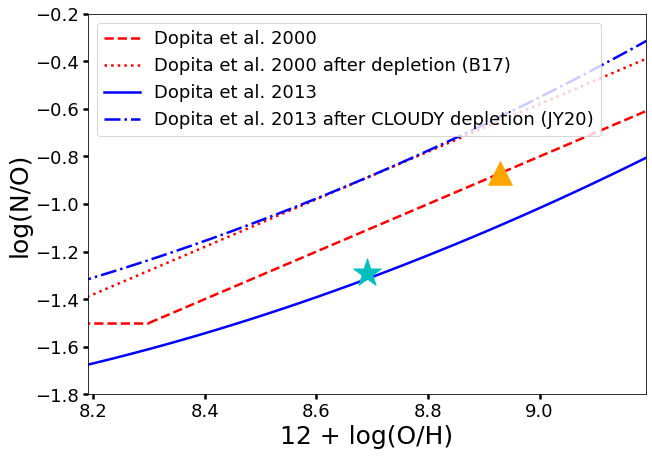}
    \includegraphics[width=0.45\textwidth]{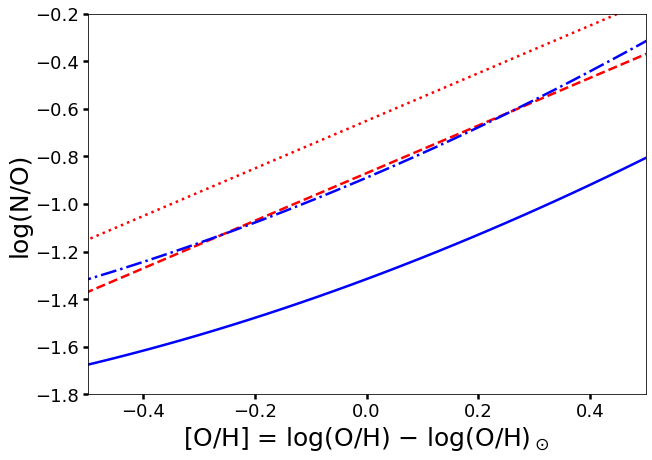}
    \caption{Comparisons of pre- and post-depletion nitrogen prescriptions for the BY17 and the JY20 models. Both the input prescription and the output prescription modified by dust depletion are shown. 
    The JY20 model adopts the same pre-depletion nitrogen prescription as \protect\cite{dopita2013} and depletes the elements using the {\sc cloudy} default depletion set. The B17 model follows \protect\cite{dopita2000} for both the pre-depletion nitrogen prescription and the depletion pattern. 
    While the left panel shows the N/O ratios as functions of the absolute oxygen abundance 12 + log(O/H), the right panel shows the N/O ratios as functions of the relative oxygen abundance [O/H], normalized by the solar O/H adopted by each model. The solar oxygen abundances, 12 + log(O/H)$_\odot$, adopted by \protect\cite{dopita2000} and \protect\cite{dopita2013} are 8.93 and 8.69 and indicated by the orange triangle and the cyan star in the plot, respectively.
    }
    \label{fig:nor}
\end{figure*}

By comparing four different SF-ionized models, we see that the model which fits the central surface of data in the multidimensional line-ratio space gives the most consistent predictions on the metallicity and ionization parameter. The location of the model surface in the line-ratio space is determined by parameters other than metallicity and ionization parameter (i.e., the secondary parameters). Variations in different secondary parameters move the model surface in different directions by different amounts, as detailed in \cite{ji2020b}. In this section we attempt to find the most relevant secondary parameters that distinguish the JY20 model from other models.

However, before we discuss the effects of the secondary parameters, we note that there are two different photoionization codes used in these models, that is to say {\sc cloudy} and {\sc mappings}. According to \cite{dagostino2019b}, the two codes result in almost identical results if the input parameters are the same. However, as pointed out by \cite{law2021b} and \cite{belfiore2021}, earlier photoionization codes do not have a reliable estimation on the dielectronic recombination rate of $S^{++}$ and use the charge-normalized mean dielectronic recombination rates of C, N, and O as an approximation \citep{ali1991}. The current version of {\sc cloudy} \citep[v17,][]{ferland2017} improves the estimation of this value \citep{badnell2015}, which results in stronger [S\,{\sc ii}] lines. Among the four models, only the JY20 model uses the latest photoionization code. Therefore, we should keep in mind that the S2 predicted by other models are underestimated by some amounts, but they could be offset by the effects of other input parameters.

Table~\ref{tab:models} lists the input parameters of the four models. We start with the effect of the density structure. The hydrogen densities of the models all lie in the range $\rm 10~cm^{-3}<n_H \leq 100~cm^{-3}$. In fact, for radiation-bounded clouds with $\rm n_H \lesssim 100~cm^{-3}$, the model-predicted line ratios are almost identical, except for some small differences at very high metallicities \citep{ji2020b}. We also tested the effects of different equations of state on the simulated cloud, and the result shows that a cloud with a constant density produces very similar results as those from an isobaric cloud with the same initial density.

The geometry of the cloud is a more complicated issue. The spherical geometry is often assumed in the dynamical modeling of \hii\ regions, while in the real world it is difficult to have perfect symmetry. Density inhomogeneities in the molecular clouds cause asymmetries and the feedback from the central star clusters could create blister \hii\ regions. In such cases, a plane-parallel geometry might be a more realistic choice \citep[e.g., the Orion nebular,][]{wen1995}. We computed models with these two different geometries while keeping other parameters the same. For the spherical model, we varied the inner radius from 3 to 10 parsecs. We found that these model surfaces cover a similar area in the line-ratio space (regardless of whether the geometry is closed or open for the spherical case) except at very high metallicities, suggesting that the geometry is not likely to be the dominant factor that causes the differences we see among the four models. 
For an ionized cloud with a high ionization parameter at the illuminated face, the \hii\ region becomes thick. If the inner radius of the \hii\ region is comparable to the thickness of the ionized layer, then the geometric dilution of the ionizing photons becomes important and the resulting emission line spectrum would be different from that produced by a plane-parallel model with the same ionization parameter. This could explain the difference we see in the models with different inner radii. In Section~\ref{sec:explanation}, we discuss whether this effect would become significant if we were to change the inner radius according to dynamical models.

The next input parameter in Table~\ref{tab:models} is the stellar SED, which determines the relative number of ionizing photons and it is quite important in setting the line ratios of H\,{\sc ii} regions. The first three models we compare all use stellar SEDs generated by {\sc starburst99}. Both the JY20 model and the D13 model assume a continuous SFH of 4 Myr. Different from the JY20 model, the D13 model assumes a Salpeter IMF \citep{salpeter1995}. However, we tested models with different IMFs and found that the IMF effect is negligible for the line ratios considered here. The remaining difference is that the D13 model uses an earlier {\sc starburst99} SED adopted by \cite{dopita2000}, which is harder compared to the newer SEDs generated by later versions of {\sc starburst99} \footnote{\cite{dopita2000} did not specify which version of the {\sc starburst99} they used. However, it should be earlier than v3.1 according to the code release information at https://www.stsci.edu/science/starburst99/docs/new.html}. A harder input SED would strengthen the low-ionization lines such as [N\,{\sc ii}] and [S\,{\sc ii}] by increasing the relative size of the partially ionized zone inside the ionized cloud. However, compared to the JY20 model, the D13 model instead shows comparable N2 ratios, but lower S2 ratios. Thus, the effect of the earlier {\sc starburst99} SED is either not large enough or is offset by the effects of some other parameters (e.g., the underestimated dielectronic recombination rate of $S^{++}$). Meanwhile, the L10 model assumes a continuous SFH of 6 Myr, which should give a very similar SED as that of the JY20 model \citep[see Figure~7 of][]{ji2020b}. Unlike the first three models, the B17 model adopts a much harder SED generated by {\sc fsps} \citep{conroy2009}, which contributes to its overpredictions of N2, S2, R3, and equivalently $P_1$. Such an SED might allow one to describe \hii\ regions undergoing intense star formation, but it is too hard for the majority of the \hii\ regions in our sample. In Figure~\ref{fig:params_com}, we plotted a model with the B17 SED set, but kept the other parameters the same as those in the JY20 model. This model shows relatively large $P_1$ values as expected.

Finally, the chemical abundance is a fundamental ingredient for setting the line ratios. The eventual chemical abundance in the cloud is determined by the adopted solar abundance, the N/O ratio, and dust depletion factors. The JY20 model and the D13 model both use the solar abundance set in \cite{grevesse2010}, while the L10 model and the B17 model use the solar abundance set in \cite{anders1989}. The abundances of relevant elements, such as nitrogen, oxygen, and sulfur, are more abundant in the latter set. However, the abundances are further altered by the N/O ratio and dust depletion factors, as we detail below.

The nitrogen is mainly created by the CNO cycles in stars. Depending on whether the carbon and oxygen that get converted to nitrogen are newly created or pre-existed in the clouds that formed the stars, the resulting nitrogen is called to have a primary origin or a secondary origin. The amount of secondary nitrogen is very sensitive to metallicity, making nitrogen-related lines good metallicity tracers \citep{alloin1979}. Theoretical metallicity calibrations that use nitrogen lines thus strongly rely on the assumed relation between the N/O ratio and metallicity (which we call the nitrogen prescription hereafter). Higher N/O ratios result in higher N2, and slightly lower S2 and R3 due to the thermal balance. \cite{ji2020b} found that the nitrogen prescription provided by \cite{dopita2013} fits the MaNGA H\,{\sc ii} regions best and they used this for their SF-ionized model. Therefore, the nitrogen prescription does not contribute to the difference between the JY20 model and the D13 model. It is unclear what kind of nitrogen prescription is used in the L10 model. Judging from the shape of the L10 model in the $P_1$-$P_2$ diagram, it could have used a nitrogen prescription that has too much primary nitrogen, causing overestimated N2 or equivalently underestimated $P_2$ at lower metallicities. The B17 model adopts the nitrogen prescription given by \cite{dopita2000}, which yields a higher N/O at a given metallicity, as is shown in Figure~\ref{fig:nor}. Interestingly, due to the effect of dust depletion and higher solar abundances, the actual N/O ratio (at a given O/H) measured in the gas phase is similar to that in the JY20 model.
However, if the relative metallicity (e.g., [O/H] or Z/Z$_\odot$) is used when comparing different models, one should be cautious about the difference in the adopted solar abundances, which we discuss in the next paragraph.

The dust depletion factors modify the gas-phase abundance in H\,{\sc ii} regions by removing part of the elements that condensed into dust grains. The depleted elements have two effects on the line ratios. First, the intensities of lines emitted by elements that are significantly depleted decrease due to the lower abundances. Second, the removed coolants influence the thermal balance of the cloud and raise the equilibrium temperature, thus strengthening lines emitted by elements that are not significantly depleted. Meanwhile, the increasing amount of dust also heats the cloud up through the photoelectric effect. The JY20 model uses the default depletion set in {\sc cloudy}, which depletes oxygen by 0.22 dex and does not deplete nitrogen at all \citep{cowie1986, jenkins1987}. In comparison, the D13 model depletes oxygen less by 0.15 dex and depletes nitrogen more by 0.05 dex. Although the difference in oxygen depletion is substantial, it is not the main cause of the apparent shift of the D13 model surface from the data at low metallicities. For comparison, we computed a model using the JY20 parameters but with the D13 depletion set. 
This model deviates from the JY20 model in the $P_1$-$P_2$ diagram in a different way, as shown by Figure~\ref{fig:params_com}.
The largest separation between this model and the JY20 model is roughly $0.1\sim 0.2$ dex along $P_1$ (with the JY20 model showing overall larger $P_1$ values) and it is negligible along $P_2$, which occurs in high metallicities rather than low metallicities.

For the B17 model, it uses the depletion set provided by \cite{dopita2000}. It depletes oxygen by 0.22 dex, just as JY20. However, it depletes 0.22 dex more nitrogen compared to the JY20 model. Its solar nitrogen abundance is 0.22 dex higher, making the gas-phase N/H the same between the two models at fixed O/H. However, B17's adopted solar oxygen abundance is 0.24 dex higher and its solar sulfur abundance is 0.09 dex higher. 
There are two ways to interpret the consequence from these differences:
(1) At the same relative pre-depletion metallicity (relative to solar), [O/H], or stellar metallicity, Z/Z$_{\odot}$ (Z$_{\odot} \approx 0.020$ for both {\sc starburst99} and {\sc fsps}), the absolute post-depletion O/H adopted by the B17 model is higher than that of the JY20 model, which also makes the corresponding post-depletion N/O larger according to the right panel of Figure~\ref{fig:nor}.
This makes the predicted R3 and N2 values larger.
(2) Alternatively, we could understand the difference as a shifted correspondence between the stellar SEDs and the absolute post-depletion metallicities. At the same post-depletion O/H, the JY20 model and the B17 model show similar post-depletion N/O as shown in the left panel of Figure~\ref{fig:nor}. However, since the solar O/H is higher in the B17 model, the corresponding [O/H] and Z/Z$_{\odot}$ are lower. Since lower metallicity SEDs are in general harder, this makes the B17 model predict larger line ratios.
We computed another model using the JY20 parameters, but with the B17 abundances and depletion set. The resulting model indeed predicts larger line ratios and it is shifted to the right in the $P_1$-$P_2$ diagram, as shown in Figure~\ref{fig:params_com}.

We note that current photoionization code does not treat the dust depletion in a totally self-consistent way.
The total mass of elements depleted does not necessarily match the total mass of dust assumed, and they  do not match in composition either \citep[see e.g.,][]{snow1996}. In addition, the depletion factors could vary in different locations inside galaxies and could depend on the environment, as noted by \cite{jenkins2009}. It also remains an open question whether or not sulfur depletes in the ISM \citep{sofia1994, jenkins2009, white2011, amayo2021}. For a more general discussion of the effect of dust depletion on predicted line ratios, readers can refer to \cite{gunasekera2022}.

In summary, there are three major factors that contribute to the differences among the models. First of all, the change in the dielectronic recombination rate of $S^{++}$ in the updated version of {\sc cloudy} makes the JY20 model predicts larger S2, resulting in some of the differences between this model and the other three models. In addition, the harder stellar SED adopted by the B17 model partly offsets the underestimation of S2, but it overestimates N2 and R3. Last but not least, the gas-phase chemical abundance plays an important role.
At a given [O/H] or Z/Z$_\odot$, the B17 model has overall higher post-depletion abundances due to a combined effect of their adopted nitrogen prescription, dust depletion set, and solar abundance set, which also contributes to its overestimation of the line ratios.
All these factors combined make the JY20 model a better representation of an average H\,{\sc ii} region in our sample. A caveat here is that there is still degeneracy between some of the secondary parameters. For example, a harder stellar SED might be able to cancel out the effect of the underestimation of elemental abundances or overestimation of depletion factors. Independent observational constraints are needed to determine the best values of these parameters. It is possible to include more emission lines and break the degeneracy into even higher dimensions, which is beyond the scope of the current paper. Nevertheless, we have seen that by tuning the model parameters within a reasonable range, we can obtain a best-fit model that lies very close to the majority of H\,{\sc ii} regions in a multidimensional line-ratio space. The true model and its predictions should not be too far from what we have presented here, as long as the adopted secondary model parameters are not significantly biased.

The best-fit model we show predicts a positive MI correlation extending from subsolar to supersolar metallicities. In the next section we discuss the physical interpretations of this correlation.

\section{Physical interpretations}
\label{sec:explanation}

\begin{figure*}
    \includegraphics[width=0.33\textwidth]{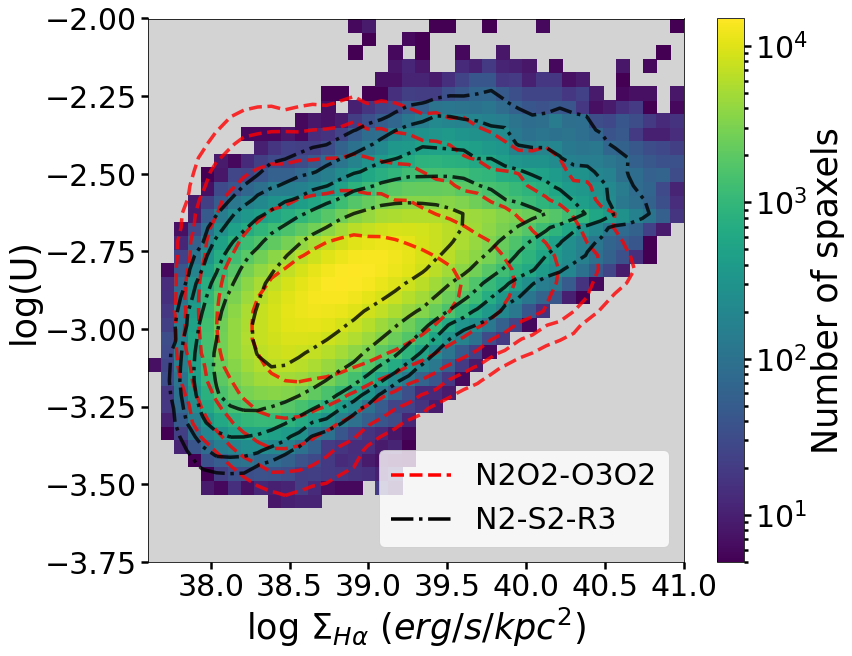}
    \includegraphics[width=0.33\textwidth]{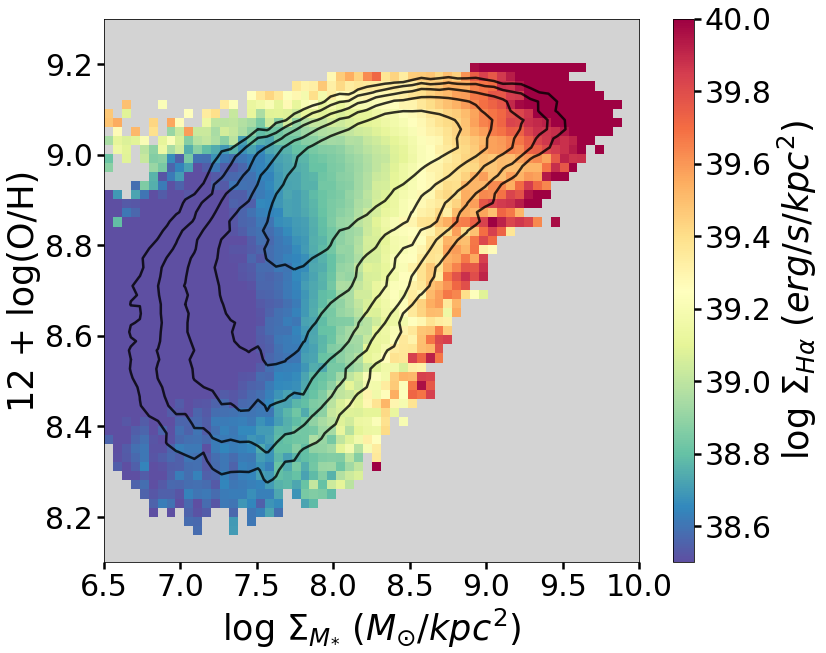}
    \includegraphics[width=0.33\textwidth]{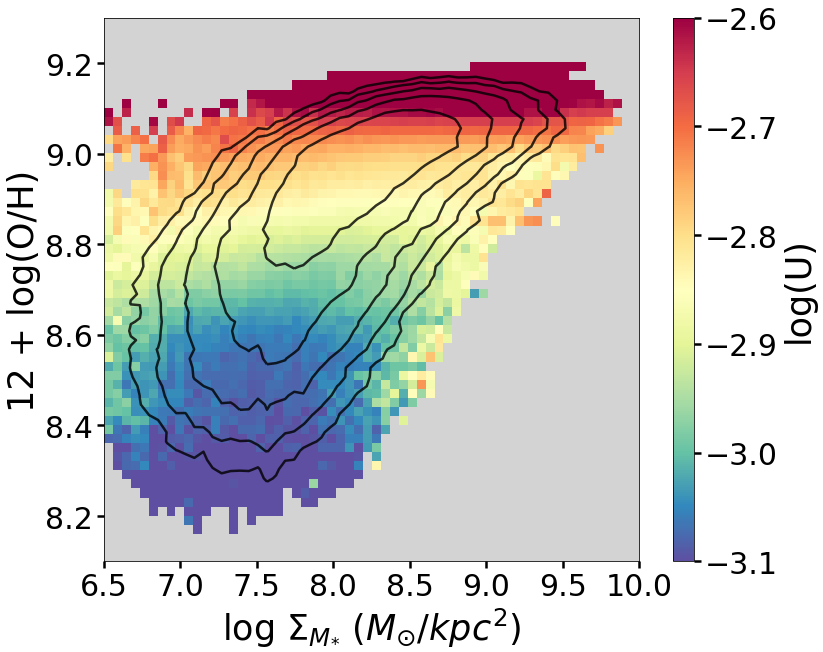}
    \caption{Correlation between the ionization parameter and H$\alpha$ surface brightness (left), and the mass-metallicity relation color coded by the H$\alpha$ surface brightness (middle) and ionization parameter (right).
    Left panel: 2D histogram of the logarithmic ionization parameter and logarithmic H$\alpha$ surface brightness. The density distribution shows the results obtained by the JY20 model using all five line ratios. The dashed red contours and the dotted-dashed black contours show the distributions obtained by using N2O2-O3O2 ratios and N2-S2-R3 ratios, respectively. The contour levels are in logarithmic scales and range from the $\rm 16^{th}$ percentile to the $\rm 84^{th}$ percentile. Middle panel: Spatially resolved mass-metallicity relation of the sample spaxels color coded by the median H$\alpha$ surface brightness inside each bin. The black contours represent the density distribution of spaxels obtained by using all five line ratios. Right panel: Same as the middle panel, but color coded by the median ionization parameter inside each bin.}
    \label{fig:surfha_logu}
\end{figure*}

\begin{figure*}
    \includegraphics[width=0.46\textwidth]{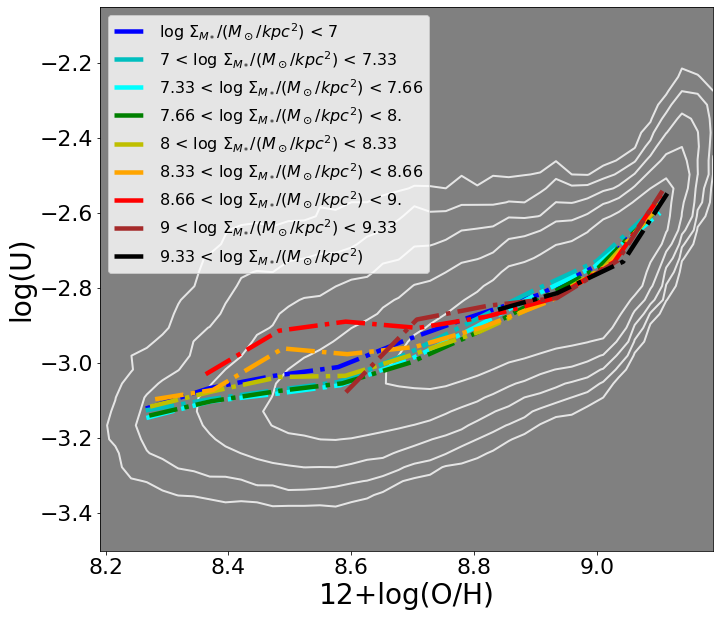}
    \includegraphics[width=0.46\textwidth]{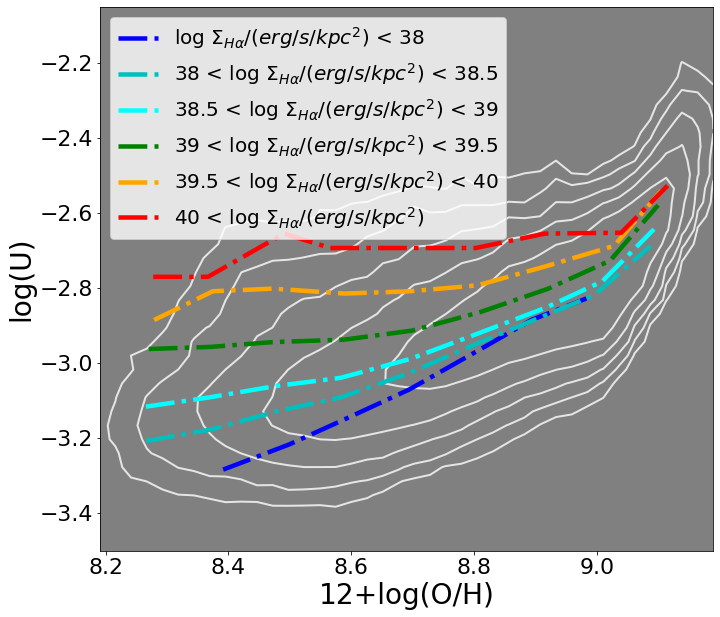}
    \caption{Dependence of the log(U) versus 12+log(O/H) relation on the stellar mass surface density (left) and H$\alpha$ surface brightness (right). The ionization parameters and metallicities shown were derived using all five line ratios. The contours represent five density levels equally spaced in the logarithmic space from the $\rm 16^{th}$ percentile to the $\rm 84^{th}$ percentile. The dotted-dashed lines indicate the median relations in different bins.}
    \label{fig:mstarsfr_dep}
\end{figure*}

The previous section has shown that according to photoionization models, the metallicity and ionization parameter are correlated.
The first theoretical analysis on how these parameters are correlated is given by D06. By studying a dynamical model of a wind-driven bubble that expands over time, D06 found $q\propto (Z/Z_{\odot})^{-0.8}$, where $q=Uc$ and $Z$ is the stellar metallicity. 
Contrary to this theoretical prediction, our best-fit photoionization model results in a positive MI correlation (assuming the stellar metallicity directly scales with the gas-phase metallicity). This positive correlation has also been noted by many authors \citep[e.g.,][]{dopita2014, poetrodjojo2018, mingozzi2020}. Specifically, D14 studied spatially resolved data of ten luminous infrared galaxies (LIRGs) and found strong MI and SFR-I correlations. Hence it is possible that the ionization parameter is related to metallicity through their common dependency on the SFR. Although the dynamical model of D06 predicts a negative correlation between the ionization parameter and metallicity, the regulation from star formation could change the statistical behavior in the data. It is also possible that the dynamical model proposed by D06 is simply not realistic enough to describe most of the observed \hii\ regions, or our photoionization model is oversimplified and introduces a bias.

In this section, we explore the potential mechanisms that lead to the positive MI correlation, including the common dependency of the metallicity and ionization parameter on a third parameter (e.g., the SFR and related parameters),
the incompleteness of the wind-driven bubble model, and the oversimplification of the photoionization model.

\subsection{Impact of star formation}

Star formation plays an important role in regulating the chemical enrichment of galaxies. At fixed stellar masses, galaxies with higher SFRs tend to have lower metallicities, leading to the fundamental relation between stellar mass, gas-phase metallicity, and SFR \citep{ellison2008, mannucci2010}. Whereas overall, the metallicity positively correlates with the SFR since more massive SF galaxies tend to have both higher SFRs and higher metallicities. In contrast, the dependence of the ionization parameter on the SFR is less clear since it is not a direct observable.

D14 explored several possibilities to relate the ionization parameter to the SFR. One plausible scenario is the effect of cluster mass. D06 observed in their dynamical model that the ionization parameter weakly depends on the cluster mass
\begin{equation}
    U\propto M_{cl}^{1/5}.
    \label{eq:cl}
\end{equation}
According to D06, this relation comes from the dependency of $U$ on the relative number density of the \hii\ region to the ambient medium (see their Equation 12). There is evidence that galaxies undergoing more intense star formation can hold a larger number of massive clusters \citep[e.g.,][]{bastian2008, powell2013}, thus connecting the ionization parameter to the SFR. However, D14 also found results that contradict observations using this relation. Their data show $U\propto \Sigma _{SFR}^{0.34}$. Assuming there are no other dependencies, we have $M_{cl}\propto \Sigma _{SFR}^{1.7}$. Combining the relation $\Sigma _{SFR} \propto \Sigma _{cl} M _{cl}$ (where $\Sigma _{SFR}$ is the SFR surface density and $\Sigma _{cl}$ is the surface number density of young clusters) and the star formation law of \cite{kennicutt1998} $\Sigma _{SFR} \propto \Sigma _{g}^{1.4}$ (where $\Sigma _{g}$ is the gas surface density), one gets $\Sigma _{cl} \propto \Sigma _g^{-0.98}$. D14 then argued that observationally speaking, the density of clusters actually increases with increasing gas density, and thus relation~\ref{eq:cl} is not a viable solution.

If the ionization parameter is truly regulated by the SFR through $U\propto \Sigma _{SFR}^{\alpha}$ with $\Sigma _{cl}$ being the medium, we must have the power-law index $\alpha <0.2$ in order to make $\Sigma _{cl}$ positively correlate with $\Sigma _{g}$. The left panel of Figure~\ref{fig:surfha_logu} shows the spatially resolved SFR-I relation of our sample derived using the J20 model. We use the extinction-corrected $\rm \log \Sigma_{H\alpha}$ to represent $\rm \log \Sigma_{SFR}$ as they differ by only a constant \citep{kennicutt2012}. We can see a weak positive correlation between the two quantities. The correlation is weakest when derived using the N2O2-O3O2 method. If we fit a linear relation to the data, the slope is below 0.2, with the N2-S2-R3 method giving the largest value and the N2O2-O3O2 method giving the smallest value. There is, however, large uncertainty in this relation due to the large intrinsic scatter in log(U). It also appears that the relation is not linear, but it flattens at high $\rm \log \Sigma_{H\alpha}$.
The Pearson correlation coefficients for the (logarithmic) MI correlation, SFR-I correlation, and SFR-M correlation are 0.70, 0.61, and 0.42, respectively.
Therefore, despite the derived slope falling into the plausible range, it is unlikely that the weaker SFR-I and SFR-M correlations are able to give rise to a stronger correlation between the metallicity and ionization parameter, with the SFR as the main driver. To produce a stronger MI correlation, the residuals in SFR-I and the residuals in SFR-M have to be correlated, which would mean whatever factor that determines the residual is a more fundamental parameter than the SFR to setup the MI correlation. This proves that the SFR cannot be the dominant factor for establishing the correlation.

D14 also proposed another potential mechanism that modifies the ionization parameter through star formation. The geometry of a realistic \hii\ region could be highly nonspherical. If the fragmented molecular clouds within an \hii\ region can survive long enough and move close to the central OB stars due to turbulent motions, they would become highly ionized and raise the overall ionization parameter we observe. D14 found that for SF regions with lower cluster masses or higher pressures, it is possible to have a molecular cloud to cross an \hii\ region within the lifetime of the OB stars. Specifically, \hii\ regions older than 1 Myr with $\rm M_{cl} \lesssim 10^4~M_{\odot}$ and $\rm n_H \gtrsim 10^2~cm^3$ can have a shorter crossing time compared to the expansion time. However, in MaNGA we found that most of the SF regions have hydrogen densities close to 14 cm$^{-3}$ \citep{ji2020a}, which means the gas pressures are not high enough for the majority of our sample. We also found no correlation between the derived ionization parameter and the density (and pressure) sensitive ratio [S\,{\sc ii}]$\lambda 6716$/[S\,{\sc ii}]$\lambda 6731$ in our sample. Unfortunately, MaNGA does not have enough resolution power for investigations of turbulent motions on subkiloparsec scales. Thus, we cannot conclude observationally on the dependency of the turbulent motion on the SFR. The theoretical calculation of \cite{joung2009} does predict an increase in the gas velocity dispersion due to supernova explosions. 
Regardless, given the weakness of the SFR-I correlation, this solution is questionable.

A related question is the dependence of the various relations we consider here on stellar masses.
Since we see the ionization parameter positively correlates with both the metallicity and SFR, whether it contradicts the fundamental relation between the stellar mass, gas-phase metallicity, and SFR is worth investigating.
In the middle and right panel of Figure~\ref{fig:surfha_logu}, we plotted the spatially resolved mass-metallicity relations color coded by $\rm \log \Sigma_{H\alpha}$ and log(U), respectively. The stellar masses are drawn from the MaNGA {\sc pipe3d} Value Added Catalogue (VAC) \citep{sanchez2016}. The fundamental relation does not seem apparent in the local scale, which has also been noted by several works \citep[e.g.,][]{sanchez2013, sanchez2017, barrera-ballesteros2017}. At a fixed stellar mass surface density, increasing the SFR surface density does not significantly lower the gas-phase metallicity on average; however, there could still be a small effect as can be seen from the color gradient. On the other hand, $\rm \Sigma_{SFR}$ appears strongly correlated with $\rm \Sigma_{M_*}$. Meanwhile, the right panel of Figure~\ref{fig:surfha_logu} shows that at a fixed stellar mass surface density, there is still a strong positive MI correlation. We further checked the median MI relations at different stellar mass surface density bins and H$\alpha$ surface brightness bins in Figure~\ref{fig:mstarsfr_dep}. One can see that the median MI relations are nearly identical in different $\rm \Sigma_{M_*}$ bins. 
This indicates that the stellar mass surface density does not drive the MI correlation.

If we investigate the MI relations in different $\rm \Sigma_{H\alpha}$ bins  instead, we see the MI relations persist in most bins. However, as $\rm \Sigma_{H\alpha}$ increases, the relation becomes flattened and elevated, and nearly vanishes at the highest $\rm \Sigma_{H\alpha}$. This is consistent with what we see in the left panel of Figure~\ref{fig:surfha_logu}: there is a large range of possible $U$ at low $\rm \Sigma_{H\alpha}$ values, but a much narrower range of $U$ at high $\rm \Sigma_{H\alpha}$ values. The ionization parameter saturates at high $\rm \Sigma_{SFR}$ and becomes independent of metallicity. It is possible that at very high $U$, the radiation pressure becomes important and prevents $U$ from getting even larger \citep{yeh2012}.
Another potential explanation is related to the diffuse ionized gas (DIG) surrounding the \hii\ regions. Leaking radiation from the \hii\ regions could ionize the DIG, which exhibits low ionization states as a result of the diluted ionizing flux. If we limit the sample to only the low SFR regions, where the DIG contribution to the emission-line spectra becomes important, we see that the MI correlation also becomes stronger. To explain the trend, however, it requires the molecular clouds with low metallicities to be more leaky compared to the high metallicity clouds, which has not been observed to our best knowledge.
Finally, the measurement bias induced by the photoionization models might contribute to this effect as well, which we detail later in \S~\ref{subsec:geometric_bias}.

In summary, there indeed exists a positive SFR-I correlation in our sample. However, the coupling between the metallicity and ionization parameter appears stronger than that between the metallicity and SFR, or between the ionization parameter and SFR. In fact, the dependence on the SFR weakens the overall MI correlation, as shown in the right panel of Figure~\ref{fig:mstarsfr_dep}.
It is thus questionable whether the main driver of the MI relation is the SFR, while the SFR-I correlation could come from the influence of the cluster mass or the turbulent motions of the molecular clouds inside \hii\ regions. This is related to the details of the dynamical models for \hii\ regions, which is the topic of the next subsection.

\subsection{Dynamical evolution of HII regions}

\begin{figure}
    \includegraphics[width=0.48\textwidth]{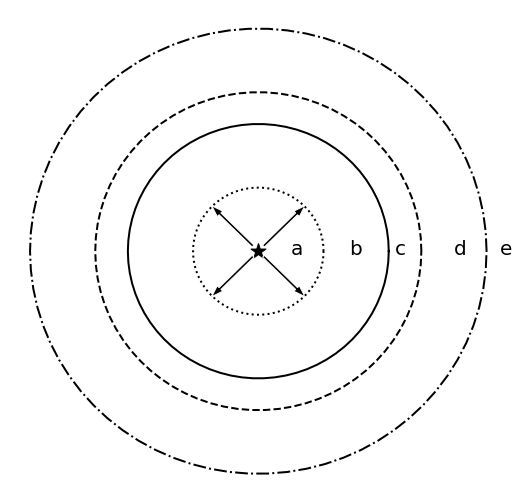}
    \caption{Schematic plot of a wind-driven bubble model. Region $a$ is the free-wind region filled with hypersonic stellar winds. Region $b$ is the shocked-wind region, which sets the inner boundary for the \hii\ region. Region $c$ is composed of shocked gas. Region $d$ is the \hii\ region. Depending on the ionization rate, region $d$ can be trapped within region $c$ or extend beyond it. Region $e$ is the ambient neutral medium. The relative sizes of the regions are not to scale.}
    \label{fig:windbubble}
\end{figure}

The structure of an \hii\ region is shaped by the feedback from the central young massive stars. The time evolution of the structure of the \hii\ region is described by the dynamical models, which in principle set the initial conditions for photoionization models. In this section, we discuss dynamical models with different assumptions and whether they lead to any correlation between the metallicity and ionization parameter.

Among the various feedback mechanisms considered in the modeling, stellar winds and photoionization are two main factors for shaping the geometry. Without stellar winds and any other source of mechanical energy, the radial structure of the ionized region is solely determined by photoionization, with the simplest case being the two-phase solution given by \cite{spitzer1978}. This solution describes a central star photoionizing a homogeneous cloud purely composed of hydrogen. During the first phase, the ionization front (IF) moves exponentially with a timescale of $\tau = 1/n_e \alpha _B$, where $n_e$ is the electron density and $\alpha _B$ is the recombination coefficient under the on-the-spot approximation. Once the IF reaches the Str\"omgren radius given by
\begin{equation}
    r_S = (\frac{3Q_0 }{4\pi n_e^2 \alpha _B })^{1/3},
    \label{eq:rs}
\end{equation}
the ionization rate equals the recombination rate and the second phase of expansion begins. A shock front is created before the IF and the compressed gas is moving toward the neutral ISM. The location of the IF  is given by
\begin{equation}
    r_i = r_S [1+\frac{7c_i (t-t_S )}{4r_S }]^{7/4},
\end{equation}
where $c_i$ is the sound speed in the ionized medium (typically $\sim 10~km/s$) and $t_S$ is the time when the IF reaches $r_S$. This solution is oversimplified and does not take important factors  into account, such as metals and dust. Meanwhile, we need to define the ionization parameter. Since it would be a function of radius, we need to specify a representative location or define an average to calculate it. One commonly adopted choice is the imagined ionization parameter at the Str\"omgren radius, which can be written as
\begin{equation}
    U_S = \frac{Q_0}{4\pi r_S^2 n_H c} \approx (\frac{Q_0 n_e \alpha _B^2}{36\pi c^3})^{1/3}.
\end{equation}
Clearly, we have $U_S = U_{VA}/3$, where $U_{VA}$ is the volume-average ionization parameter inside the Str\"omgren sphere. 
We note that this expression does not describe the actual ionization parameter measured at the Str\"omgren radius, as the number of ionizing photons decreases with increasing depth into the cloud.

Given that $\alpha _B \approx 2.56\times 10^{-13} T_4^{-0.83}~cm^3 s^{-1}$ \citep{draine2011}, we have
\begin{equation}
    U_S \propto Q_0^{1/3}n_e^{1/3}T^{-0.55}.
    \label{eq:us}
\end{equation}
If we now consider adding metals to the simple model, $T$  drops due to cooling. In the meantime, dust abundances increase with increasing metal abundances \citep{draine2011}, which would heat the gas  up due to photoelectric heating. Despite this competing effect, the net result of scaling the metallicity and dust abundances up is a drop in $T$, which causes an increase in $U_S$ defined above. However, we note that the actual radius within which the recombination happens would be smaller than $r_S$ given in Equation~\ref{eq:rs} since metals and dust grains also absorb part of the ionizing photons. Therefore, a better way to calculate $U_S$ when metals and dust grains are included is to simply define $r_S$ at the location where half of the hydrogen is ionized (i.e., $n(H+)/n(H)\sim 0.5$). The whole H\,{\sc ii} region would become more compact, which results in a larger $U_S$, as shown by \cite{haworth2015}. On the other hand, according to the {\sc starburst99} models, $Q_0$ provided by the ionizing stars drops as metallicity increases, provided that all models are normalized to the same SFR. 

We can check these dependencies with our measurements. Figure~\ref{fig:mstarsfr_dep} shows that log(U) roughly increases by 0.2 dex as [O/H] changes from 0 to 0.3 (solar to double-solar value). 
We generated a series of photoionization models corresponding to spherical \hii\ regions with no inner cavity and with $Q_0$ ranging from $10^{47.5}~s^{-1}$ to $10^{49}~s^{-1}$. The other conditions were set to be consistent with our fiducial model. For the model with the largest $Q_0$, the change in log($\rm U_S$) with increasing metallicity is comparable to what we see in the observed data\footnote{In fact, the ionization parameter measured by our fiducial model could be different from $U_S$ for thick spherical \hii\ regions. This effect associated with geometry is detailed in Section~\ref{subsec:geometric_bias}.}. However, there are two difficulties in explaining the observed MI correlation with this scenario. First, at each metallicity, $U_S$ returned by the model is nearly 6 times the median $U$ in the observed data. Lowering $Q_0$ can lower $U_S$, but it would flatten the MI relation significantly. Also, the ionizing luminosity of the central stars should decrease with increasing stellar metallicity, which further flattens the MI relation predicted by the model. Second, observations of nearby \hii\ regions show that the ionizing stars are separated from the bulk of the ionized clouds by hot diffuse gas \citep[e.g.,][]{pellegrini2007, gudel2008, pellegrini2011}. This is because stellar wind feedback from young massive stars creates shocked-wind bubbles within their birth clouds. If we include non-negligible inner cavities (with radii comparable to $r_S$) in \hii\ region models, the MI relation becomes much flatter. The strength of the stellar wind feedback, however, is also a function of metallicity. This adds more complexity to the dynamical modeling of \hii\ regions, as we detail in the following.

Once the stellar winds are included, the structure of the ISM around the central star cluster changes substantially. Figure~\ref{fig:windbubble} shows the structure of a spherical wind-blown cloud. A wind-driven bubble is created and gradually sweeps the outer H\,{\sc ii} region. Due to their high speed (typically $1500\sim 2500~km/s$ for O-type stars), stellar winds drive shocks into the H\,{\sc ii} region and produce a region with a very high temperature and low density (region $b$). For the dynamical models containing stellar winds, the geometry of the H\,{\sc ii} region is set by the inner (shock) radius, $r_{in}$, and the Str\"omgren radius. When the stellar winds are important, the relative thickness of the H\,{\sc ii} region is determined by the ionization parameter at $r_{in}$ and is generally small if $U < 10^{-2.5\sim -2}$ \citep[][but the exact value depends on assumptions of the dynamical model]{2006ApJ...647..244D}.
If the shocked-wind radius is small and/or the ionization parameter is high, the \hii\ region becomes thick. In this situation, the geometric dilution of the ionization photons would become important. Measuring the ionization parameter of such an \hii\ region with a photoionization model that has a mismatched geometry would introduce a bias.
We discuss a thin \hii\ region first, and investigate the geometric bias for measuring a thick \hii\ region in the following section.

\cite{weaver1977} gave an analytical solution for such a wind-driven bubble. In their model, the H\,{\sc ii} region is a thin isobaric shell.
The position of the shell is given by the adiabatic solution
\begin{equation}
    r_{shell} = (\frac{250}{308\pi })^{1/5}L^{1/5}_w \rho ^{-1/5} t^{3/5} \approx r_{in},
    \label{eq:wind}
\end{equation}
where $L_w = \frac{1}{2}\dot{M}v_w^2$ is the mechanical power of the stellar winds and $\rho $ is the density of the ambient medium, that is the H\,{\sc ii} region beyond the shock front. The pressure of the H\,{\sc ii} region is
\begin{equation}
    P = \frac{7}{(3850\pi )^{2/5}}L^{2/5}_w \rho ^{3/5} t^{-4/5}.
\end{equation}
Eliminating $t$ and using $\rho \propto n$, we have the following:
\begin{equation}
    P \propto (\frac{L_w}{n r_{in}^2})^{2/3}n = [4\pi cU(\frac{L_w}{Q_0})]^{2/3}n,
    \label{eq:wind_u}
\end{equation}
where we use $n$ to represent the hydrogen density of the H\,{\sc ii} region. The shell would keep expanding following Equation~\ref{eq:wind} until its internal pressure equals the ambient pressure (i.e., the stall condition). After that, the shell moves in a momentum-conserving way and is finally destroyed by turbulence \citep{oey1997, dopita2005}. Combining Equation~\ref{eq:wind_u} with $P = nkT$, we have
\begin{equation}
    U \propto (\frac{Q_0}{L_w}) T^{3/2}.
    \label{eq:wind_u0}
\end{equation}
This equation is a function of time. According to this equation, increasing the gas-phase metallicity would lower $T$, thus decreasing $U$. Meanwhile, increasing the stellar metallicity would generally lower the $Q_0 /L_w $ ratio as the stellar winds and the stellar atmosphere become more opaque. The combined effect is a decrease in $U$, which gives an anti-correlation between the ionization parameter and metallicity. For a single massive star, the mass loss rate $\dot{M}$ roughly scales with the stellar metallicity following $\dot{M} \propto Z^{0.5\sim 0.85}$ \citep{puls2008}. Based on the {\sc starburst99} models, D06 estimated the anti-correlation to roughly follow $U \propto Z^{-0.8}$. It is noteworthy that the mechanical energy from the stellar winds have to be manually scaled down by roughly 1 dex to reproduce the observed range of $U$ \citep{dopita2005}. The same discrepancy has been noted by \cite{naze2001} and \cite{harper2009} when comparing the observed sizes and expanding velocities of the interstellar bubbles with theoretical predictions. Hence it is likely that there are mechanisms in reality that lower the efficiency of the power output from stars, which could be related to the fact that the stellar winds are clumpy \citep{evans2004,bouret2005,fullerton2006,puls2006, puls2008}. In addition to the inefficiency of stellar winds, \cite{harper2009} suggested that the leakage of the hot shocked gas from \hii\ regions could also lower the radius and increase the ionization parameter by lowering the internal pressure. Whether and how these hyper parameters depend on the metallicity is unknown and requires further investigation.

The above adiabatic solution is based on the assumption that the shocked stellar winds do not have enough time to cool. However, as shown by \cite{maclow1988}, the cooling time of the wind bubble is typically short in dense molecular clouds, with $t_{cool}\lesssim 10^4 yr$. In such cases, the pressure directly comes from the momentum of the winds,
\begin{equation}
    P = \frac{\dot{p}_w}{4\pi r_{in}^2} = \frac{(2\dot{M}L_w )^{1/2}}{4\pi r_{in}^2}.
\end{equation}
Where $\dot{p}_w$ is the force imposed by the wind momentum and $\dot{M}$ is the mass loss rate of the central star. Equating the above equation with $P=nkT$, we obtained
\begin{equation}
    U = \frac{k}{2^{1/2}c}\frac{Q_0}{(\dot{M}L_w)^{1/2}}T.
\end{equation}
If the metallicity of the cloud is increased, $T$ and $Q_0$ would decrease, while $(\dot{M}L_w)^{1/2}$ would increase. Therefore, the model still predicts an anti-correlation between the ionization parameter and metallicity, despite being in a different form.

Thus far, we have discussed dynamical models with and without stellar winds. A related question is whether the stellar winds are important in H\,{\sc ii} regions in general. \cite{geen2020} studied this problem and conclude that dynamically speaking, stellar winds do not play an important role in most H\,{\sc ii} regions. However, stellar winds do shape the geometry of the H\,{\sc ii} regions even when they are dynamically unimportant. In other words, a considerably large shocked wind bubble can exist when the location of the IF is barely affected by the presence of stellar winds.

We have seen that none of these dynamical models are able to explain the positive correlation we found between $U$ and $Z$. The complicating factor in this problem is that $U$ is not a direct observable and it has to rely on certain assumptions of the geometry of the \hii\ regions. In reality, the geometries of \hii\ regions are complicated and can be far from spherical, and $U$ also varies with locations inside \hii\ regions. It is also unclear how stellar winds shape the shocked bubble when the ambient medium is highly asymmetric and inhomogeneous. The mysterious inefficiency of stellar winds in the wind-driven bubble model could be the key to resolving the problem, which we intend to investigate in future work. Quite surprisingly, despite the complicated picture shown here, the resulting correlation between $U$ and $Z$ is relatively clear and strong. This raises the question about the robustness and reliability of our derivations of these quantities based on photoionization models. In the following section, we discuss this point in detail.

\subsection{Geometric bias}
\label{subsec:geometric_bias}

\begin{figure}
    \includegraphics[width=0.48\textwidth]{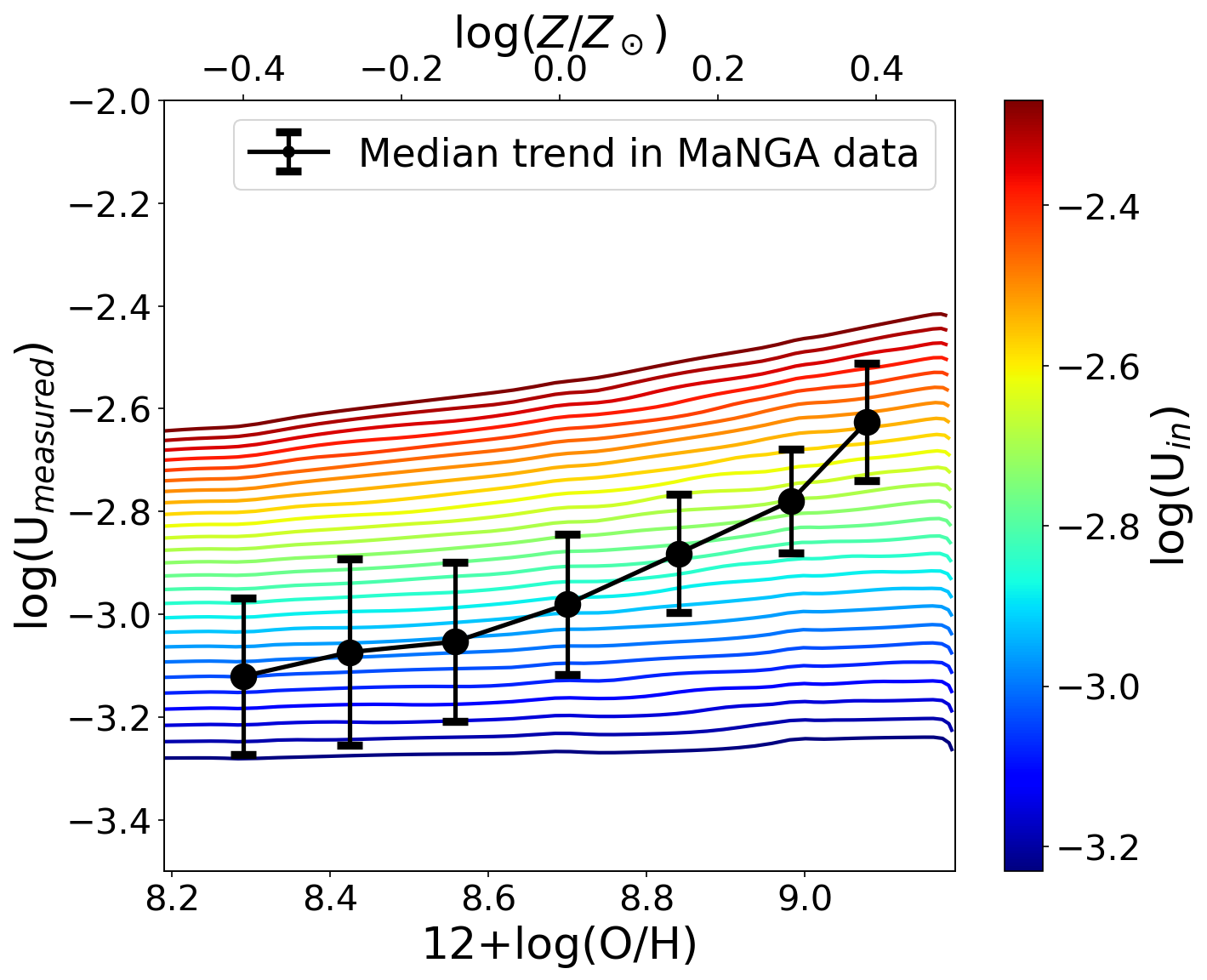}
    \caption{Measured ionization parameters and metallicities for a spherically ionized cloud of which the inner radius varies with the metallicity accoridng to Equation~\ref{eq:inner_r}. Each colored line corresponds to \hii\ regions with the same ionization parameter at their inner radii. The black line shows the median trend we measured in the MaNGA data, with the error bars indicating the standard derivations of log(U) in individual metallicity bins. The measurements were performed using N2, S2, R3, N2O2, and O3O2 line ratios.}
    \label{fig:geo_bias}
\end{figure}

\begin{figure}
    \includegraphics[width=0.49\textwidth]{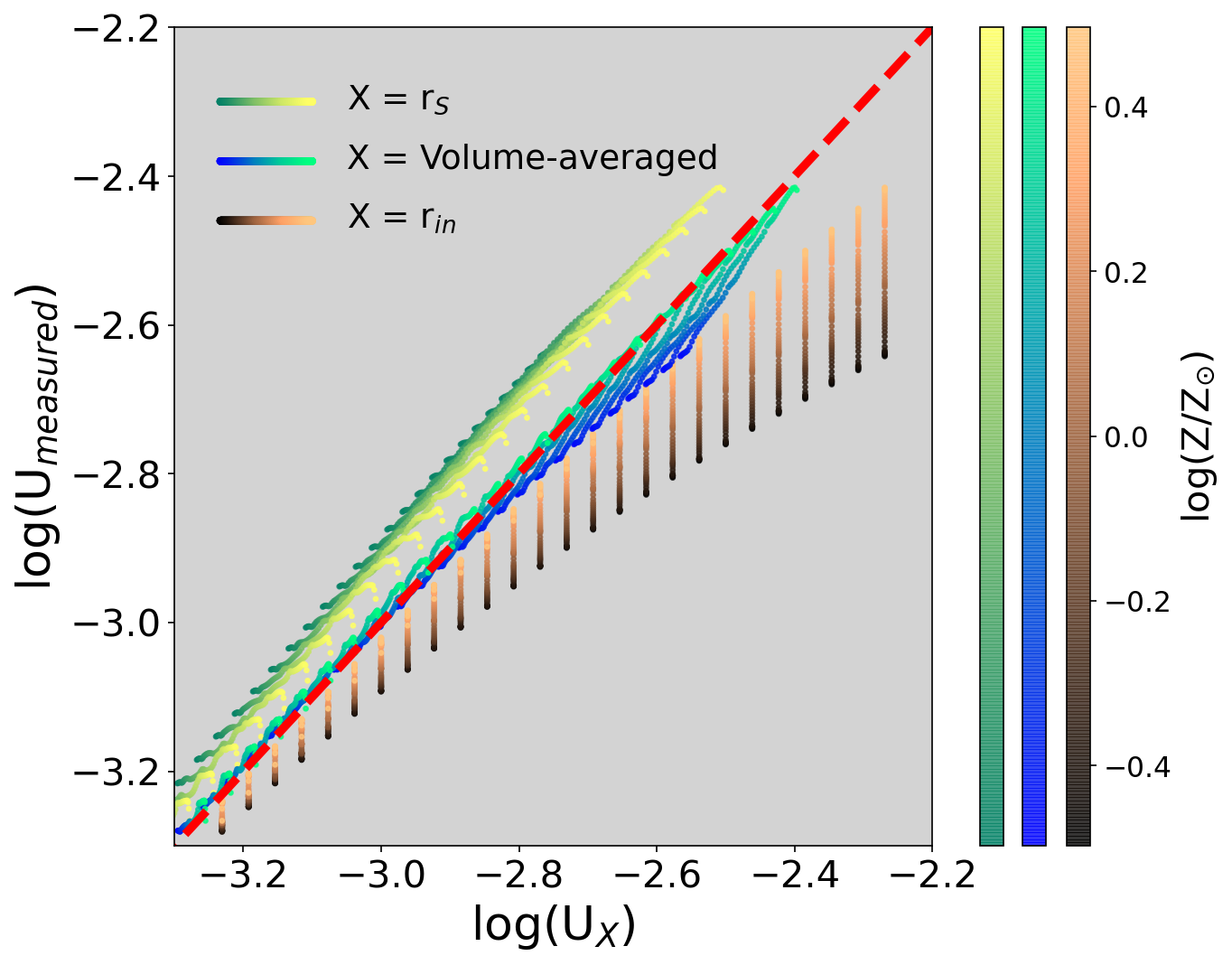}
    \caption{Comparisons between the ionization parameters measured by our plane-parallel fiducial model ($U_{measured}$) and the ionization parameters under different definitions for a spherically ionized cloud ($U_X$). The inner radius of the cloud was set to vary with metallicity according to Equation~\ref{eq:inner_r}. The ionization parameters of the spherical cloud are defined using the value at the inner radius ($r_{in}$), the value at the Str\"omgren radius ($r_S$), and the volume-averaged value, respectively.} The dashed red line is a diagonal line for reference.
    \label{fig:usinva}
\end{figure}

When measuring the ionization parameter of a given \hii\ region, one needs to define at which location the ionization parameter is measured. In addition to choosing the imagined $U$ at the Str\"omgren radius, choosing the $U$ at the inner radius is common and easy to carry out in photoionization modeling. Thus, this definition has been adopted by many photoionization models \citep[e.g.,][]{levesque2010,dopita2013,byler2017}. However, the definition of the ionization parameter inevitably introduces an uncertainty associated with the actual geometry of the \hii\ region, or more precisely, the relative thickness of the \hii\ region (i.e., $r_{in}/r_{i}$).

For a spherical \hii\ region, the ionizing flux decreases more rapidly with radius due to the geometric dilution. Hence, the spherical cloud would have a thinner ionized layer compared to a plane-parallel one with the same inner $U$. Their ionization structures are also different. Roughly speaking, the average location at which we measure the average line ratios would exhibit a lower $U$ for a spherical model since the ionizing flux dilutes more.
In other words, suppose we measure a spherical \hii\ region with a photoionization model with a plane-parallel geometry, the measured ionization parameter would be lower compared to the ionization parameter defined at the inner radius. Similarly, if we use a spherical model with a larger inner radius to measure a spherical \hii\ region with a smaller radius, the resulting $U$ would also be underestimated.

The amount of the bias depends on the relative thickness of the \hii\ region. If $r_{in}/r_{i}$ is small, the \hii\ region is thick and the effect of geometric dilution becomes important. Both the ionization parameter and metallicity affect the thickness of the \hii\ region. For a higher ionization parameter, there is a larger number of ionizing photons and thus the ionized layer becomes thicker. A higher metallicity, on the other hand, would have two effects. First of all, $r_{i}$ deceases due to both a lower equilibrium temperature and a reduction in the number of ionizing photons, which reduces the thickness of the \hii\ region. In addition, according to the wind-driven
bubble model by \cite{weaver1977}, the inner radius would increase as a result of stronger stellar winds (see Equation~\ref{eq:wind}). If the inner $U$ is held fixed, there would be less geometric dilution as the \hii\ region becomes more plane-parallel-like. The combined effect is a reduced underestimation of the inner $U$ for higher metallicities. Therefore, a spurious MI correlation may be induced by the geometric bias, as the ionization parameters of higher metallicity \hii\ regions are less underestimated. The final slope of the MI relation depends on both the slope of the intrinsic relation and the amount of the geometric bias. It might also explain the saturation of the MI relation at a high SFR. If a cluster has a higher SFR and thus holds more stars, both the number of ionizing photons and the power of stellar winds increase, but their ratio remains roughly the same. This keeps the inner $U$ unchanged (see Equation~\ref{eq:wind_u0}), but increases the inner radius. If the SFR is high enough, the metallicity dependence of the geometric bias would become negligible as the geometry becomes closer to plane-parallel overall, producing an apparent saturation of $U$. Still this depends on whether the intrinsic MI relation is flat or not, which is the weak point of this argument as we discuss shortly.

Figure~\ref{fig:geo_bias} shows an example of the geometric bias. Here we measure the metallicities and ionization parameters of a series of spherical \hii\ region models with the inner radii given by
\begin{equation}
    r_{in} = r_0~(Z/Z_{\odot})^{0.17} = 10~(Z/Z_{\odot})^{0.17}~pc,
    \label{eq:inner_r}
\end{equation}
with our best-fit plane-parallel model. We used Equation~\ref{eq:wind} and assumed the most extreme scaling relation for the mass loss rate, that is, $\dot{M}\propto (Z/Z_{\odot})^{0.85}$ \citep{puls2008}. We chose the normalization to be $r_0 = 10~pc$, which is typical for a late-stage wind-driven bubble with $\rm t_{age} \gtrsim 1~Myr$ \citep[e.g.,][]{dopita2005, dale2014, geen2020}. 
We treated the predicted line ratios from the spherical models as observations and added uncertainties typically found in MaNGA data to the line ratios.
As expected, for \hii\ regions with low ionization parameters at the inner radii, the measured $U$ is closer to $U_{in}$ and it barely depends on the metallicity. On the other hand, for \hii\ regions with high $U_{in}$, there are positive MI correlations caused by the geometric dilution. Still the bias-induced correlation is insufficient to explain the MI correlation we measured in the MaNGA data, assuming the intrinsic MI relation is flat. Scaling down $r_0$ would steepen the slopes of the constant $U_{in}$ lines, but even after reducing $r_0$ to 3 pc (which is too small for common \hii\ regions with densities of $\rm 10\sim 100~cm^{-3}$), we still cannot explain the median trend in the MaNGA data with a single constant $U_{in}$ line. Making the metallicity dependence of the inner radius stronger would also yield a steeper slope, but it would require the power-law index in Equation~\ref{eq:inner_r} to be on the order of 1, which is not evident in any dynamical model. Furthermore, once we consider the inner radius as a function of the metallicity, the intrinsic MI correlation is unlikely to be flat anymore. If we follow D06's argument and use $U_{in} \propto (Z/Z_{\odot})^{-0.8}$, it is even harder to reconcile with the results from the MaNGA data.

To summarize, the geometric bias could potentially influence the MI trend observed at high ionization parameters.\ However, it is insufficient to explain the whole MI relation.

One might wonder if we can use a better definition of $U$ to avoid this geometric bias. In Figure~\ref{fig:usinva} we compare the measured ionization parameters of the same set of spherical models described by Equation~\ref{eq:inner_r} (using our fiducial model) with the ionization parameters computed at their inner radii ($U_{in}$), at their Str\"omgren radii ($U_{S}$), and using the volume-averaged values ($U_{VA}$), respectively. At nearly all input ionization parameters and metallicities, we have $U_{S} < U_{measured} \approx U_{VA} < U_{in}$. The variation in $U_{measured}$ (due to metallicity variation) at fixed $U_{S}$ or fixed $U_{VA}$ is much smaller than that at fixed $U_{in}$. Therefore, the geometric bias would be much weaker if we replace $U_{in}$ with $U_{S}$ or $U_{VA}$. This simple test shows that our plane-parallel fiducial model is actually roughly measuring the volume-averaged ionization parameter of these spherical models, although we have defined its ionization parameter at the inner surface of the ionized cloud. Even so, the definition of $U_{in}$ has its own advantage. First of all, it is easy to set $U_{in}$ as the input parameter for a photoionization model, while $U_{S}$ and $U_{VA}$ have to be calculated after the model is computed\footnote{\cite{stasinska2015} suggest defining a shape factor, $f_S \equiv r_{in} /r_S$, and using it to calculate $U_{VA}$ as input for photoionization models. However, this approach still relies on the assumption that metals and dust grains do not absorb a significant number of hydrogen ionizing photons.}. In addition, the dynamical models we compare also explicitly define the ionization parameter at the inner radii of \hii\ regions \citep[see][]{dopita2005, 2006ApJ...647..244D}.
For the purpose of comparison, it is more straightforward to use the same definition.
Furthermore, our test shows that the bias in the MI relation is not significant even when we consider an extreme scaling relation for the size of the wind-driven bubble in \hii\ regions. Future works with more sophisticated combinations of dynamical models and photoionization models of \hii\ regions will be helpful to provide a clearer picture on which definition is most useful.

We have seen that there is no simple physical picture explaining the entire MI relation in our data. However, we have not investigated the uncertainties associated with the self-consistency of the photoionization model, the method to derive the parameters, and the sample so far. We explore these points in the next section.

\section{Discussions}
\label{sec:discuss}

As we have seen in Figure~\ref{fig:fitting}, the derived correlation between the ionization parameter and metallicity strongly depends on the choice of photoionization models. The idea behind our derivations is that there exists a best-fit photoionization model surface sitting in the center of the data distribution in the high dimensional line-ratio space. This implicitly assumes that the ionization parameter and metallicity are two primary parameters that determine a 2D manifold embedded in high dimensions, and observed data tend to cluster around this manifold. The secondary parameters, on the other hand, contribute to the scatters around the manifold. Starting from this point, a few questions have yet to be answered.

First, it remains to be checked whether our best-fit model is still good enough if we include more emission lines. We already see that our model outperforms others in a 3D line-ratio space. In higher dimensions, there are more constraints on model parameters. Models that appear to fit the data in lower dimensions do not necessarily work in higher dimensions, which we have already seen in the case of 2D diagnostic diagrams. Second, given the existence of the intrinsic scatters induced by the secondary parameters, it is important to check whether our Bayesian approach is the right choice to capture the true distributions of the primary parameters. In many similar practices, people have invoked priors in their derivations. We need to carefully interpret the effect of priors. Finally, the choice of the sample certainly affects the derivations as it could change the mean values of the secondary parameters. Our best-fit model is based on MaNGA data. 
We need to check whether the MI correlation is still the same if it is derived from other samples.
In what follows, we investigate the above points.

\subsection{Offset between the observed O2 and model predictions}
\label{subsec:o2_corr}

\begin{figure*}
    \includegraphics[width=0.95\textwidth]{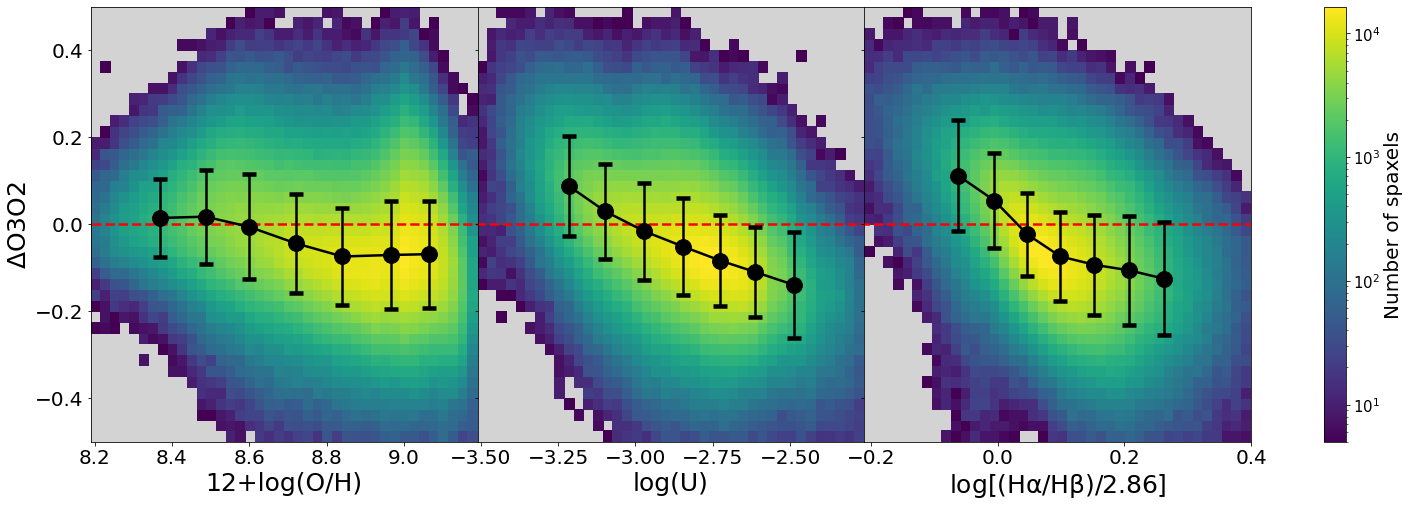}
    \caption{$\rm O3O2_{data} - O3O2_{model}$ as a function of the gas-phase metallicity (left panel), ionization parameter (middle panel), and Balmer decrement (right panel). The median trend in each panel is shown as the black line, with the standard deviations indicated by the error bars.}
    \label{fig:o2_corr}
\end{figure*}

In Figure~\ref{fig:hist}, one can see that the ionization parameters predicted by the N2-S2-R3 method are systematically higher than those predicted by the N2O2-O3O2 method. This implies that the inclusion of the [O\,{\sc ii}] doublet no longer makes our model optimal in describing the data distribution. To check how much the model surface is offset from the data along the axes containing [O\,{\sc ii}] in high dimensions, we can define the logarithmic emission-line ratio difference as
\begin{equation}
    \rm \Delta \log ER = \log ER_{data} - \log ER_{model} [(O/H)_{3D}, U_{3D}].
\end{equation}
For a given data point, $\rm ER_{data}$ is its extinction-corrected emission-line ratio (involving [O\,{\sc ii}]), and $\rm ER_{model} [(O/H)_{3D}, U_{3D}]$ is the same emission-line ratio predicted by the model with the metallicity and ionization parameter given by the N2-S2-R3 method. The left and middle panels of Figure~\ref{fig:o2_corr} show the O3O2 difference as a function of $\rm {12+log(O/H)}_{3D}$ and $\rm U_{3D}$.
The median difference in O3O2 reaches $\sim 0.1$ dex at high metallicities and ionization parameters.
In comparison, the median O3O2 (not the difference, but the value itself) in our sample changes roughly by 0.4 dex as 12+log(O/H) changes from 8.3 to 9.1.
There is a clear trend of decreasing O3O2 as $\rm U_{3D}$ increases, indicating a systematic bias in the model-predicted O3O2. At high ionization parameters and high metallicities, our model tends to overestimate O3O2, or underestimate O2.

However, there is a caveat for this interpretation. The O3O2 used here has an extra uncertainty from the extinction correction. It is observed that the overall Balmer decrement (and thus the extinction) is positively correlated with the metallicity. Therefore, it is possible that the extinction correction is increasingly overestimated as the metallicity (and thus the ionization parameter) increases. The right panel of Figure~\ref{fig:o2_corr} shows how the O3O2 difference changes as the Balmer decrement increases. One can see a similar relation as shown in the middle panel. The extinction correction we used is based on Balmer decrements and a \cite{fitzpatrick1999} extinction curve with $\rm R_V = 3.1$ being adopted. If we instead use a \cite{cardelli1989} extinction curve, the trend still remains. Interestingly, the absolute O3O2 difference would become smaller if we do not apply the extinction correction (but this time the difference is positive), which again implies that the extinction correction is overestimated.

In summary, there could be a systematic offset in both the model predictions concerning O2 and the estimation of the extinction based on Balmer decrements.
Although the relative offset in O3O2 is not significant, it is correlated with the predicted ionization parameter and the Balmer decrements. We investigate the model predicted O2 and the potential systematic uncertainties associated with the extinction correction in detail in a future paper.

It is unsurprising that the model already becomes "imperfect" in a 4D space. Although photoionization models work well in predicting a large number of emission lines for individual well-observed \hii\ regions \citep[e.g.,][]{baldwin1991, jamet2005, morisset2013}, finding a model that reproduces all strong emission lines for the general population of \hii\ regions is notoriously difficult (see e.g., Section 6 of \citealp{law2021b}, Section 3.2 of \citealp{mingozzi2020}, and references therein). As a compromise, when using photoionization models with the Bayesian inference, people usually assume some nonflat priors for metallicities or ionization parameters, and adding terms representing the intrinsic scatters not reflected by the models \citep[e.g.,][]{perezmontero2014, blanc2015, mingozzi2020}. In the next section, we investigate the effects of these methods and their interpretations.

\subsection{Nonflat priors and intrinsic scatters}

The introduction of nonflat priors changes the form of the posteriors. Using the likelihood we introduced in Equation~\ref{eq:likelihood}, the posterior is given by
\begin{equation}
    p(\theta |D,M) = \frac{p(\theta |M)p(D|M,\theta )}{p(D|M)},
\end{equation}
where $p(\theta |M)$ is the prior and $p(D|M)$ is the normalization factor. Priors come in different forms, which usually assume the distribution of the parameter to follow a certain form within a given range, or correlate with some observables. \cite{mingozzi2020}, for example, used [S\,{\sc iii}]$\lambda \lambda 9068, 9532$/[S\,{\sc ii}]$\lambda \lambda 6716, 6731$ to set a prior for the ionization parameter. Photoionization models predict a very tight correlation between [S\,{\sc iii}]/[S\,{\sc ii}] and the ionization parameter, with little dependence on the metallicity \citep{2019ARA&A..57..511K}. \cite{mingozzi2020} adopted the relation given by \cite{diaz1991} and found that the smaller peak in the metallicity distribution predicted by the D13 model vanishes (see the lower left panel of Figure~\ref{fig:hist}). Meanwhile, the correlation between the ionization parameter and metallicity seemed to disappear at fixed stellar masses (although the overall correlation still exists). Despite the seeming advantage of using such a prior for "correcting" the model predictions, one caveat needs to be taken into account. Most of the current photoionization models (including ours) fail to predict the observed range of the [S\,{\sc iii}]/[S\,{\sc ii}] ratios. Under this circumstance, using the correlations between [S\,{\sc iii}]/[S\,{\sc ii}] and $U$ set by photoionization models themselves as priors can introduce a bias. There is also a risk concerning self-consistency if the relation derived from a model is applied to correct the predictions of another model with different model assumptions. In this work, we have seen that the double-peak features in the metallicity distribution are likely a result of the deviation of the model surface from the dense region of the data distribution in the line-ratio space. The effect of priors is to regulate the mappings between the model parameters and the data positions, which can also be achieved by adjusting the secondary model parameters and make the model surface match the data. We think the latter approach is more physically motivated. Even so, it would be interesting to see how the photoionization models for \hii\ regions can be improved to correctly reproduce observed [S\,{\sc iii}]/[S\,{\sc ii}] and update the ionization parameter calibration, which we leave for future investigations.

Another potentially useful treatment in the Bayesian inference is the inclusion of model uncertainties. As we have mentioned, the variations in the secondary model parameters manifest themselves as intrinsic scatters in the data distribution. Since each photoionization model uses a single set of secondary model parameters, including terms of constant uncertainties for the model-predicted line ratios might help improve the fitting \citep{blanc2015}. Indeed, if we add a systematic uncertainty of 0.1 dex for each line ratio, the agreement between the N2-S2-R3 method and the N2O2-O3O2 method becomes slightly better, and their predicted distributions for the ionization parameter are more consistent. This might be because the 2D method is more susceptible to intrinsic scatters as there are fewer degrees of freedom. Whereas we need to be cautious that the manually added uncertainties could hide the true discrepancy between different line ratios. An alternative way is to include some of the secondary parameters in the modeling as free parameters. In principle, this would reduce the impact of intrinsic scatters. However, more degeneracy would occur as more free parameters are considered. It would be necessary to add more emission lines and go to higher dimensions, which in turn requires models to accurately simulate more lines. No doubt such a self-consistent treatment is complicated and may be computationally expensive, but it might be worth doing if one wants to understand subtle correlations between different parameters.

\subsection{Sample selection effect}

\begin{figure}
    \includegraphics[width=0.48\textwidth]{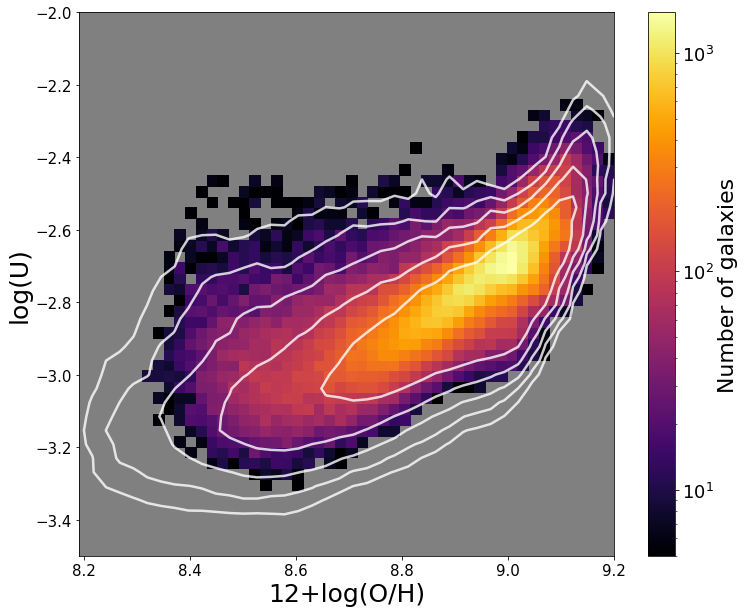}
    \caption{ MI correlation derived from the SF galaxies in the DR7 of SDSS. The counters represent the density distribution of the MaNGA data, with five density levels equally spaced in the logarithmic space from the $\rm 16^{th}$ percentile to the $\rm 84^{th}$ percentile.}
    \label{fig:sdss}
\end{figure}

Our derived positive correlation between the metallicity and ionization parameter is entirely based on the spatially resolved MaNGA \hii\ regions. In this section, we discuss the impact of the sample selection on our result.

MaNGA's sample galaxies have a relatively flat mass distribution (see \S~\ref{sec:data_models}), and this explains why a large fraction of our sample spaxels show high metallicities. The data we used are spatially resolved spaxels. For individual spaxels, the measured spectra are light-weighted, while the density distribution of the spaxels are heavily influenced by the on-sky areas of the \hii\ regions in galaxies. This makes the MaNGA data more weighted toward the outer SF regions in galaxies compared to single fiber spectroscopic data. \cite{2019ARA&A..57..511K} suspect that the MI correlation depends on the scales of observations, with the single fiber spectroscopic data yielding anti-correlations as expected by the wind-driven bubble models, while the spatially resolved data give positive or no correlations. To check this possibility, we applied our method to the SF galaxies in the DR7 of SDSS \citep{2000AJ....120.1579Y, 2009ApJS..182..543A}. We obtained the emission line measurements using the updated code of \cite{yan2006} with flux calibrations and zero-point corrections to the equivalent widths for emission lines \citep{yan2011,yan2018}. The result is shown in Figure~\ref{fig:sdss}. Clearly, there is a similar positive trend similar to what we see in the spatially resolved case. We thereby conclude that the discrepant MI correlations seen in previous works are not due to the difference between single-fiber observations and IFS observations or the difference between the centers of galaxies and outskirts of galaxies. They are more likely a result of different metallicity and ionization parameter calibrators.

Even so, the spatial resolution of MaNGA is not enough to sample individual \hii\ regions. The full width at half maximum (FWHM) of MaNGA's point spread function (PSF) is roughly $2^{\prime \prime}_. 5$ \citep{law2015}, which corresponds to $1\sim 2$ kpc at the typical redshifts of MaNGA galaxies. As suggested by \cite{sanders2020} and \cite{mannucci2021}, the observed line ratios from \hii\ regions could change considerably if individual \hii\ regions are resolved. There are two effects involved. First, the emission from the diffuse ionized gas (DIG) around \hii\ regions can be blended with that from within \hii\ regions. The DIG emission has enhanced low-ionization lines and can impact the determination of metallicities \citep{zhang2017}. The DIG impact can largely be removed by applying a cut to the equivalent width of the H$\alpha$ line, the surface brightness of the H$\alpha$ line, or the fractional contribution to the total luminosity by young stars \cite[e.g.,][]{2011MNRAS.413.1687C, sanchez2014, zhang2017, espinosa-ponce2020}. The majority of our sample spaxels have a large enough EW(H$\alpha$). Thus, the DIG is not likely to be dominant in our case. Second, within a single \hii\ region, the emission line spectrum changes with the aperture size as well. The inner region would be featured by stronger high ionization lines but weaker low ionization lines. How exactly the spectrum changes would depend on the geometry of the \hii\ region. The effective ionization parameter, being sensitive to geometry, would also change with the aperture size. The derived ionization parameter of the entire \hii\ region should thus be related to its internal structure. As we have mentioned, the ionization parameter is not a direct observable most of the time and it is not well defined if the geometry of the \hii\ region is unclear. Therefore, it is vital to have detailed analyses on well-resolved nearby \hii\ regions to understand how well the geometry is reflected by the emission line measurements, which could be addressed by future surveys such as the Affordable Multi-Aperture Spectroscopy Explorer \citep[AMASE,][]{yan2020} and SDSS-V/Local Volume Mapper \citep[LVM,][]{kollmeier2017}.

Last but not least, the redshift of the \hii\ region sample could also impact the result. MaNGA's sample is mainly composed of low-redshift galaxies (0.01 < $z$ < 0.14). \cite{sanders2016} and \cite{sanders2020} found no evidence of evolution in the IM relation for their high redshift sample from the MOSDEF survey \citep{kriek2015}, but the range of the MI relation they could measure is limited. Star-forming galaxies at higher redshifts form an offset sequence in the line-ratio space compared to the local ones \citep[e.g.,][]{shapley2005, erb2006,steidel2014,shapley2015}. They could have higher electron densities \citep{kewley2013,sanders2016}, elevated N/O versus O/H relations \citep{shapley2015, sanders2016}, or harder ionizing spectra \citep{steidel2014}, which would generally result in larger N2, S2, and R3 line ratios. The density effect on line ratios of \hii\ regions is small compared to the changing SED effect \citep{ji2020b}. To fit a high redshift sample, one needs to consider the average changes in all secondary parameters and the potential model degeneracy associated with the changes. The accuracy of the model predictions depends on how well these secondary parameters can be constrained. This requires a better understanding of the mappings between different secondary parameters and observed line ratios in \hii\ regions. Precise and self-consistent photoionization modeling as well as carefully chosen visualizations of data distribution in multidimensional line-ratio space would be the key to solving the problem.

\section{Conclusions}
\label{sec:conclude}

In this work we study the correlation between the gas-phase metallicity and ionization parameter (MI correlation) in general \hii\ regions and investigate its physical origin. We measured the gas-phase metallicities and ionization parameters for \hii\ regions in MaNGA data released in SDSS DR15 using four different photoionization models (L10, D13, B17, and JY20). We performed the measurements with Bayesian inference and calculated the weighted average values assuming flat priors. A total of five emission line ratios were used for the measurements, including N2, S2, R3, N2O2, and O3O2. We compared the results from different models and different combinations of line ratios. Our conclusions are summarized as the following.
\begin{enumerate}
    \item When compared with the data distribution in the 3D line-ratio space spanned by N2, S2, and R3, our updated model (JY20 model) provides a better fit to the location and the shape of the central surface of the data.
    \item The goodness of fit in the 3D space can reflect the consistency of the model predictions over different line ratios.
    This is confirmed by the measurements of metallicities and ionization parameters using different subsets of line ratios. For each model, we carried out three sets of measurements using two line ratios (N2O2 and O3O2), three line ratios (N2, S2, and R3), and five line ratios (N2O2, O3O2, N2, S2, and R3), respectively.
    Compared to the other three models considered in this work (L10, D13, and B17), our model provides the most consistent results no matter which set of line ratios is used.
    \item With our best-fit photoionization model, we found a positive MI correlation, which is in contrast to the prediction of a wind-driven bubble model frequently used for describing the dynamical evolution of \hii\ regions.
    \item In order to solve this discrepancy, we investigated the potential influence from star formation. We found that star formation activities alone are not enough to explain the MI correlation. Our results indicate that the variation in the SFR actually contributes to the scatter in the MI correlation.
    \item We also study the bias brought by the assumed geometry in the photoionization model. Since the relative thickness of a spherical \hii\ region would change with the metallicity due to changes in temperature and stellar wind power, measuring it using a photoionization model with a fixed geometry introduces a bias correlated with the metallicity. We found that this geometric bias can result in a slight positive MI correlation, but it is still not enough to fully explain the MI correlation in the MaNGA data.
\end{enumerate}

As a concluding remark, the discrepancy between the measurements from photoionization models and the prediction from the dynamical model is still an open question. Both a better photoionization model that could self-consistently reproduce more emission line ratios, and a better dynamical model that could accurately describe the impact from stellar wind as well as other feedback mechanisms are vital for future studies. Equally important is the detailed analysis of the structures of the nearby \hii\ regions. A future data release from the Affordable Multi-Aperture Spectroscopy Explorer \citep[AMASE,][]{yan2020} and SDSS-V/Local Volume Mapper \citep[LVM,][]{kollmeier2017} would be useful for resolving this issue.

\begin{acknowledgements}
We thank the anonymous referee, whose thoughtful suggestions improved the clarity of this work.
We acknowledge support by NSF AST-1715898 and NASA grant 80NSSC20K0436 subaward S000353.
RY acknowledges support by the Hong Kong Global STEM Scholar scheme and the Direct Grant of CUHK Faculty of Science.
Funding for the Sloan Digital Sky Survey IV has been provided by the Alfred P. Sloan Foundation, the U.S. Department of Energy Office of Science, and the Participating Institutions. SDSS acknowledges support and resources from the Center for High-Performance Computing at the University of Utah. The SDSS web site is www.sdss.org.

SDSS is managed by the Astrophysical Research Consortium for the Participating Institutions of the SDSS Collaboration including the Brazilian Participation Group, the Carnegie Institution for Science, Carnegie Mellon University, the Chilean Participation Group, the French Participation Group, Harvard-Smithsonian Center for Astrophysics, Instituto de Astrof\'isica de Canarias, The Johns Hopkins University, Kavli Institute for the Physics and Mathematics of the Universe (IPMU) / University of Tokyo, the Korean Participation Group, Lawrence Berkeley National Laboratory, Leibniz Institut f\"ur Astrophysik Potsdam (AIP), Max-Planck-Institut f\"ur Astronomie (MPIA Heidelberg), Max-Planck-Institut f\"ur Astrophysik (MPA Garching), Max-Planck-Institut f\"ur Extraterrestrische Physik (MPE), National Astronomical Observatories of China, New Mexico State University, New York University, University of Notre Dame, Observat\'orio Nacional / MCTI, The Ohio State University, Pennsylvania State University, Shanghai Astronomical Observatory, United Kingdom Participation Group, Universidad Nacional Aut\'onoma de M\'exico, University of Arizona, University of Colorado Boulder, University of Oxford, University of Portsmouth, University of Utah, University of Virginia, University of Washington, University of Wisconsin, Vanderbilt University, and Yale University.
\end{acknowledgements}

%
   \bibliographystyle{aa} 
   \bibliography{ref} 

\begin{thebibliography}{140}
\expandafter\ifx\csname natexlab\endcsname\relax\def\natexlab#1{#1}\fi

\bibitem[{{Abazajian} {et~al.}(2009){Abazajian}, {Adelman-McCarthy},
  {Ag{\"u}eros}, {Allam}, {Allende Prieto}, {An}, {Anderson}, {Anderson},
  {Annis}, {Bahcall}, {Bailer-Jones}, {Barentine}, {Bassett}, {Becker},
  {Beers}, {Bell}, {Belokurov}, {Berlind}, {Berman}, {Bernardi}, {Bickerton},
  {Bizyaev}, {Blakeslee}, {Blanton}, {Bochanski}, {Boroski}, {Brewington},
  {Brinchmann}, {Brinkmann}, {Brunner}, {Budav{\'a}ri}, {Carey}, {Carliles},
  {Carr}, {Castander}, {Cinabro}, {Connolly}, {Csabai}, {Cunha}, {Czarapata},
  {Davenport}, {de Haas}, {Dilday}, {Doi}, {Eisenstein}, {Evans}, {Evans},
  {Fan}, {Friedman}, {Frieman}, {Fukugita}, {G{\"a}nsicke}, {Gates},
  {Gillespie}, {Gilmore}, {Gonzalez}, {Gonzalez}, {Grebel}, {Gunn},
  {Gy{\"o}ry}, {Hall}, {Harding}, {Harris}, {Harvanek}, {Hawley}, {Hayes},
  {Heckman}, {Hendry}, {Hennessy}, {Hindsley}, {Hoblitt}, {Hogan}, {Hogg},
  {Holtzman}, {Hyde}, {Ichikawa}, {Ichikawa}, {Im}, {Ivezi{\'c}}, {Jester},
  {Jiang}, {Johnson}, {Jorgensen}, {Juri{\'c}}, {Kent}, {Kessler}, {Kleinman},
  {Knapp}, {Konishi}, {Kron}, {Krzesinski}, {Kuropatkin}, {Lampeitl},
  {Lebedeva}, {Lee}, {Lee}, {French Leger}, {L{\'e}pine}, {Li}, {Lima}, {Lin},
  {Long}, {Loomis}, {Loveday}, {Lupton}, {Magnier}, {Malanushenko},
  {Malanushenko}, {Mand elbaum}, {Margon}, {Marriner}, {Mart{\'\i}nez-Delgado},
  {Matsubara}, {McGehee}, {McKay}, {Meiksin}, {Morrison}, {Mullally}, {Munn},
  {Murphy}, {Nash}, {Nebot}, {Neilsen}, {Newberg}, {Newman}, {Nichol},
  {Nicinski}, {Nieto-Santisteban}, {Nitta}, {Okamura}, {Oravetz}, {Ostriker},
  {Owen}, {Padmanabhan}, {Pan}, {Park}, {Pauls}, {Peoples}, {Percival}, {Pier},
  {Pope}, {Pourbaix}, {Price}, {Purger}, {Quinn}, {Raddick}, {Re Fiorentin},
  {Richards}, {Richmond}, {Riess}, {Rix}, {Rockosi}, {Sako}, {Schlegel},
  {Schneider}, {Scholz}, {Schreiber}, {Schwope}, {Seljak}, {Sesar}, {Sheldon},
  {Shimasaku}, {Sibley}, {Simmons}, {Sivarani}, {Allyn Smith}, {Smith},
  {Smol{\v{c}}i{\'c}}, {Snedden}, {Stebbins}, {Steinmetz}, {Stoughton},
  {Strauss}, {SubbaRao}, {Suto}, {Szalay}, {Szapudi}, {Szkody}, {Tanaka},
  {Tegmark}, {Teodoro}, {Thakar}, {Tremonti}, {Tucker}, {Uomoto}, {Vanden
  Berk}, {Vandenberg}, {Vidrih}, {Vogeley}, {Voges}, {Vogt}, {Wadadekar},
  {Watters}, {Weinberg}, {West}, {White}, {Wilhite}, {Wonders}, {Yanny},
  {Yocum}, {York}, {Zehavi}, {Zibetti}, \& {Zucker}}]{2009ApJS..182..543A}
{Abazajian}, K.~N., {Adelman-McCarthy}, J.~K., {Ag{\"u}eros}, M.~A., {et~al.}
  2009, \apjs, 182, 543

\bibitem[{{Ali} {et~al.}(1991){Ali}, {Blum}, {Bumgardner}, {Cranmer},
  {Ferland}, {Haefner}, \& {Tiede}}]{ali1991}
{Ali}, B., {Blum}, R.~D., {Bumgardner}, T.~E., {et~al.} 1991, \pasp, 103, 1182

\bibitem[{{Alloin} {et~al.}(1979){Alloin}, {Collin-Souffrin}, {Joly}, \&
  {Vigroux}}]{alloin1979}
{Alloin}, D., {Collin-Souffrin}, S., {Joly}, M., \& {Vigroux}, L. 1979, \aap,
  78, 200

\bibitem[{{Amayo} {et~al.}(2021){Amayo}, {Delgado-Inglada}, \&
  {Stasi{\'n}ska}}]{amayo2021}
{Amayo}, A., {Delgado-Inglada}, G., \& {Stasi{\'n}ska}, G. 2021, \mnras, 505,
  2361

\bibitem[{{Anders} \& {Grevesse}(1989)}]{anders1989}
{Anders}, E. \& {Grevesse}, N. 1989, \gca, 53, 197

\bibitem[{{Badnell} {et~al.}(2015){Badnell}, {Ferland}, {Gorczyca},
  {Nikoli{\'c}}, \& {Wagle}}]{badnell2015}
{Badnell}, N.~R., {Ferland}, G.~J., {Gorczyca}, T.~W., {Nikoli{\'c}}, D., \&
  {Wagle}, G.~A. 2015, \apj, 804, 100

\bibitem[{{Baldwin} {et~al.}(1991){Baldwin}, {Ferland}, {Martin}, {Corbin},
  {Cota}, {Peterson}, \& {Slettebak}}]{baldwin1991}
{Baldwin}, J.~A., {Ferland}, G.~J., {Martin}, P.~G., {et~al.} 1991, \apj, 374,
  580

\bibitem[{{Baldwin} {et~al.}(1981){Baldwin}, {Phillips}, \&
  {Terlevich}}]{1981PASP...93....5B}
{Baldwin}, J.~A., {Phillips}, M.~M., \& {Terlevich}, R. 1981, \pasp, 93, 5

\bibitem[{{Barrera-Ballesteros} {et~al.}(2017){Barrera-Ballesteros},
  {S{\'a}nchez}, {Heckman}, {Blanc}, \& {MaNGA Team}}]{barrera-ballesteros2017}
{Barrera-Ballesteros}, J.~K., {S{\'a}nchez}, S.~F., {Heckman}, T., {Blanc},
  G.~A., \& {MaNGA Team}. 2017, \apj, 844, 80

\bibitem[{{Bastian} {et~al.}(2008){Bastian}, {Gieles}, {Goodwin}, {Trancho},
  {Smith}, {Konstantopoulos}, \& {Efremov}}]{bastian2008}
{Bastian}, N., {Gieles}, M., {Goodwin}, S.~P., {et~al.} 2008, \mnras, 389, 223

\bibitem[{{Belfiore} {et~al.}(2021){Belfiore}, {Santoro}, {Groves},
  {Schinnerer}, {Kreckel}, {Glover}, {Klessen}, {Emsellem}, {Blanc}, {Congiu},
  {Barnes}, {Boquien}, {Chevance}, {Dale}, {Kruijssen}, {Leroy}, {Pan},
  {Pessa}, {Schruba}, \& {Williams}}]{belfiore2021}
{Belfiore}, F., {Santoro}, F., {Groves}, B., {et~al.} 2021, arXiv e-prints,
  arXiv:2111.14876

\bibitem[{{Belfiore} {et~al.}(2019){Belfiore}, {Westfall}, {Schaefer},
  {Cappellari}, {Ji}, {Bershady}, {Tremonti}, {Law}, {Yan}, {Bundy}, {Shetty},
  {Drory}, {Thomas}, {Emsellem}, \& {S{\'a}nchez}}]{belfiore2019}
{Belfiore}, F., {Westfall}, K.~B., {Schaefer}, A., {et~al.} 2019, \aj, 158, 160

\bibitem[{{Blanc} {et~al.}(2015){Blanc}, {Kewley}, {Vogt}, \&
  {Dopita}}]{blanc2015}
{Blanc}, G.~A., {Kewley}, L., {Vogt}, F. P.~A., \& {Dopita}, M.~A. 2015, \apj,
  798, 99

\bibitem[{{Blanton} {et~al.}(2017){Blanton}, {Bershady}, {Abolfathi},
  {Albareti}, {Allende Prieto}, {Almeida}, {Alonso-Garc{\'{\i}}a}, {Anders},
  {Anderson}, {Andrews}, \& et~al.}]{blanton2017}
{Blanton}, M.~R., {Bershady}, M.~A., {Abolfathi}, B., {et~al.} 2017, \aj, 154,
  28

\bibitem[{{Bouret} {et~al.}(2005){Bouret}, {Lanz}, \& {Hillier}}]{bouret2005}
{Bouret}, J.~C., {Lanz}, T., \& {Hillier}, D.~J. 2005, \aap, 438, 301

\bibitem[{{Bundy} {et~al.}(2015){Bundy}, {Bershady}, {Law}, {Yan}, {Drory},
  {MacDonald}, {Wake}, {Cherinka}, {S{\'a}nchez-Gallego}, {Weijmans}, {Thomas},
  {Tremonti}, {Masters}, {Coccato}, {Diamond-Stanic}, {Arag{\'o}n-Salamanca},
  {Avila-Reese}, {Badenes}, {Falc{\'o}n-Barroso}, {Belfiore}, {Bizyaev},
  {Blanc}, {Bland-Hawthorn}, {Blanton}, {Brownstein}, {Byler}, {Cappellari},
  {Conroy}, {Dutton}, {Emsellem}, {Etherington}, {Frinchaboy}, {Fu}, {Gunn},
  {Harding}, {Johnston}, {Kauffmann}, {Kinemuchi}, {Klaene}, {Knapen},
  {Leauthaud}, {Li}, {Lin}, {Maiolino}, {Malanushenko}, {Malanushenko}, {Mao},
  {Maraston}, {McDermid}, {Merrifield}, {Nichol}, {Oravetz}, {Pan}, {Parejko},
  {Sanchez}, {Schlegel}, {Simmons}, {Steele}, {Steinmetz}, {Thanjavur},
  {Thompson}, {Tinker}, {van den Bosch}, {Westfall}, {Wilkinson}, {Wright},
  {Xiao}, \& {Zhang}}]{bundy2015}
{Bundy}, K., {Bershady}, M.~A., {Law}, D.~R., {et~al.} 2015, \apj, 798, 7

\bibitem[{{Byler} {et~al.}(2017){Byler}, {Dalcanton}, {Conroy}, \&
  {Johnson}}]{byler2017}
{Byler}, N., {Dalcanton}, J.~J., {Conroy}, C., \& {Johnson}, B.~D. 2017, \apj,
  840, 44

\bibitem[{{Cappellari}(2017)}]{cappellari2017}
{Cappellari}, M. 2017, \mnras, 466, 798

\bibitem[{{Cappellari} \& {Emsellem}(2004)}]{cappellari2004}
{Cappellari}, M. \& {Emsellem}, E. 2004, \pasp, 116, 138

\bibitem[{{Cardelli} {et~al.}(1989){Cardelli}, {Clayton}, \&
  {Mathis}}]{cardelli1989}
{Cardelli}, J.~A., {Clayton}, G.~C., \& {Mathis}, J.~S. 1989, \apj, 345, 245

\bibitem[{{Charlot} \& {Longhetti}(2001)}]{charlot2001}
{Charlot}, S. \& {Longhetti}, M. 2001, \mnras, 323, 887

\bibitem[{{Cherinka} {et~al.}(2019){Cherinka}, {Andrews},
  {S{\'a}nchez-Gallego}, {Brownstein}, {Argudo-Fern{\'a}ndez}, {Blanton},
  {Bundy}, {Jones}, {Masters}, {Law}, {Rowlands}, {Weijmans}, {Westfall}, \&
  {Yan}}]{cherinka2019}
{Cherinka}, B., {Andrews}, B.~H., {S{\'a}nchez-Gallego}, J., {et~al.} 2019,
  \aj, 158, 74

\bibitem[{{Choi} {et~al.}(2016){Choi}, {Dotter}, {Conroy}, {Cantiello},
  {Paxton}, \& {Johnson}}]{choi2016}
{Choi}, J., {Dotter}, A., {Conroy}, C., {et~al.} 2016, \apj, 823, 102

\bibitem[{{Cid Fernandes} {et~al.}(2011){Cid Fernandes}, {Stasi{\'n}ska},
  {Mateus}, \& {Vale Asari}}]{2011MNRAS.413.1687C}
{Cid Fernandes}, R., {Stasi{\'n}ska}, G., {Mateus}, A., \& {Vale Asari}, N.
  2011, \mnras, 413, 1687

\bibitem[{{Conroy} {et~al.}(2009){Conroy}, {Gunn}, \& {White}}]{conroy2009}
{Conroy}, C., {Gunn}, J.~E., \& {White}, M. 2009, \apj, 699, 486

\bibitem[{{Cowie} \& {Songaila}(1986)}]{cowie1986}
{Cowie}, L.~L. \& {Songaila}, A. 1986, \araa, 24, 499

\bibitem[{{D'Agostino} {et~al.}(2019){D'Agostino}, {Kewley}, {Groves}, {Byler},
  {Sutherland}, {Nicholls}, {Leitherer}, \& {Stanway}}]{dagostino2019b}
{D'Agostino}, J.~J., {Kewley}, L.~J., {Groves}, B., {et~al.} 2019, \apj, 878, 2

\bibitem[{{Dale} {et~al.}(2014){Dale}, {Ngoumou}, {Ercolano}, \&
  {Bonnell}}]{dale2014}
{Dale}, J.~E., {Ngoumou}, J., {Ercolano}, B., \& {Bonnell}, I.~A. 2014, \mnras,
  442, 694

\bibitem[{{Diaz} {et~al.}(1991){Diaz}, {Terlevich}, {Vilchez}, {Pagel}, \&
  {Edmunds}}]{diaz1991}
{Diaz}, A.~I., {Terlevich}, E., {Vilchez}, J.~M., {Pagel}, B. E.~J., \&
  {Edmunds}, M.~G. 1991, \mnras, 253, 245

\bibitem[{{Dopita} \& {Evans}(1986)}]{1986ApJ...307..431D}
{Dopita}, M.~A. \& {Evans}, I.~N. 1986, \apj, 307, 431

\bibitem[{{Dopita} {et~al.}(2006){Dopita}, {Fischera}, {Sutherland}, {Kewley},
  {Tuffs}, {Popescu}, {van Breugel}, {Groves}, \&
  {Leitherer}}]{2006ApJ...647..244D}
{Dopita}, M.~A., {Fischera}, J., {Sutherland}, R.~S., {et~al.} 2006, \apj, 647,
  244

\bibitem[{{Dopita} {et~al.}(2005){Dopita}, {Groves}, {Fischera}, {Sutherland},
  {Tuffs}, {Popescu}, {Kewley}, {Reuland}, \& {Leitherer}}]{dopita2005}
{Dopita}, M.~A., {Groves}, B.~A., {Fischera}, J., {et~al.} 2005, \apj, 619, 755

\bibitem[{{Dopita} {et~al.}(2000){Dopita}, {Kewley}, {Heisler}, \&
  {Sutherland}}]{dopita2000}
{Dopita}, M.~A., {Kewley}, L.~J., {Heisler}, C.~A., \& {Sutherland}, R.~S.
  2000, \apj, 542, 224

\bibitem[{{Dopita} {et~al.}(2014){Dopita}, {Rich}, {Vogt}, {Kewley}, {Ho},
  {Basurah}, {Ali}, \& {Amer}}]{dopita2014}
{Dopita}, M.~A., {Rich}, J., {Vogt}, F. P.~A., {et~al.} 2014, \apss, 350, 741

\bibitem[{{Dopita} {et~al.}(2013){Dopita}, {Sutherland}, {Nicholls}, {Kewley},
  \& {Vogt}}]{dopita2013}
{Dopita}, M.~A., {Sutherland}, R.~S., {Nicholls}, D.~C., {Kewley}, L.~J., \&
  {Vogt}, F.~P.~A. 2013, \apjs, 208, 10

\bibitem[{{Dors} {et~al.}(2011){Dors}, {Krabbe}, {H{\"a}gele}, \&
  {P{\'e}rez-Montero}}]{2011MNRAS.415.3616D}
{Dors}, O.~L., J., {Krabbe}, A., {H{\"a}gele}, G.~F., \& {P{\'e}rez-Montero},
  E. 2011, \mnras, 415, 3616

\bibitem[{{Dotter}(2016)}]{dotter2016}
{Dotter}, A. 2016, \apjs, 222, 8

\bibitem[{{Draine}(2011)}]{draine2011}
{Draine}, B.~T. 2011, {Physics of the Interstellar and Intergalactic Medium}
  (Princeton, NJ: Princeton Univ. Press)

\bibitem[{{Drory} {et~al.}(2015){Drory}, {MacDonald}, {Bershady}, {Bundy},
  {Gunn}, {Law}, {Smith}, {Stoll}, {Tremonti}, {Wake}, {Yan}, {Weijmans},
  {Byler}, {Cherinka}, {Cope}, {Eigenbrot}, {Harding}, {Holder}, {Huehnerhoff},
  {Jaehnig}, {Jansen}, {Klaene}, {Paat}, {Percival}, \& {Sayres}}]{drory2015}
{Drory}, N., {MacDonald}, N., {Bershady}, M.~A., {et~al.} 2015, \aj, 149, 77

\bibitem[{{Edmunds} \& {Pagel}(1984)}]{edmunds1984}
{Edmunds}, M.~G. \& {Pagel}, B.~E.~J. 1984, \mnras, 211, 507

\bibitem[{{Ellison} {et~al.}(2008){Ellison}, {Patton}, {Simard}, \&
  {McConnachie}}]{ellison2008}
{Ellison}, S.~L., {Patton}, D.~R., {Simard}, L., \& {McConnachie}, A.~W. 2008,
  \apjl, 672, L107

\bibitem[{{Erb} {et~al.}(2006){Erb}, {Shapley}, {Pettini}, {Steidel}, {Reddy},
  \& {Adelberger}}]{erb2006}
{Erb}, D.~K., {Shapley}, A.~E., {Pettini}, M., {et~al.} 2006, \apj, 644, 813

\bibitem[{{Espinosa-Ponce} {et~al.}(2020){Espinosa-Ponce}, {S{\'a}nchez},
  {Morisset}, {Barrera-Ballesteros}, {Galbany}, {Garc{\'\i}a-Benito},
  {Lacerda}, \& {Mast}}]{espinosa-ponce2020}
{Espinosa-Ponce}, C., {S{\'a}nchez}, S.~F., {Morisset}, C., {et~al.} 2020,
  \mnras, 494, 1622

\bibitem[{{Evans} {et~al.}(2004){Evans}, {Crowther}, {Fullerton}, \&
  {Hillier}}]{evans2004}
{Evans}, C.~J., {Crowther}, P.~A., {Fullerton}, A.~W., \& {Hillier}, D.~J.
  2004, \apj, 610, 1021

\bibitem[{{Ferland} {et~al.}(2017){Ferland}, {Chatzikos}, {Guzm{\'a}n},
  {Lykins}, {van Hoof}, {Williams}, {Abel}, {Badnell}, {Keenan}, {Porter}, \&
  {Stancil}}]{ferland2017}
{Ferland}, G.~J., {Chatzikos}, M., {Guzm{\'a}n}, F., {et~al.} 2017, \rmxaa, 53,
  385

\bibitem[{{Fitzpatrick}(1999)}]{fitzpatrick1999}
{Fitzpatrick}, E.~L. 1999, \pasp, 111, 63

\bibitem[{{Fullerton} {et~al.}(2006){Fullerton}, {Massa}, \&
  {Prinja}}]{fullerton2006}
{Fullerton}, A.~W., {Massa}, D.~L., \& {Prinja}, R.~K. 2006, \apj, 637, 1025

\bibitem[{{Geen} {et~al.}(2020){Geen}, {Pellegrini}, {Bieri}, \&
  {Klessen}}]{geen2020}
{Geen}, S., {Pellegrini}, E., {Bieri}, R., \& {Klessen}, R. 2020, \mnras, 492,
  915

\bibitem[{{Grevesse} {et~al.}(2010){Grevesse}, {Asplund}, {Sauval}, \&
  {Scott}}]{grevesse2010}
{Grevesse}, N., {Asplund}, M., {Sauval}, A.~J., \& {Scott}, P. 2010, \apss,
  328, 179

\bibitem[{{G{\"u}del} {et~al.}(2008){G{\"u}del}, {Briggs}, {Montmerle},
  {Audard}, {Rebull}, \& {Skinner}}]{gudel2008}
{G{\"u}del}, M., {Briggs}, K.~R., {Montmerle}, T., {et~al.} 2008, Science, 319,
  309

\bibitem[{{Gunasekera} {et~al.}(2022){Gunasekera}, {Ji}, {Chatzikos}, {Yan}, \&
  {Ferland}}]{gunasekera2022}
{Gunasekera}, C., {Ji}, X., {Chatzikos}, M., {Yan}, R., \& {Ferland}, G. 2022,
  arXiv e-prints, arXiv:2201.02882

\bibitem[{{Gunn} {et~al.}(2006){Gunn}, {Siegmund}, {Mannery}, {Owen}, {Hull},
  {Leger}, {Carey}, {Knapp}, {York}, {Boroski}, {Kent}, {Lupton}, {Rockosi},
  {Evans}, {Waddell}, {Anderson}, {Annis}, {Barentine}, {Bartoszek}, {Bastian},
  {Bracker}, {Brewington}, {Briegel}, {Brinkmann}, {Brown}, {Carr},
  {Czarapata}, {Drennan}, {Dombeck}, {Federwitz}, {Gillespie}, {Gonzales},
  {Hansen}, {Harvanek}, {Hayes}, {Jordan}, {Kinney}, {Klaene}, {Kleinman},
  {Kron}, {Kresinski}, {Lee}, {Limmongkol}, {Lindenmeyer}, {Long}, {Loomis},
  {McGehee}, {Mantsch}, {Neilsen}, {Neswold}, {Newman}, {Nitta}, {Peoples},
  {Pier}, {Prieto}, {Prosapio}, {Rivetta}, {Schneider}, {Snedden}, \&
  {Wang}}]{gunn2006}
{Gunn}, J.~E., {Siegmund}, W.~A., {Mannery}, E.~J., {et~al.} 2006, \aj, 131,
  2332

\bibitem[{{Harper-Clark} \& {Murray}(2009)}]{harper2009}
{Harper-Clark}, E. \& {Murray}, N. 2009, \apj, 693, 1696

\bibitem[{{Haworth} {et~al.}(2015){Haworth}, {Harries}, {Acreman}, \&
  {Bisbas}}]{haworth2015}
{Haworth}, T.~J., {Harries}, T.~J., {Acreman}, D.~M., \& {Bisbas}, T.~G. 2015,
  \mnras, 453, 2277

\bibitem[{{Hillier} \& {Miller}(1998)}]{hillier1998}
{Hillier}, D.~J. \& {Miller}, D.~L. 1998, \apj, 496, 407

\bibitem[{{Hogg} {et~al.}(2010){Hogg}, {Bovy}, \& {Lang}}]{hogg2010}
{Hogg}, D.~W., {Bovy}, J., \& {Lang}, D. 2010, arXiv e-prints, arXiv:1008.4686

\bibitem[{{Jamet} {et~al.}(2005){Jamet}, {Stasi{\'n}ska}, {P{\'e}rez},
  {Gonz{\'a}lez Delgado}, \& {V{\'\i}lchez}}]{jamet2005}
{Jamet}, L., {Stasi{\'n}ska}, G., {P{\'e}rez}, E., {Gonz{\'a}lez Delgado},
  R.~M., \& {V{\'\i}lchez}, J.~M. 2005, \aap, 444, 723

\bibitem[{{Jenkins}(1987)}]{jenkins1987}
{Jenkins}, E.~B. 1987, in Astrophysics and Space Science Library, Vol. 134,
  Interstellar Processes, ed. D.~J. {Hollenbach} \& H.~A. {Thronson}, Jr.,
  533--559

\bibitem[{{Jenkins}(2009)}]{jenkins2009}
{Jenkins}, E.~B. 2009, \apj, 700, 1299

\bibitem[{{Ji} \& {Yan}(2020)}]{ji2020b}
{Ji}, X. \& {Yan}, R. 2020, \mnras, 499, 5749

\bibitem[{{Ji} {et~al.}(2020){Ji}, {Yan}, {Riffel}, {Drory}, \&
  {Zhang}}]{ji2020a}
{Ji}, X., {Yan}, R., {Riffel}, R., {Drory}, N., \& {Zhang}, K. 2020, \mnras

\bibitem[{{Joung} {et~al.}(2009){Joung}, {Mac Low}, \& {Bryan}}]{joung2009}
{Joung}, M.~R., {Mac Low}, M.-M., \& {Bryan}, G.~L. 2009, \apj, 704, 137

\bibitem[{{Kennicutt}(1998)}]{kennicutt1998}
{Kennicutt}, Robert~C., J. 1998, \apj, 498, 541

\bibitem[{{Kennicutt} \& {Evans}(2012)}]{kennicutt2012}
{Kennicutt}, R.~C. \& {Evans}, N.~J. 2012, \araa, 50, 531

\bibitem[{{Kewley} \& {Dopita}(2002)}]{kewley2002}
{Kewley}, L.~J. \& {Dopita}, M.~A. 2002, \apjs, 142, 35

\bibitem[{{Kewley} {et~al.}(2013){Kewley}, {Dopita}, {Leitherer}, {Dav{\'e}},
  {Yuan}, {Allen}, {Groves}, \& {Sutherland}}]{kewley2013}
{Kewley}, L.~J., {Dopita}, M.~A., {Leitherer}, C., {et~al.} 2013, \apj, 774,
  100

\bibitem[{{Kewley} {et~al.}(2001){Kewley}, {Dopita}, {Sutherland}, {Heisler},
  \& {Trevena}}]{2001ApJ...556..121K}
{Kewley}, L.~J., {Dopita}, M.~A., {Sutherland}, R.~S., {Heisler}, C.~A., \&
  {Trevena}, J. 2001, \apj, 556, 121

\bibitem[{{Kewley} {et~al.}(2019){Kewley}, {Nicholls}, \&
  {Sutherland}}]{2019ARA&A..57..511K}
{Kewley}, L.~J., {Nicholls}, D.~C., \& {Sutherland}, R.~S. 2019, \araa, 57, 511

\bibitem[{{Kimura} {et~al.}(2003){Kimura}, {Mann}, \&
  {Jessberger}}]{kimura2003}
{Kimura}, H., {Mann}, I., \& {Jessberger}, E.~K. 2003, \apj, 582, 846

\bibitem[{{Kobulnicky} \& {Kewley}(2004)}]{kobulnicky2004}
{Kobulnicky}, H.~A. \& {Kewley}, L.~J. 2004, \apj, 617, 240

\bibitem[{{Kollmeier} {et~al.}(2017){Kollmeier}, {Zasowski}, {Rix}, {Johns},
  {Anderson}, {Drory}, {Johnson}, {Pogge}, {Bird}, {Blanc}, {Brownstein},
  {Crane}, {De Lee}, {Klaene}, {Kreckel}, {MacDonald}, {Merloni}, {Ness},
  {O'Brien}, {Sanchez-Gallego}, {Sayres}, {Shen}, {Thakar}, {Tkachenko},
  {Aerts}, {Blanton}, {Eisenstein}, {Holtzman}, {Maoz}, {Nandra}, {Rockosi},
  {Weinberg}, {Bovy}, {Casey}, {Chaname}, {Clerc}, {Conroy}, {Eracleous},
  {G{\"a}nsicke}, {Hekker}, {Horne}, {Kauffmann}, {McQuinn}, {Pellegrini},
  {Schinnerer}, {Schlafly}, {Schwope}, {Seibert}, {Teske}, \& {van
  Saders}}]{kollmeier2017}
{Kollmeier}, J.~A., {Zasowski}, G., {Rix}, H.-W., {et~al.} 2017, arXiv
  e-prints, arXiv:1711.03234

\bibitem[{{Kreckel} {et~al.}(2019){Kreckel}, {Ho}, {Blanc}, {Groves},
  {Santoro}, {Schinnerer}, {Bigiel}, {Chevance}, {Congiu}, {Emsellem}, {Faesi},
  {Glover}, {Grasha}, {Kruijssen}, {Lang}, {Leroy}, {Meidt}, {McElroy}, {Pety},
  {Rosolowsky}, {Saito}, {Sandstrom}, {Sanchez-Blazquez}, \&
  {Schruba}}]{kreckel2019}
{Kreckel}, K., {Ho}, I.~T., {Blanc}, G.~A., {et~al.} 2019, \apj, 887, 80

\bibitem[{{Kriek} {et~al.}(2015){Kriek}, {Shapley}, {Reddy}, {Siana}, {Coil},
  {Mobasher}, {Freeman}, {de Groot}, {Price}, {Sanders}, {Shivaei}, {Brammer},
  {Momcheva}, {Skelton}, {van Dokkum}, {Whitaker}, {Aird}, {Azadi}, {Kassis},
  {Bullock}, {Conroy}, {Dav{\'e}}, {Kere{\v{s}}}, \& {Krumholz}}]{kriek2015}
{Kriek}, M., {Shapley}, A.~E., {Reddy}, N.~A., {et~al.} 2015, \apjs, 218, 15

\bibitem[{{Kroupa}(2001)}]{kroupa2001}
{Kroupa}, P. 2001, \mnras, 322, 231

\bibitem[{{Law} {et~al.}(2016){Law}, {Cherinka}, {Yan}, {Andrews}, {Bershady},
  {Bizyaev}, {Blanc}, {Blanton}, {Bolton}, {Brownstein}, {Bundy}, {Chen},
  {Drory}, {D'Souza}, {Fu}, {Jones}, {Kauffmann}, {MacDonald}, {Masters},
  {Newman}, {Parejko}, {S{\'a}nchez-Gallego}, {S{\'a}nchez}, {Schlegel},
  {Thomas}, {Wake}, {Weijmans}, {Westfall}, \& {Zhang}}]{law2016}
{Law}, D.~R., {Cherinka}, B., {Yan}, R., {et~al.} 2016, \aj, 152, 83

\bibitem[{{Law} {et~al.}(2021{\natexlab{a}}){Law}, {Ji}, {Belfiore},
  {Bershady}, {Cappellari}, {Westfall}, {Yan}, {Bizyaev}, {Brownstein},
  {Drory}, \& {Andrews}}]{law2021b}
{Law}, D.~R., {Ji}, X., {Belfiore}, F., {et~al.} 2021{\natexlab{a}}, \apj, 915,
  35

\bibitem[{{Law} {et~al.}(2021{\natexlab{b}}){Law}, {Westfall}, {Bershady},
  {Cappellari}, {Yan}, {Belfiore}, {Bizyaev}, {Brownstein}, {Chen}, {Cherinka},
  {Drory}, {Lazarz}, \& {Shetty}}]{law2021a}
{Law}, D.~R., {Westfall}, K.~B., {Bershady}, M.~A., {et~al.}
  2021{\natexlab{b}}, \aj, 161, 52

\bibitem[{{Law} {et~al.}(2015){Law}, {Yan}, {Bershady}, {Bundy}, {Cherinka},
  {Drory}, {MacDonald}, {S{\'a}nchez-Gallego}, {Wake}, {Weijmans}, {Blanton},
  {Klaene}, {Moran}, {Sanchez}, \& {Zhang}}]{law2015}
{Law}, D.~R., {Yan}, R., {Bershady}, M.~A., {et~al.} 2015, \aj, 150, 19

\bibitem[{{Leitherer} {et~al.}(1999){Leitherer}, {Schaerer}, {Goldader},
  {Delgado}, {Robert}, {Kune}, {de Mello}, {Devost}, \&
  {Heckman}}]{leitherer1999}
{Leitherer}, C., {Schaerer}, D., {Goldader}, J.~D., {et~al.} 1999, \apjs, 123,
  3

\bibitem[{{Levesque} {et~al.}(2010){Levesque}, {Kewley}, \&
  {Larson}}]{levesque2010}
{Levesque}, E.~M., {Kewley}, L.~J., \& {Larson}, K.~L. 2010, \aj, 139, 712

\bibitem[{{Mac Low} \& {McCray}(1988)}]{maclow1988}
{Mac Low}, M.-M. \& {McCray}, R. 1988, \apj, 324, 776

\bibitem[{{Maier} {et~al.}(2006){Maier}, {Lilly}, {Carollo}, {Meisenheimer},
  {Hippelein}, \& {Stockton}}]{2006ApJ...639..858M}
{Maier}, C., {Lilly}, S.~J., {Carollo}, C.~M., {et~al.} 2006, \apj, 639, 858

\bibitem[{{Mannucci} {et~al.}(2021){Mannucci}, {Belfiore}, {Curti}, {Cresci},
  {Maiolino}, {Marasco}, {Marconi}, {Mingozzi}, {Tozzi}, \&
  {Amiri}}]{mannucci2021}
{Mannucci}, F., {Belfiore}, F., {Curti}, M., {et~al.} 2021, \mnras, 508, 1582

\bibitem[{{Mannucci} {et~al.}(2010){Mannucci}, {Cresci}, {Maiolino}, {Marconi},
  \& {Gnerucci}}]{mannucci2010}
{Mannucci}, F., {Cresci}, G., {Maiolino}, R., {Marconi}, A., \& {Gnerucci}, A.
  2010, \mnras, 408, 2115

\bibitem[{{Marino} {et~al.}(2013){Marino}, {Rosales-Ortega}, {S{\'a}nchez},
  {Gil de Paz}, {V{\'\i}lchez}, {Miralles-Caballero}, {Kehrig},
  {P{\'e}rez-Montero}, {Stanishev}, {Iglesias-P{\'a}ramo}, {D{\'\i}az},
  {Castillo-Morales}, {Kennicutt}, {L{\'o}pez-S{\'a}nchez}, {Galbany},
  {Garc{\'\i}a-Benito}, {Mast}, {Mendez-Abreu}, {Monreal-Ibero}, {Husemann},
  {Walcher}, {Garc{\'\i}a-Lorenzo}, {Masegosa}, {Del Olmo Orozco},
  {Mour{\~a}o}, {Ziegler}, {Moll{\'a}}, {Papaderos},
  {S{\'a}nchez-Bl{\'a}zquez}, {Gonz{\'a}lez Delgado}, {Falc{\'o}n-Barroso},
  {Roth}, {van de Ven}, \& {Califa Team}}]{marino2013}
{Marino}, R.~A., {Rosales-Ortega}, F.~F., {S{\'a}nchez}, S.~F., {et~al.} 2013,
  \aap, 559, A114

\bibitem[{{Mingozzi} {et~al.}(2020){Mingozzi}, {Belfiore}, {Cresci}, {Bundy},
  {Bershady}, {Bizyaev}, {Blanc}, {Boquien}, {Drory}, {Fu}, {Maiolino},
  {Riffel}, {Schaefer}, {Storchi-Bergmann}, {Telles}, {Tremonti}, {Zakamska},
  \& {Zhang}}]{mingozzi2020}
{Mingozzi}, M., {Belfiore}, F., {Cresci}, G., {et~al.} 2020, \aap, 636, A42

\bibitem[{{Morisset} {et~al.}(2016){Morisset}, {Delgado-Inglada},
  {S{\'a}nchez}, {Galbany}, {Garc{\'\i}a-Benito}, {Husemann}, {Marino}, {Mast},
  \& {Roth}}]{2016A&A...594A..37M}
{Morisset}, C., {Delgado-Inglada}, G., {S{\'a}nchez}, S.~F., {et~al.} 2016,
  \aap, 594, A37

\bibitem[{{Nagao} {et~al.}(2006){Nagao}, {Maiolino}, \&
  {Marconi}}]{2006A&A...459...85N}
{Nagao}, T., {Maiolino}, R., \& {Marconi}, A. 2006, \aap, 459, 85

\bibitem[{{Naz{\'e}} {et~al.}(2001){Naz{\'e}}, {Chu}, {Points}, {Danforth},
  {Rosado}, \& {Chen}}]{naze2001}
{Naz{\'e}}, Y., {Chu}, Y.-H., {Points}, S.~D., {et~al.} 2001, \aj, 122, 921

\bibitem[{{Oey} \& {Clarke}(1997)}]{oey1997}
{Oey}, M.~S. \& {Clarke}, C.~J. 1997, \mnras, 289, 570

\bibitem[{{Pagel} {et~al.}(1979){Pagel}, {Edmunds}, {Blackwell}, {Chun}, \&
  {Smith}}]{pagel1979}
{Pagel}, B.~E.~J., {Edmunds}, M.~G., {Blackwell}, D.~E., {Chun}, M.~S., \&
  {Smith}, G. 1979, \mnras, 189, 95

\bibitem[{{Pagel} {et~al.}(1980){Pagel}, {Edmunds}, \& {Smith}}]{pagel1980}
{Pagel}, B.~E.~J., {Edmunds}, M.~G., \& {Smith}, G. 1980, \mnras, 193, 219

\bibitem[{{Pauldrach} {et~al.}(2001){Pauldrach}, {Hoffmann}, \&
  {Lennon}}]{pauldrach2001}
{Pauldrach}, A.~W.~A., {Hoffmann}, T.~L., \& {Lennon}, M. 2001, \aap, 375, 161

\bibitem[{{Pellegrini} {et~al.}(2007){Pellegrini}, {Baldwin}, {Brogan},
  {Hanson}, {Abel}, {Ferland}, {Nemala}, {Shaw}, \& {Troland}}]{pellegrini2007}
{Pellegrini}, E.~W., {Baldwin}, J.~A., {Brogan}, C.~L., {et~al.} 2007, \apj,
  658, 1119

\bibitem[{{Pellegrini} {et~al.}(2011){Pellegrini}, {Baldwin}, \&
  {Ferland}}]{pellegrini2011}
{Pellegrini}, E.~W., {Baldwin}, J.~A., \& {Ferland}, G.~J. 2011, \apj, 738, 34

\bibitem[{{Pellegrini} {et~al.}(2020){Pellegrini}, {Rahner}, {Reissl},
  {Glover}, {Klessen}, {Rousseau-Nepton}, \&
  {Herrera-Camus}}]{2020MNRAS.496..339P}
{Pellegrini}, E.~W., {Rahner}, D., {Reissl}, S., {et~al.} 2020, \mnras, 496,
  339

\bibitem[{{P{\'e}rez-Montero}(2014)}]{perezmontero2014}
{P{\'e}rez-Montero}, E. 2014, \mnras, 441, 2663

\bibitem[{{Pettini} \& {Pagel}(2004)}]{pettini2004}
{Pettini}, M. \& {Pagel}, B. E.~J. 2004, \mnras, 348, L59

\bibitem[{{Pilyugin} \& {Thuan}(2005)}]{pilyugin2005}
{Pilyugin}, L.~S. \& {Thuan}, T.~X. 2005, \apj, 631, 231

\bibitem[{{Poetrodjojo} {et~al.}(2018){Poetrodjojo}, {Groves}, {Kewley},
  {Medling}, {Sweet}, {van de Sande}, {Sanchez}, {Bland-Hawthorn}, {Brough},
  {Bryant}, {Cortese}, {Croom}, {L{\'o}pez-S{\'a}nchez}, {Richards}, {Zafar},
  {Lawrence}, {Lorente}, {Owers}, \& {Scott}}]{poetrodjojo2018}
{Poetrodjojo}, H., {Groves}, B., {Kewley}, L.~J., {et~al.} 2018, \mnras, 479,
  5235

\bibitem[{{Powell} {et~al.}(2013){Powell}, {Bournaud}, {Chapon}, \&
  {Teyssier}}]{powell2013}
{Powell}, L.~C., {Bournaud}, F., {Chapon}, D., \& {Teyssier}, R. 2013, \mnras,
  434, 1028

\bibitem[{{Puls} {et~al.}(2006){Puls}, {Markova}, {Scuderi}, {Stanghellini},
  {Taranova}, {Burnley}, \& {Howarth}}]{puls2006}
{Puls}, J., {Markova}, N., {Scuderi}, S., {et~al.} 2006, \aap, 454, 625

\bibitem[{{Puls} {et~al.}(2008){Puls}, {Vink}, \& {Najarro}}]{puls2008}
{Puls}, J., {Vink}, J.~S., \& {Najarro}, F. 2008, \aapr, 16, 209

\bibitem[{{Salpeter}(1955)}]{salpeter1995}
{Salpeter}, E.~E. 1955, \apj, 121, 161

\bibitem[{{S{\'a}nchez} {et~al.}(2017){S{\'a}nchez}, {Barrera-Ballesteros},
  {S{\'a}nchez-Menguiano}, {Walcher}, {Marino}, {Galbany}, {Bland-Hawthorn},
  {Cano-D{\'\i}az}, {Garc{\'\i}a-Benito}, {L{\'o}pez-Cob{\'a}}, {Zibetti},
  {Vilchez}, {Igl{\'e}sias-P{\'a}ramo}, {Kehrig}, {L{\'o}pez S{\'a}nchez},
  {Duarte Puertas}, \& {Ziegler}}]{sanchez2017}
{S{\'a}nchez}, S.~F., {Barrera-Ballesteros}, J.~K., {S{\'a}nchez-Menguiano},
  L., {et~al.} 2017, \mnras, 469, 2121

\bibitem[{{S{\'a}nchez} {et~al.}(2016){S{\'a}nchez}, {P{\'e}rez},
  {S{\'a}nchez-Bl{\'a}zquez}, {Gonz{\'a}lez}, {Ros{\'a}les-Ortega},
  {Cano-D{\'\i}az}, {L{\'o}pez-Cob{\'a}}, {Marino}, {Gil de Paz}, {Moll{\'a}},
  {L{\'o}pez-S{\'a}nchez}, {Ascasibar}, \& {Barrera-Ballesteros}}]{sanchez2016}
{S{\'a}nchez}, S.~F., {P{\'e}rez}, E., {S{\'a}nchez-Bl{\'a}zquez}, P., {et~al.}
  2016, \rmxaa, 52, 21

\bibitem[{{S{\'a}nchez} {et~al.}(2014){S{\'a}nchez}, {Rosales-Ortega},
  {Iglesias-P{\'a}ramo}, {Moll{\'a}}, {Barrera-Ballesteros}, {Marino},
  {P{\'e}rez}, {S{\'a}nchez-Blazquez}, {Gonz{\'a}lez Delgado}, {Cid Fernandes},
  {de Lorenzo-C{\'a}ceres}, {Mendez-Abreu}, {Galbany}, {Falcon-Barroso},
  {Miralles-Caballero}, {Husemann}, {Garc{\'\i}a-Benito}, {Mast}, {Walcher},
  {Gil de Paz}, {Garc{\'\i}a-Lorenzo}, {Jungwiert}, {V{\'\i}lchez},
  {J{\'\i}lkov{\'a}}, {Lyubenova}, {Cortijo-Ferrero}, {D{\'\i}az}, {Wisotzki},
  {M{\'a}rquez}, {Bland-Hawthorn}, {Ellis}, {van de Ven}, {Jahnke},
  {Papaderos}, {Gomes}, {Mendoza}, \& {L{\'o}pez-S{\'a}nchez}}]{sanchez2014}
{S{\'a}nchez}, S.~F., {Rosales-Ortega}, F.~F., {Iglesias-P{\'a}ramo}, J.,
  {et~al.} 2014, \aap, 563, A49

\bibitem[{{S{\'a}nchez} {et~al.}(2013){S{\'a}nchez}, {Rosales-Ortega},
  {Jungwiert}, {Iglesias-P{\'a}ramo}, {V{\'\i}lchez}, {Marino}, {Walcher},
  {Husemann}, {Mast}, {Monreal-Ibero}, {Cid Fernandes}, {P{\'e}rez},
  {Gonz{\'a}lez Delgado}, {Garc{\'\i}a-Benito}, {Galbany}, {van de Ven},
  {Jahnke}, {Flores}, {Bland-Hawthorn}, {L{\'o}pez-S{\'a}nchez}, {Stanishev},
  {Miralles-Caballero}, {D{\'\i}az}, {S{\'a}nchez-Blazquez}, {Moll{\'a}},
  {Gallazzi}, {Papaderos}, {Gomes}, {Gruel}, {P{\'e}rez}, {Ruiz-Lara},
  {Florido}, {de Lorenzo-C{\'a}ceres}, {Mendez-Abreu}, {Kehrig}, {Roth},
  {Ziegler}, {Alves}, {Wisotzki}, {Kupko}, {Quirrenbach}, {Bomans}, \& {Califa
  Collaboration}}]{sanchez2013}
{S{\'a}nchez}, S.~F., {Rosales-Ortega}, F.~F., {Jungwiert}, B., {et~al.} 2013,
  \aap, 554, A58

\bibitem[{{Sanders} {et~al.}(2020){Sanders}, {Jones}, {Shapley}, {Reddy},
  {Kriek}, {Coil}, {Siana}, {Mobasher}, {Shivaei}, {Price}, {Freeman}, {Azadi},
  {Leung}, {Fetherolf}, {Zick}, {de Groot}, {Barro}, \&
  {Fornasini}}]{sanders2020}
{Sanders}, R.~L., {Jones}, T., {Shapley}, A.~E., {et~al.} 2020, \apjl, 888, L11

\bibitem[{{Sanders} {et~al.}(2016){Sanders}, {Shapley}, {Kriek}, {Reddy},
  {Freeman}, {Coil}, {Siana}, {Mobasher}, {Shivaei}, {Price}, \& {de
  Groot}}]{sanders2016}
{Sanders}, R.~L., {Shapley}, A.~E., {Kriek}, M., {et~al.} 2016, \apj, 816, 23

\bibitem[{{Shapley} {et~al.}(2005){Shapley}, {Coil}, {Ma}, \&
  {Bundy}}]{shapley2005}
{Shapley}, A.~E., {Coil}, A.~L., {Ma}, C.-P., \& {Bundy}, K. 2005, \apj, 635,
  1006

\bibitem[{{Shapley} {et~al.}(2015){Shapley}, {Reddy}, {Kriek}, {Freeman},
  {Sanders}, {Siana}, {Coil}, {Mobasher}, {Shivaei}, {Price}, \& {de
  Groot}}]{shapley2015}
{Shapley}, A.~E., {Reddy}, N.~A., {Kriek}, M., {et~al.} 2015, \apj, 801, 88

\bibitem[{{Smee} {et~al.}(2013){Smee}, {Gunn}, {Uomoto}, {Roe}, {Schlegel},
  {Rockosi}, {Carr}, {Leger}, {Dawson}, {Olmstead}, {Brinkmann}, {Owen},
  {Barkhouser}, {Honscheid}, {Harding}, {Long}, {Lupton}, {Loomis}, {Anderson},
  {Annis}, {Bernardi}, {Bhardwaj}, {Bizyaev}, {Bolton}, {Brewington}, {Briggs},
  {Burles}, {Burns}, {Castander}, {Connolly}, {Davenport}, {Ebelke}, {Epps},
  {Feldman}, {Friedman}, {Frieman}, {Heckman}, {Hull}, {Knapp}, {Lawrence},
  {Loveday}, {Mannery}, {Malanushenko}, {Malanushenko}, {Merrelli}, {Muna},
  {Newman}, {Nichol}, {Oravetz}, {Pan}, {Pope}, {Ricketts}, {Shelden},
  {Sandford}, {Siegmund}, {Simmons}, {Smith}, {Snedden}, {Schneider},
  {SubbaRao}, {Tremonti}, {Waddell}, \& {York}}]{smee2013}
{Smee}, S.~A., {Gunn}, J.~E., {Uomoto}, A., {et~al.} 2013, \aj, 146, 32

\bibitem[{{Snow} \& {Witt}(1996)}]{snow1996}
{Snow}, T.~P. \& {Witt}, A.~N. 1996, \apjl, 468, L65

\bibitem[{{Sofia} {et~al.}(1994){Sofia}, {Cardelli}, \& {Savage}}]{sofia1994}
{Sofia}, U.~J., {Cardelli}, J.~A., \& {Savage}, B.~D. 1994, \apj, 430, 650

\bibitem[{{Spitzer}(1978)}]{spitzer1978}
{Spitzer}, L. 1978, {Physical processes in the interstellar medium} (New York:
  Wiley)

\bibitem[{{Stasi{\'n}ska} {et~al.}(2006){Stasi{\'n}ska}, {Cid Fernandes},
  {Mateus}, {Sodr{\'e}}, \& {Asari}}]{stasinska2006}
{Stasi{\'n}ska}, G., {Cid Fernandes}, R., {Mateus}, A., {Sodr{\'e}}, L., \&
  {Asari}, N.~V. 2006, \mnras, 371, 972

\bibitem[{{Stasi{\'n}ska} {et~al.}(2015){Stasi{\'n}ska}, {Izotov}, {Morisset},
  \& {Guseva}}]{stasinska2015}
{Stasi{\'n}ska}, G., {Izotov}, Y., {Morisset}, C., \& {Guseva}, N. 2015, \aap,
  576, A83

\bibitem[{{Stasi{\'n}ska} {et~al.}(2013){Stasi{\'n}ska}, {Morisset},
  {Sim{\'o}n-D{\'\i}az}, {Bresolin}, {Schaerer}, \& {Brandl}}]{morisset2013}
{Stasi{\'n}ska}, G., {Morisset}, C., {Sim{\'o}n-D{\'\i}az}, S., {et~al.} 2013,
  \aap, 551, A82

\bibitem[{{Steidel} {et~al.}(2014){Steidel}, {Rudie}, {Strom}, {Pettini},
  {Reddy}, {Shapley}, {Trainor}, {Erb}, {Turner}, {Konidaris}, {Kulas}, {Mace},
  {Matthews}, \& {McLean}}]{steidel2014}
{Steidel}, C.~C., {Rudie}, G.~C., {Strom}, A.~L., {et~al.} 2014, \apj, 795, 165

\bibitem[{{Telford} {et~al.}(2016){Telford}, {Dalcanton}, {Skillman}, \&
  {Conroy}}]{2016ApJ...827...35T}
{Telford}, O.~G., {Dalcanton}, J.~J., {Skillman}, E.~D., \& {Conroy}, C. 2016,
  \apj, 827, 35

\bibitem[{{Thomas} {et~al.}(2019){Thomas}, {Kewley}, {Dopita}, {Groves},
  {Hopkins}, \& {Sutherland}}]{Thomas19}
{Thomas}, A.~D., {Kewley}, L.~J., {Dopita}, M.~A., {et~al.} 2019, \apj, 874,
  100

\bibitem[{{Tremonti} {et~al.}(2004){Tremonti}, {Heckman}, {Kauffmann},
  {Brinchmann}, {Charlot}, {White}, {Seibert}, {Peng}, {Schlegel}, \&
  {Uomoto}}]{2004ApJ...613..898T}
{Tremonti}, C.~A., {Heckman}, T.~M., {Kauffmann}, G., {et~al.} 2004, \apj, 613,
  898

\bibitem[{{van Zee} {et~al.}(1998){van Zee}, {Salzer}, {Haynes}, {O'Donoghue},
  \& {Balonek}}]{vanzee1998}
{van Zee}, L., {Salzer}, J.~J., {Haynes}, M.~P., {O'Donoghue}, A.~A., \&
  {Balonek}, T.~J. 1998, \aj, 116, 2805

\bibitem[{{Veilleux} \& {Osterbrock}(1987)}]{1987ApJS...63..295V}
{Veilleux}, S. \& {Osterbrock}, D.~E. 1987, \apjs, 63, 295

\bibitem[{{Wake} {et~al.}(2017){Wake}, {Bundy}, {Diamond-Stanic}, {Yan},
  {Blanton}, {Bershady}, {S{\'a}nchez-Gallego}, {Drory}, {Jones}, \&
  {Kauffmann}}]{wake2017}
{Wake}, D.~A., {Bundy}, K., {Diamond-Stanic}, A.~M., {et~al.} 2017, \aj, 154,
  86

\bibitem[{{Weaver} {et~al.}(1977){Weaver}, {McCray}, {Castor}, {Shapiro}, \&
  {Moore}}]{weaver1977}
{Weaver}, R., {McCray}, R., {Castor}, J., {Shapiro}, P., \& {Moore}, R. 1977,
  \apj, 218, 377

\bibitem[{{Wen} \& {O'dell}(1995)}]{wen1995}
{Wen}, Z. \& {O'dell}, C.~R. 1995, \apj, 438, 784

\bibitem[{{Westfall} {et~al.}(2019){Westfall}, {Cappellari}, {Bershady},
  {Bundy}, {Belfiore}, {Ji}, {Law}, {Schaefer}, {Shetty}, {Tremonti}, {Yan},
  {Andrews}, {Brownstein}, {Cherinka}, {Coccato}, {Drory}, {Maraston},
  {Parikh}, {S{\'a}nchez-Gallego}, {Thomas}, {Weijmans}, {Barrera-Ballesteros},
  {Du}, {Goddard}, {Li}, {Masters}, {Ibarra Medel}, {S{\'a}nchez}, {Yang},
  {Zheng}, \& {Zhou}}]{westfall2019}
{Westfall}, K.~B., {Cappellari}, M., {Bershady}, M.~A., {et~al.} 2019, \aj,
  158, 231

\bibitem[{{White} \& {Sofia}(2011)}]{white2011}
{White}, B. \& {Sofia}, U.~J. 2011, in American Astronomical Society Meeting
  Abstracts, Vol. 218, American Astronomical Society Meeting Abstracts \#218,
  129.23

\bibitem[{{Yan}(2011)}]{yan2011}
{Yan}, R. 2011, \aj, 142, 153

\bibitem[{{Yan}(2018)}]{yan2018}
{Yan}, R. 2018, \mnras, 481, 467

\bibitem[{{Yan} {et~al.}(2020){Yan}, {Bershady}, {Smith}, {MacDonald},
  {Bizyaev}, {Bundy}, {Chattopadhyay}, {Gunn}, {Westfall}, \& {Wolf}}]{yan2020}
{Yan}, R., {Bershady}, M.~A., {Smith}, M.~P., {et~al.} 2020, in Society of
  Photo-Optical Instrumentation Engineers (SPIE) Conference Series, Vol. 11447,
  Society of Photo-Optical Instrumentation Engineers (SPIE) Conference Series,
  114478Y

\bibitem[{{Yan} {et~al.}(2016{\natexlab{a}}){Yan}, {Bundy}, {Law}, {Bershady},
  {Andrews}, {Cherinka}, {Diamond-Stanic}, {Drory}, {MacDonald},
  {S{\'a}nchez-Gallego}, {Thomas}, {Wake}, {Weijmans}, {Westfall}, {Zhang},
  {Arag{\'o}n-Salamanca}, {Belfiore}, {Bizyaev}, {Blanc}, {Blanton},
  {Brownstein}, {Cappellari}, {D'Souza}, {Emsellem}, {Fu}, {Gaulme}, {Graham},
  {Goddard}, {Gunn}, {Harding}, {Jones}, {Kinemuchi}, {Li}, {Li}, {Maiolino},
  {Mao}, {Maraston}, {Masters}, {Merrifield}, {Oravetz}, {Pan}, {Parejko},
  {Sanchez}, {Schlegel}, {Simmons}, {Thanjavur}, {Tinker}, {Tremonti}, {van den
  Bosch}, \& {Zheng}}]{yan2016b}
{Yan}, R., {Bundy}, K., {Law}, D.~R., {et~al.} 2016{\natexlab{a}}, \aj, 152,
  197

\bibitem[{{Yan} {et~al.}(2006){Yan}, {Newman}, {Faber}, {Konidaris}, {Koo}, \&
  {Davis}}]{yan2006}
{Yan}, R., {Newman}, J.~A., {Faber}, S.~M., {et~al.} 2006, \apj, 648, 281

\bibitem[{{Yan} {et~al.}(2016{\natexlab{b}}){Yan}, {Tremonti}, {Bershady},
  {Law}, {Schlegel}, {Bundy}, {Drory}, {MacDonald}, {Bizyaev}, {Blanc},
  {Blanton}, {Cherinka}, {Eigenbrot}, {Gunn}, {Harding}, {Hogg},
  {S{\'a}nchez-Gallego}, {S{\'a}nchez}, {Wake}, {Weijmans}, {Xiao}, \&
  {Zhang}}]{yan2016a}
{Yan}, R., {Tremonti}, C., {Bershady}, M.~A., {et~al.} 2016{\natexlab{b}}, \aj,
  151, 8

\bibitem[{{Yeh} \& {Matzner}(2012)}]{yeh2012}
{Yeh}, S. C.~C. \& {Matzner}, C.~D. 2012, \apj, 757, 108

\bibitem[{{York} {et~al.}(2000){York}, {Adelman}, {Anderson}, {Anderson},
  {Annis}, {Bahcall}, {Bakken}, {Barkhouser}, {Bastian}, {Berman}, {Boroski},
  {Bracker}, {Briegel}, {Briggs}, {Brinkmann}, {Brunner}, {Burles}, {Carey},
  {Carr}, {Castander}, {Chen}, {Colestock}, {Connolly}, {Crocker}, {Csabai},
  {Czarapata}, {Davis}, {Doi}, {Dombeck}, {Eisenstein}, {Ellman}, {Elms},
  {Evans}, {Fan}, {Federwitz}, {Fiscelli}, {Friedman}, {Frieman}, {Fukugita},
  {Gillespie}, {Gunn}, {Gurbani}, {de Haas}, {Haldeman}, {Harris}, {Hayes},
  {Heckman}, {Hennessy}, {Hindsley}, {Holm}, {Holmgren}, {Huang}, {Hull},
  {Husby}, {Ichikawa}, {Ichikawa}, {Ivezi{\'c}}, {Kent}, {Kim}, {Kinney},
  {Klaene}, {Kleinman}, {Kleinman}, {Knapp}, {Korienek}, {Kron}, {Kunszt},
  {Lamb}, {Lee}, {Leger}, {Limmongkol}, {Lindenmeyer}, {Long}, {Loomis},
  {Loveday}, {Lucinio}, {Lupton}, {MacKinnon}, {Mannery}, {Mantsch}, {Margon},
  {McGehee}, {McKay}, {Meiksin}, {Merelli}, {Monet}, {Munn}, {Narayanan},
  {Nash}, {Neilsen}, {Neswold}, {Newberg}, {Nichol}, {Nicinski}, {Nonino},
  {Okada}, {Okamura}, {Ostriker}, {Owen}, {Pauls}, {Peoples}, {Peterson},
  {Petravick}, {Pier}, {Pope}, {Pordes}, {Prosapio}, {Rechenmacher}, {Quinn},
  {Richards}, {Richmond}, {Rivetta}, {Rockosi}, {Ruthmansdorfer}, {Sand ford},
  {Schlegel}, {Schneider}, {Sekiguchi}, {Sergey}, {Shimasaku}, {Siegmund},
  {Smee}, {Smith}, {Snedden}, {Stone}, {Stoughton}, {Strauss}, {Stubbs},
  {SubbaRao}, {Szalay}, {Szapudi}, {Szokoly}, {Thakar}, {Tremonti}, {Tucker},
  {Uomoto}, {Vanden Berk}, {Vogeley}, {Waddell}, {Wang}, {Watanabe},
  {Weinberg}, {Yanny}, {Yasuda}, \& {SDSS Collaboration}}]{2000AJ....120.1579Y}
{York}, D.~G., {Adelman}, J., {Anderson}, John~E., J., {et~al.} 2000, \aj, 120,
  1579

\bibitem[{{Zhang} {et~al.}(2017){Zhang}, {Yan}, {Bundy}, {Bershady}, {Haffner},
  {Walterbos}, {Maiolino}, {Tremonti}, \& {et al.}}]{zhang2017}
{Zhang}, K., {Yan}, R., {Bundy}, K., {et~al.} 2017, \mnras, 466, 3217

\bibitem[{{Zinchenko} {et~al.}(2019){Zinchenko}, {Dors}, {H{\"a}gele},
  {Cardaci}, \& {Krabbe}}]{zinchenko2019}
{Zinchenko}, I.~A., {Dors}, O.~L., {H{\"a}gele}, G.~F., {Cardaci}, M.~V., \&
  {Krabbe}, A.~C. 2019, \mnras, 483, 1901

\end{thebibliography}
%

\end{document}